\documentclass[fleqn,useAMS,usenatbib]{mnras}
\usepackage{txfonts}
\usepackage{natbib}
\usepackage{graphicx}% Include Figure files
\usepackage{dcolumn}% Align Table columns on decimal point
\usepackage{bm}% bold math
\usepackage{amssymb}	% Extra maths symbols
\usepackage{hyperref}

\pubyear{2021}
\date{\today}

\title[Acceleration and clustering of cosmic dust]{Acceleration and clustering of cosmic dust in a gravoturbulent  gas\\ I. Numerical simulation of the nearly Jeans-unstable case.}

\author[Mattsson \& Hedvall]{Lars Mattsson$^{1}$\thanks{E-mail: lars.mattsson@nordita.org}, Robert Hedvall$^{1,2}$\\
$^1$Nordita, KTH Royal Institute of Technology and Stockholm University, Hannes Alfv\'ens v\"ag 12, SE-106 91 Stockholm, Sweden\\
$^2$Department of Physics, Stockholm University, Roslagstullsbacken 23, SE-106 91 Stockholm, Sweden}

\begin{document}
\label{firstpage}
\pagerange{\pageref{firstpage}--\pageref{lastpage}} 

\maketitle

\begin{abstract} 
We investigate the dynamics of interstellar dust particles in moderately high resolution ($512^3$ grid points) simulations of forced compressible transonic turbulence including self-gravity of the gas. Turbulence is induced by stochastic compressive forcing which is delta-correlated in time. By considering the nearly Jeans-unstable case, where the scaling of the simulation is such that a statistical steady state without any irreversible collapses is obtained, we obtain a randomly varying potential, acting as a second stochastic forcing. We show that, in this setting, low-inertia grains follow the gas flow and cluster in much the same way as in a case of statistical steady-state turbulence without self-gravity. Large, high-inertia grains, however, are accelerated to much higher mean velocities in the presence of self-gravity. Grains of intermediate size also show an increased degree of clustering. We conclude that self-gravity effects can play an important role for aggregation/coagulation of dust even in a turbulent system which is not Jeans-unstable. In particular, the collision rate of large grains in the interstellar medium can be much higher than predicted by previous work.
\end{abstract}


\section{Introduction}
The interstellar medium (ISM) of galaxies is known to be turbulent and highly compressible, like almost all gaseous media in astrophysics. Depending on the temperature, which defines the sound speed, the turbulence of the ISM can be anything from nearly incompressible to hypersonic; interstellar gas flows can have mean Mach numbers ranging from 0.1 to 20. The warm diffuse ISM is transonic and display non-thermal motions of the order 10\,km~s$^{-1}$ \citep{Heiles03}. Moreover, according to the empirical \citet{Larson81} relations, the turbulent velocity dispersion $\sigma_{\rm u}$ in molecular clouds (MCs) is scale dependent, such that small-scale systems have smaller $\sigma_{\rm u}$ than large-scale systems. This also implies that locally, the turbulence in an MCs may be transonic (rather than hyper- or supersonic), although MCs have a high mean Mach number on a global scale.

Typical average densities in the diffuse ISM range between one and a few hydrogen atoms per cm$^{3}$, which means the kinetic drag on dust particles due to collisions with gas particles is relatively weak and the interaction timescale (stopping time) is relatively long. In MCs densities can be several orders of magnitude higher, which suggest tighter coupling between gas and dust.  But the interstellar gas is also self-gravitating and concentrations of gas may form and generate potential wells which can accelerate grains independently of the drag force. This has been studied quite extensively for planetary discs \citep[see, e.g.,][]{Rice04,Shi16,Baehr19}, but must apply to other astrophysical contexts as well. In the ISM we could expect there is a stochastic background potential resulting from continuous formation and disruption of gas concentrations. \citet{Hedvall19} hypothesised, based on simulations of supersonic turbulence with dust, that sufficiently large (decoupled) grains will be accelerated to higher velocities on average in the presence of self-gravity in the form of a steady-state stochastic potential. They also suggested that this effect should occur for smaller grains at lower Mach numbers (their study considered $\mathcal{M}_{\rm rms}\approx 3.6$) due to the lower gas-flow velocities and longer frictional timescales. 

Regardless of the self-gravity effects, small inertial particles, like interstellar dust grains, can show fractal clustering on scales smaller than the typical length scale of the turbulent flow. Such {\it small-scale clustering}, a.k.a. preferential concentration, of particles is well-studied for incompressible flows, for which it has been shown that the centrifuging of particles away from vortex cores leads to the accumulation of particles in convergence zones  \citep{Maxey87,Squires91,Eaton94,Bec05,Yavuz18}. Vorticity and the inertia of the particles is therefore decisive for the amount of clustering \citep[see][]{Toschi09}. This phenomenon has been much studied and simulated for incompressible turbulence \citep[see, e.g.,][]{Sundaram97,Hogan99,Hogan01,Bec03,Bec07,Bec07b,Bec10,Bhatnagar18}, but has only in the last decade become a subject for numerical experiments in highly compressible turbulence \citep[e.g.,][]{Pan11,Laibe12a,Laibe12b,Hopkins16,Tricco17,Mattsson19a,Mattsson19c}. As for the incompressible case, small grains (with radii $\lesssim 0.1\,\mu$m) tend to cluster and follow the gas to first approximation, while larger grains ($\gtrsim 1\,\mu$m) tend to decouple from the gas flow and not cluster notably. At which grain size the clustering maximum occurs and how the fractal {\it correlation dimension} depends on physical grain size seem to change with the degree of compressibility (Mach number) due to clustering via shock-shock interaction \citep{Haugen21}. 

It is well known grain clustering and relative grain velocities play important roles for coagulation, i.e., the formation of dust aggregates \citep[see, e.g.,][]{Pumir16}, and it has recently been shown that compressible turbulence can accelerate coagulation significantly \citep{Li21}. Adding to this the results by \citet{Hedvall19}, which indicate that fractal clustering and velocity statistics of large grains in a self-gravitating gas may differ from the case of pure kinetic-drag acceleration, exploration of self-gravity effects appear needed to understand grain growth in the ISM. Large grains ($\sim 1\,\mu$m in radius) are needed to overcome the ``coagulation bottle-neck effect'', i.e., grains must also grow to a certain average size for coagulation to be efficient \citep{Mattsson16}.  Depletion patterns in interstellar gas abundances suggest accretion of molecules onto dust in MCs may be efficient and a viable mechanism to form large grains \citep[see, e.g.,][]{Jenkins09,DeCia13,DeCia16,Mattsson19b}. This picture is also supported by the fact that late-type galaxies seem to have steeper dust-to-gas gradients than metallicity gradients along their discs \citep{Mattsson12a,Mattsson12b,Mattsson14b,Vilchez19}. But in a dilute medium such as the ISM, low velocities may form a second ``bottle neck'', which cannot be overcome by turbulent gas drag alone. 

In the present paper we aim to show, by means of numerical simulations of hydrodynamic turbulence with dust treated as Lagrangian inertial particles, how dust grains are affected by self-gravity effects in a nearly Jeans-unstable setting and discuss a few of its consequences.

\section{Basic equations and theory}
\label{equations}
\subsection{Fluid equations}
The basic equations governing the dynamics of the ISM are the equations of fluid dynamics. For a compressible fluid/gas the density is given by the continuity equation,
\begin{equation}
{{\rm d\rho}\over {\rm d}t} =
{\partial \rho\over \partial t} + \nabla\cdot(\rho\,\mathbfit{u})=0,
\end{equation}
where $\rho$ is density and $\mathbfit{u}$ is the velocity field. The velocity field is governed by the equation of motion (a.k.a. momentum equation),
\begin{equation}
\label{EOM}
\rho{{\rm d\mathbfit{u}}\over {\rm d}t} =
\rho\left({\partial \mathbfit{u}\over \partial t} + \mathbfit{u}\cdot\nabla\mathbfit{u}\right)= -{\nabla P} +  {\mathbfit{f}_{\rm visc}} + {\mathbfit{f}_{\rm force}},
\end{equation}
in which $P$ is (gas) pressure, $\mathbfit{f}_{\rm visc} = \nabla \cdot \,\left(2\nu\,\rho\,\mathbfss{S} _{0}\right)+\nabla \cdot \,\left(3\zeta\, \mathbfss{C} \right)$ represents viscous forces, where $\nu$ is the kinematic viscosity, $\mathbfss{C}={1\over 3}\left(\nabla \!\cdot \!\mathbfit{u} \right)\mathbfss{I} $ is the compression tensor, $\mathbfss{S}_0 = \mathbfss{S} - \mathbfss{C}$ is the rate-of-strain tensor, and $\mathbfss{S} = {1\over 2}\left[\nabla \mathbfit{u} +\left(\nabla \mathbfit{u} \right)^{T}\right]$. The physical viscous forces are complemented with an artificial (shock) viscosity to ensure numerical stability. The last term in Eq. (\ref{EOM}), $\mathbfit{f}_{\rm force}$, is external forcing, which in the present case represents stochastic driving of turbulence and self-gravity. The self-gravity terms in eq. (\ref{EOM}) is obtained by adding the Poisson equation,
\begin{equation}
\label{Poisson}
\nabla^2 \Phi = 4\pi\,G\,\rho,
\end{equation}
where $\Phi$ is the gravitational potential.

To obtain closure of the equations above we must also introduce a coupling between pressure and density, which is here just the isothermal condition $P=c_{\rm s}^2\,\rho$, with $c_{\rm s}$ the isothermal sound speed. 

\subsection{Dust equation of motion}
Interstellar dust shall in our simulations be treated as ideal Lagrangian inertial particles, i.e., spherical massive particles in an $N$-body system. Such particles, suspended in a gaseous medium, will show a delayed response to kinetic drag and will therefore have their own equation of motion and not just trace the motion of the carrier (gas). Each particle $i$ will obey
\begin{equation}
\label{eom_dust}
m_{{\rm gr},\,i}{{\rm d}\mathbfit{v}_i\over {\rm dt}} = \mathbfit{F}_{{\rm drag},\,i} + \mathbfit{F}_{\rm grav},
\end{equation}
where $m_{{\rm gr},\,i}$ is the mass of the particle and  $\mathbfit{F}_{{\rm drag},\,i}$, $\mathbfit{F}_{\rm grav}$ denote the kinetic drag force imposed by the gas and the gravitational force created due to the self-gravity of the gas (and calculated using Poisson's equation as previously explained). The latter is a consequence of the gas being highly compressible, an important fact that should not be disregarded in an astrophysical context. The kinetic drag force is
\begin{equation}
\label{stokeseq}
\mathbfit{F}_{{\rm drag},\,i} = {m_{{\rm gr},\,i} \over \tau_{{\rm s},\,i}}\,(\mathbfit{u}-\mathbfit{v}_i),
\end{equation}
where $\tau_{{\rm s},\,i}$ is the stopping time, i.e., the timescale of acceleration (or deceleration) of grains, which will be discussed in more detail below.

\subsection{Analytical predictions and hypotheses}
Before we go on to describe our simulations, we will first have a look at some analytic results which can be derived based on the equations presented above. It will be valuable to have these results in mind when we later interpret the simulation results.

\subsubsection{Dust fluid}
\label{sec:dustfluid}
In order to derive analytical results on the acceleration of dust particles in the presence of a background gravitational potential, we will introduce an equation of motion for a ``dust fluid'' which has grain radius $a$ as a scalar variable, i.e.,
\begin{equation}
\label{eom_dust2}
{{\rm d} \mathbfit{v}\over {\rm d}t} = {\mathbfit{u}- \mathbfit{v}(a)\over \tau_{\rm s}} - \nabla\Phi.
\end{equation}
The root-mean-square velocities for gas and dust are given by $u_{\rm rms}^2 = \langle \mathbfit{u}\cdot \mathbfit{u} \rangle = \langle \mathbfit{u}^2\rangle$ and $v_{\rm rms}^2 = \langle \mathbfit{v}\cdot\mathbfit{v} \rangle = \langle \mathbfit{v}^2\rangle$, respectively, where the brackets denote volume means, which are the same as ensemble means in this case (see Section \ref{sec:enseblemean}).

\subsubsection{Mean equation of motion}
\label{sec:MEOM}
To obtain an analytically tractable equation, we will assume $u_{\rm rms}$ remains constant and that equilibrium drag/drift applies once a statistical steady state for the gas is reached\footnote{The assumption of equilibrium drag/drift requires that we treat the dust as a fluid.}. Equilibrium drag/drift means ${d \mathbfit{v}/ dt} = {d \mathbfit{u}/ dt}$, which is a condition that can be assumed to be fulfilled on average in homogeneous isotropic steady-state turbulence. The stopping time $\tau_{\rm s}$ will here be treated as a constant, defined as
\begin{equation}
 \label{eq:conststoptime}
\tau_{\rm s} = \sqrt{\upi\over 8} {\rho_{\rm gr}\over \langle \rho\rangle}{a\over  c_{\rm s}}.
\end{equation}
Adopting the dimensionless variables $\mathcal{V} \equiv (v_{\rm rms}/u_{\rm rms})^2$, $\mathcal{V}_{\rm G}\equiv u_{\rm rms}^{-2}\tau_{\rm s}\,\langle\mathbfit{v}\cdot \nabla\Phi \rangle$, $\vartheta \equiv t/\tau_{\rm s}$ and assuming that the acceleration of dust particles mainly occurs after a statistical steady state for the gas has been obtained, we obtain an equation for $\mathcal{V}$ (see Appendix \ref{apx:dustacceq} for a derivation),
\begin{equation}
\label{acc_eq}
    {1\over 2}{{\rm d} \mathcal{V}\over {\rm d}\vartheta} = 1 - (\mathcal{V} + \mathcal{V}_{\rm G}),
\end{equation}
which can be seen as a ``dust acceleration equation'' describing how the kinetic energy of the gas and the potential energy of the gravitational field is transferred to the dust.
This equation has a simple analytical solution $\mathcal{V}(\vartheta) = 1-{\rm e}^{-2\,\vartheta}$ for the case $\mathcal{V}_{\rm G} = 0$ (no self-gravity). This solution seem to imply that the final mean velocity is always that of the gas, but that is not the case. For a turbulent system, there exists a finite time of acceleration $\Delta t_{\rm acc}$, which is related to the integral (or large-eddy) timescale of the system. Dust acceleration will follow $\mathcal{V}(\vartheta) = 1-{\rm e}^{-2\,\vartheta}$ for a certain time and then the overall acceleration stops and a statistical steady state emerges. 

\subsubsection{Grain-size dependence}
If we let $\Delta t_{\rm acc}$ be the typical time it takes to accelerate the dust, we can write $v_{\rm rms}$ as a function of $a$ by using equation (\ref{eq:conststoptime}),
\begin{equation}
\label{vrmsa_eq}
    \mathcal{V}(\vartheta=\Delta\vartheta, a) = \left[1-\exp\left(-{a_0\over a}\right) \right],
\end{equation}
where
\begin{equation}
a_0 \equiv \sqrt{32\over \upi}{\langle\rho \rangle\over \rho_{\rm gr}}\,c_{\rm s}\tau_{\rm s}\,\Delta \vartheta, \quad \Delta\vartheta = {\Delta t_{\rm acc}\over \tau_{\rm s}}.
\end{equation}
With $\Delta t_{\rm acc} \sim \tau_{L}^{\rm d} = L\,(v_{\rm rms})^{-1}$, where $L$ is the characteristic energy-injection scale, we have that $\Delta t_{\rm acc}$ depends on grain size and $a_0$ is effectively a function of $a$. If we make a power-law ansatz, $\Delta t_{\rm acc} \propto a^\beta$, one can easily show by series expansion of equation (\ref{vrmsa_eq}) that the power index seems to converge to $\beta \approx 0.3$. This is a relatively weak dependence and in the following subsection we will assume $a_0$ is a constant. Note also that the integral/eddy-turnover timescale is shorter for higher Mach number and thus $a_0$ will be smaller.

\subsubsection{Adding self-gravity of the gas}
\label{sec:addselfgrav}
To include the effect of a self-gravitating carrier (gas), we have to estimate the contribution from the $\mathcal{V}_{\rm G}$ term in equation (\ref{acc_eq}). Let us first define another dimensionless variable: $\mathcal{G} \equiv u_{\rm rms}^{-2}\tau_{\rm s}^2\,\langle(\nabla\Phi)^2\rangle$. Taking the scalar product of both sides of equation (\ref{eom_dust2}) with $\nabla\Phi$ and then spatial average of each term yields
\begin{equation}
\label{vrms_phi}
\left\langle \nabla\Phi\cdot {{\rm d} \mathbfit{v}\over {\rm d}t}\right\rangle = -\left[{ \langle\nabla\Phi\cdot\mathbfit{v}\rangle\over \tau_{\rm s}} + \langle(\nabla\Phi)^2\rangle\right] = -u_{\rm rms}^2 \,(\mathcal{V}_{\rm G} + \mathcal{G}),
\end{equation}
where we have used the fact that the carrier is ``incompressible on average'', i.e., $\langle \nabla\cdot \mathbfit{u}\rangle = 0 \Leftrightarrow\langle\nabla\Phi\cdot\mathbfit{u}\rangle = 0$ (see Appendix \ref{apx:sssgas}). In the low-inertia limit, the right-hand-side term is vanishing, since $\mathcal{V}_{\rm G} \to 0$, $\mathcal{G} \to 0$. The right-hand-side is zero also if equilibrium drag holds. Provided equilibrium drag holds on average, we may then adopt $\mathcal{V}_{\rm G} \approx - \mathcal{G}$ as a general approximation. However, it breaks down for large-inertia dust, since in that limit ${\rm d}\mathbfit{v}/{\rm d}t \to -\nabla\Phi$ and therefore $\mathcal{V}_{\rm G}\ll 1$. But since the fluid approximation of dust also breaks down in the large-inertia limit, we shall not go into details about what happens for very high-inertia dust particles. In the range where the fluid approximation more or less works we find that gravitational acceleration of dust particles can be included in a modified solution of the form $\mathcal{V}(\vartheta) = (1+\mathcal{G})(1-e^{-2\vartheta})$,
which also corresponds to
\begin{equation}
\label{vrmsa_eq_grav2}
    \mathcal{V}(\vartheta=\Delta \vartheta, a) = \bigg[1 + \mathcal{G}_0\,\left({a\over a_0} \right)^2\bigg] \,\bigg[1-\exp\left(-{a_0\over a}\right) \bigg].
\end{equation}
where
$\mathcal{G}_0 \equiv {\tau_{\rm s,\,0}^{-2}\, u_{\rm rms}^{2}}\,\langle(\nabla\Phi)^2\rangle$, $\tau_{\rm s}  = \tau_{\rm s,\,0}\,{a/a_0}$.
Note that $\mathcal{G}_0$ is a constant, given the assumptions made above. Numerical analysis of the extrema of the solution (\ref{vrmsa_eq_grav2}), given by
\begin{equation}
    0= {{\rm d}\mathcal{V}\over {\rm d}x} = {1\over x^2}\left[2\,\mathcal{G}_0\,x^3(1-e^{-1/x}) - e^{-1/x}(1+\mathcal{G}_0\,x^2) \right],
\end{equation}
where we have used $x = {a/a_0}$ for convenience, yields that $\mathcal{V}$ has local minima $x_{\rm min} > 0$ for $\mathcal{G}_0$ values in the range $0<\mathcal{G}_0\lesssim 0.65$, i.e., for $\mathcal{G}_0$ values in this range, $\mathcal{V}$ reaches a minimum and begins to increase again for larger grains. 
For $x = a/a_0 \gg 1$, equation (\ref{vrmsa_eq_grav2}) suggests $\mathcal{V} = \mathcal{G}_0\,x$, which is a false prediction and to some degree a consequence of the breakdown of the dust-fluid approximation. \citet{Hedvall19} suggested that $\mathcal{V}$ should satisfy
\begin{equation}
\label{eq:HMeq}
\mathcal{V} = 1+\mathcal{G} - \mathcal{A}, \quad {\rm with} \quad \mathcal{A} \equiv u_{\rm rms}^{-2}\tau_{\rm s}^2\,\left\langle\,\left({{\rm d}\mathbfit{v}\over {\rm d}t}\right)^2 \right\rangle.
\end{equation}
$\mathcal{A}$ is the dimensionless mean square of the acceleration of dust particles, which should rapidly approach a nearly constant value (statistical steady state). Equation (\ref{eq:HMeq}) holds more generally than equation (\ref{vrmsa_eq_grav2}) and implies $\mathcal{V}=1$ for very large, completely decoupled grains, which of course seems more reasonable.

\subsubsection{A generalised parametric model}
\label{sec:genformula}
If we assume $\Delta t_{\rm acc} \sim a^\beta$, as suggested above, a natural generalisation of equation (\ref{vrmsa_eq_grav2}) is
 \begin{equation}
 \label{vrms_fit}
 \mathcal{V}(\mathcal{G}_0,b,\alpha_0;\alpha) =
    \bigg[1 + \mathcal{G}_0\,\left({\alpha\over \alpha_0} \right)^{2b}\bigg] \,\bigg[1-\exp\left\{- \left({\alpha_0\over \alpha}\right)^b\right\}\bigg],
\end{equation}
where $b$ is a parameter related to $\beta$ ($b\approx 1-\beta$). This formula should be sufficiently general to be used as a fitting function in an attempt to model the output of detailed simulations. We will return to this in Section \ref{results}, where we present results of the numerical simulations to be described in Section \ref{sec:numerical}. 

\subsubsection{Summary of analytical predictions}
Based on equations (\ref{eq:HMeq}) and (\ref{vrmsa_eq_grav2}), we make the following important observations: 
\begin{enumerate}
    \item[(1)] in the tracer-particle limit, where $\mathcal{V} = 1$ ($\mathbfit{v}=\mathbfit{u}$) and $\tau_{\rm s}\to 0$, we have $\mathcal{G}= \mathcal{A}$; 
    \item[(2)] there exist a minimum in $\mathcal{V}$ in the range $0 <a/a_0 \lesssim 1/2$;
    \item[(3)] in the high-inertia limit (complete decoupling), where the dust-fluid approximation breaks down, kinetic drag is negligible and $d\mathbfit{v}/dt = -\nabla\Phi$, we should again have $\mathcal{G}= \mathcal{A}$ and $\mathcal{V}= 1$;
    \item[(4)] in the regime where $\mathcal{V}_{\rm G} \sim \mathcal{G}$,  we may have $\mathcal{A} < \mathcal{G}$, so that $v_{\rm rms} > u_{\rm rms}$.
\end{enumerate}
The dynamics of low-inertia particles, which are well correlated with the gas flow, is governed by the same equation regardless of whether self-gravity is included or not. Thus, such particles should follow the usual declining trend for small particles. In the extreme large-inertia limit $\tau_{\rm s}\tau_{\rm L} \gg 1$, which means the fluid approximation cannot hold. Consequently, we must take the result (3) above with a pinch salt. In this limit, according to equation (\ref{vrms_phi}), $\mathcal{V}_{\rm G} = 0$, which is inconsistent with ${\rm d}\mathbfit{v}/{\rm d}t = -\nabla\Phi$. Particles with more moderately large inertia are essentially uncorrelated with the smaller features of the gas flow, but still experience a kinetic drag on larger scales which is not negligible. In such a case, the fluid approximation is more reasonable, and the prediction of an upturn in $\mathcal{V}$ vs. $a/a_0$ is a real effect, as well as the possibility that $\mathcal{V}> 1$.

\section{Numerical simulation methods}
\label{sec:numerical}
\subsection{Simulation setup}
We simulate a turbulent gas flow aimed at describing the dynamics of the ISM. We set up a moderately high-resolution ($512^3$) three-dimensional (3D) periodic-boundary box with sides $L_{\rm box} = 2\upi$ (simulation units). We then solve the standard hydrodynamic equations as described in section \ref{equations} with a constant sound speed set to $c_{\rm s} = 1$, which is the velocity unit in the simulations. We refer to \citet{Mattsson19a} for further details about scaling and simulation units. Dust particles are included as inertial particles in size bins with $10^6$ particles in each. To solve the equations of the model, we use the {\sc Pencil Code}, which is a non-conservative, high-order, finite-difference code (sixth order in space and third order in time). For a more detailed description of the code, see \citet{Brandenburg02} and \citet{PCC21} as well as the GitHub repository (see Data Availability statement at the end of this paper) which include extensive documentation of the code.

        \begin{table*}
  \begin{center}
  \caption{\label{simulations}  Basic properties and time-averaged physical parameters of the simulations. All simulation have the mean gas density and isothermal sound speed set to unity, i.e., $\langle\rho\rangle=c_{\rm s}=1$.}
  \begin{tabular}{l|lllllllllll}
  %\hline
  \hline\\[-3mm]
  \rule[-0.2cm]{0mm}{0.6cm}
  & $f$  & $\log(\rho_{\rm min})$ & $\log(\rho_{\rm max})$  &$\mathcal{M}_{\rm rms}$ & $\mathcal{M}_{\rm max}$ & $\omega_{\rm rms}$ &Re & $\nu_{\rm shock}$& $L_{\rm box}$ & $\lambda_{\rm  J,\,0}/L_{\rm box}$\\[2mm]
  \hline\\[-1mm]
  S3G & $1.0$ & $-4.86\pm 0.57$ & $8.13\pm 0.12$ & $1.15\pm 0.046$& $4.26\pm 0.35$ & $4.61\pm 0.24$ & $\sim 700$ & 1.0 & $2\upi$ & 1.0\\
  S3    & $1.0$ & $-4.54\pm 0.55$ & $8.08\pm 0.13$ & $1.14\pm 0.048$& $4.26\pm 0.37$ & $4.44\pm 0.22$ & $\sim 700$ & 1.0 & $2\upi$ & --\\[2mm]
  %\hline
  \hline
  \end{tabular}
  \end{center}
  \end{table*}

\subsection{External forces}
\subsubsection{Driving of turbulence}
A steady-state turbulence simulation requires some kind of external forcing. We use a white-in-time stochastic forcing term with both solenoidal (rotational) and compressible components. The forcing is applied at low wave-numbers in Fourier space and is delta-correlated in time. More precisely, we apply an Ornstein-Uhlenbeck process to excite turbulence and the projections in Fourier space necessary to separate into compressive and solenoidal force components.  Here, we consider only purely compressive forcing, since it is reasonable to assume that transonic turbulence is dominated by acoustic-type forcing from, e.g., supernovae and stellar wind bubbles. Moreover, compressive modes in the flow will be of greater importance in a self-gravity context \citep{Federrath12} and we are thus maximising the effect of introducing self-gravity. The forcing is then controlled by only one parameter, the forcing amplitude $f_0$, which we set to $f_0 = 1$. This yields a mean Mach number $\mathcal{M}_{\rm rms}\sim 1$.

\subsubsection{Kinetic drag force}
\label{dustflow}
The dust grains experience kinetic drag as well as gravitational pull (see section \ref{sec:selfgrav}) from the gas. The timescale governing the drag experienced by a grain $i$ is the stopping time $\tau_{{\rm s},\,i}$, which in a compressible flow is inverse proportional to the gas density $\rho$. Assuming the dust is accelerated by a turbulent gas flow via an \citet{Epstein24} drag law, $\tau_{{\rm s},\,i}$ depends on the size and density of the grain as well as the gas density and the relative Mach number $\mathcal{W}_{{\rm s},\,i} = |\mathbfit{u}-\mathbfit{v}|/c_{\rm s}$ \citep{Schaaf63}. 
The full expression for $\tau_{{\rm s},\,i}$ is complicated and tedious to use, but there exist a convenient formula which is sufficiently accurate for our purposes \citep[see][]{Kwok75,Draine79},
\begin{equation}
\label{stoppingtime}
\tau_{{\rm s},\,i} = \sqrt{\upi\over 8}{\rho_{\rm gr}\over\rho}{a_i\over  c_{\rm s}} \left(1 + {9\upi\over 128}{|\mathbfit{u}-\mathbfit{v}_i|^2\over c_{\rm s}^2 } \right)^{-1/2},
\end{equation}
where $a_i$ is the grain radius (assuming spherical grains), $\rho_{\rm gr}$ is the bulk material density of the grain and the isothermal sound speed $c_{\rm s}$ replaces the thermal mean speed of molecules. 
The second term inside the parenthesis can be seen as a correction for supersonic flow velocities and compression.

Non-inertial particles, a.k.a. tracer particles, will have $\mathbfit{v}_i = \mathbfit{u}$ and maintain position coupling with the medium in which they reside. It is often assumed that the approximation $\mathbfit{v}_i \approx \mathbfit{u}$ is justified when $\tau_{{\rm s},\,i}/\tau_{\rm L} \ll 1$, which is indeed the case for very small interstellar dust particles. 

However, as the ISM is highly compressible and show a wide range of mean Mach numbers, even small amount of inertia can be important.

\subsubsection{Self-gravity and size of the simulation box}
\label{sec:selfgrav}
We include self-gravity of the gas is by adding the Poisson equation (\ref{Poisson}) and corresponding acceleration terms in the equations of motion. The ``Poisson constant'', $\kappa = 4\pi\,G$, is then an additional model parameter and its value will set either an intrinsic length scale or a fundamental (mean) density, even if the simulation is performed in dimensionless units. Simulations with self-gravity are therefore not scale free, even if isothermal conditions are adopted for simplicity. In the present study we seek to understand self-gravity effects on scales where no irreversible collapses occur. Using the Jeans criterion \citep{Jeans02} and the fact that we assume that forcing wavelength is $\lambda_{\rm f} \approx L_{\rm box}/2$  it is easy to show that this situation corresponds to $L_{\rm box}\sim\lambda_{\rm J,\,0}$, where $\lambda_{\rm J,\,0}$ is the Jeans wavelength at the initial equilibrium state. This in turn corresponds to $\kappa = 1$, since the mean density is $\langle \rho\rangle = 1$, i.e., the unit density is $\langle \rho\rangle$. 

  \begin{figure}
      \resizebox{\hsize}{!}{
	\includegraphics{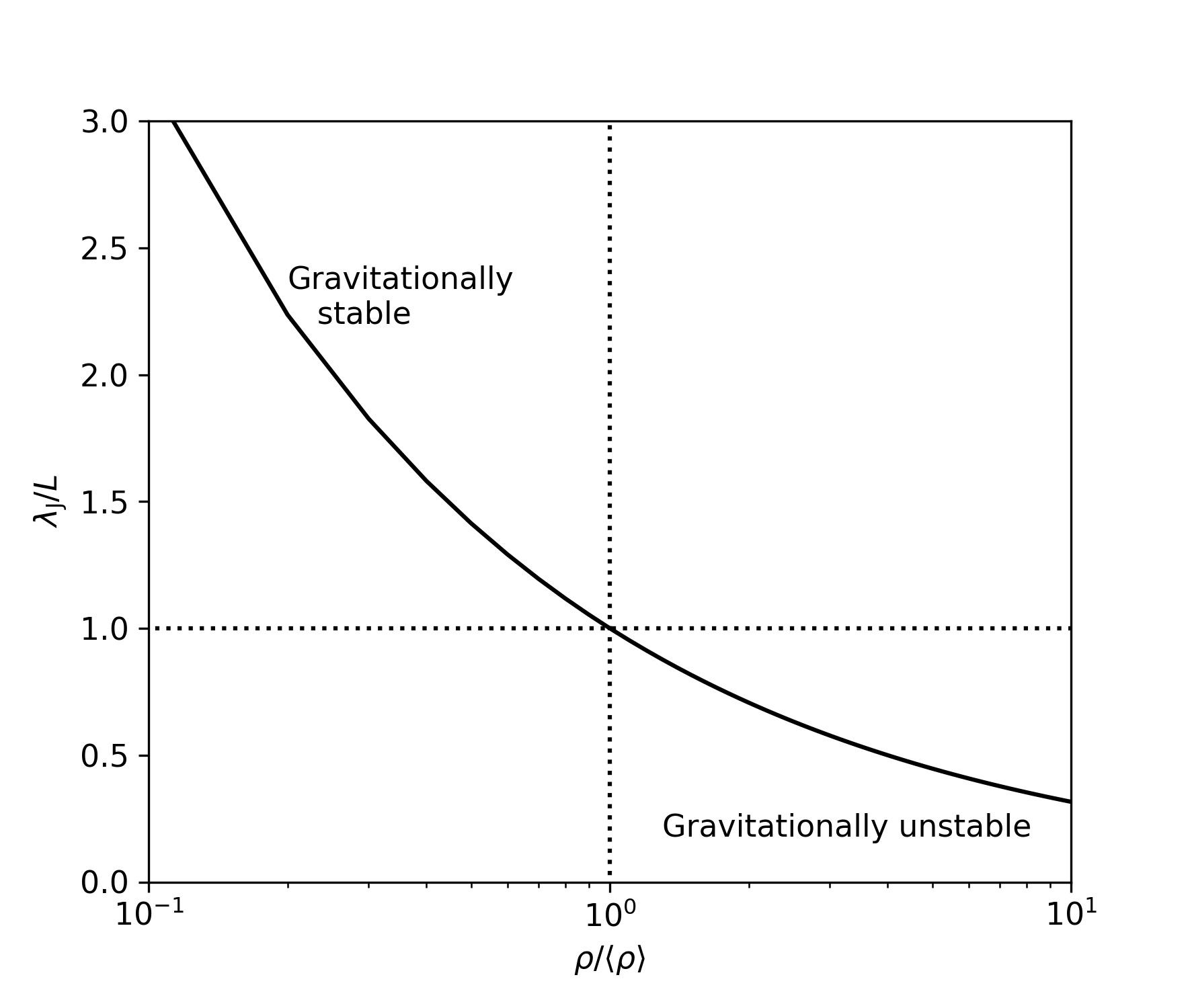}
    }
  \caption{\label{fig:lambda_J} Dependence of the Jeans wavelength $\lambda_{\rm J}$ in units of the simulation-box size $L_{\rm box}$ on the gas density $\rho$.}
  \end{figure} 

We can estimate the physical scale in pc by means of the formula
\begin{equation}
\lambda _{\rm {J}}\approx 0.4\,\bigg({\frac {c_{s}}{0.2{\mbox{ km s}}^{-1}}}\bigg)\, \bigg({\frac {n_{\rm gas}}{10^{3}{\mbox{ cm}}^{-3}}}\bigg)^{-1/2}{\mbox{ pc}},    
\end{equation}
where $n_{\rm gas}$ is the number density of gas particles (essentially hydrogen atoms/molecules). Thus, the equilibrium Jeans wavelength $\lambda_{\rm J}$ in periodic boundary box with a homogeneous distribution of gas with $\langle n_{\rm gas}\rangle =1$~cm$^{-3}$ is $\lambda_{\rm J,\,0} \sim 10$~pc. In Fig. \ref{fig:lambda_J} we show how $L_{\rm box}$ must scale with $\rho_0 = \langle \rho\rangle$ in order to maintain $\lambda_{\rm J,\,0} = L$. The volume of the simulation box $V = L_{\rm box}^3$ must be decreased whenever $\langle \rho\rangle$ is increased, because the stability of the system depends on the total mass.

\subsection{Physical viscosity and Reynolds number}
Astrophysical flows can often be treated as essentially inviscid. That is, they have very large Reynolds numbers (Re) and in principle we would like to target the limit Re~$\to\infty$. Solving the equations of fluid dynamics numerically usually require a finite Re, however. We define Re as
\begin{equation}
{\rm Re} \simeq  {L_{\rm box}\over 2}\,{u_{\rm rms}\over \nu},
\end{equation}
where $\nu$ is the kinematic viscosity. With this definition of Re, we can easily reach Re~$\sim 1000$ using the {\sc Pencil Code} with the present setup and spatial resolution. We set $\nu = 5.0\cdot 10^{-3}$ (in dimensionless simulation units) for both simulations, with and without a self-gravitating gas (labelled S3 and S3G, respectively), which yields Re~$\sim 700$ in both cases (see Table \ref{simulations}).

\subsection{Artificial viscosity and shock capturing}
In the equations actually solved in the \textsc{Pencil Code}, the viscous force (see equation \ref{EOM}) includes also a shock-capturing term, proportional to $\nabla\cdot\mathbfit{u}$ (see the \textsc{Pencil Code} User Manual for further details), where a dimensionless artificial viscosity factor $\chi_{\rm s}$ is introduced to soften strong shocks and avoid unresolved jumps in the velocity field, which would otherwise cause the code to eventually crash. We use a standard choice of $\chi_{\rm s} = 1$ in both S3 and S3G (see Table \ref{simulations}). The simulation without self-gravity of the gas (S3) would run without artificial viscosity (the chosen resolution of $512^3$ grid points is normally sufficient to resolve transonic and moderately supersonic shocks), but as it is occasionally needed when self-gravity is included (simulation S3G) and we want the two simulations to be directly comparable, we use $\chi_{\rm s} = 1.0$ for both S3 and S3G.

\subsection{Grain size and Stokes number}
We use a dimensionless grain-size parameter as defined by \citet{Hopkins16},
\begin{equation}
\alpha = {\rho_{\rm gr}\over\langle \rho\rangle}{a\over L}.
\end{equation}
Because the total mass of a simulation box and the mass of a grain of a given radius $a$ are constants, the quantity $\alpha$ must also be a constant regardless of characteristics of the simulated flow.  The physical size of the grains can be estimated from
\begin{equation}
a = 0.04\,\alpha\, \left({L\over 10\,{\rm pc}} \right)\left({\langle n_{\rm gas}\rangle \over 1\,{\rm cm}^{-3}} \right) \left({\rho_{\rm gr}\over 2.4\,{\rm g\,cm}^{-3}} \right)^{-1}\,\mu{\rm m},
\end{equation}
where $\langle n_{\rm gas}\rangle$ is the average number density of gas particles (molecules). 

The integral-scale Stokes number ${\rm St}_{\rm int}$ can be estimated as the ratio of the stopping time $\tau_{\rm s}$ and the forcing timescale of the flow $\tau_{\rm f} \propto (k_{\rm f}\,u_{\rm rms})^{-1}$, where $k_{\rm f}$ is the forcing wavenumber. For small and moderate Mach numbers $\mathcal{M}_{\rm rms} \lesssim 1$, the parameter $\alpha$ is essentially the ratio of ${\rm St}_{\rm int}$ and $\mathcal{M}_{\rm rms}$, i.e.,
\begin{equation}
\label{eq:stint}
    \langle{\rm St}_{\rm int}\rangle \approx {\langle\tau_{\rm s}\rangle\over \tau_{\rm f}} \sim \alpha\,\mathcal{M}_{\rm rms}.
\end{equation}
We will in the remainder of this paper use $\alpha$ rather than ${\rm St}_{\rm int}$, since for compressible turbulence $\alpha$ is a better indicator of the characteristic physical grain size than the average Stokes number $\langle {\rm St}_{\rm int}\rangle$ \citep{Mattsson19c}. However, we emphasise that, for subsonic and transonic turbulence, $\langle {\rm St}_{\rm int}\rangle$ is still a useful global diagnostic for the dust-gas flow interaction. In such a case, maximal clustering should always occur around a characteristic $\langle {\rm St}_{\rm int}\rangle \sim 1$ \citep[see, e.g.,][]{Bec07}.
We note that this may explain the linear shift of grain-clustering effects towards lower $\alpha$ seen in simulations of supersonic turbulence, compered to the nearly incompressible case \citep[see Figs. 4 and 5 in][]{Mattsson19c}. Moreover, we shall remember that at which $\langle {\rm St}_{\rm int}\rangle$ maximal clustering occurs, may depend on the type and strength of forcing used to sustain the turbulence \citep{Mattsson19c,Haugen21}.

\subsection{Clustering of dust}
Clustering of particles can be measured in several ways. 
 We can measure the clustering of grains in terms of nearest-neighbour statistics (NNS), the correlation dimension $d_2$ -- a kind of fractal dimension -- or just using density-based measures of clustering \citep{Monchaux12}. Power spectra of the spatial particle distribution can also be used as a multi-scale diagnostic \citep{Haugen21}, but will not be considered here.

\subsubsection{Nearest-neighbour statistics}
The NNS is a common and straight forward diagnostic of clustering, usually represented by the average nearest-neighbour distance ratio $R_{\rm ANN}$ \citep[see][]{Monchaux12,Mattsson19a,Mattsson19c}, which measures clustering relative to the non-clustered case (Poisson process). 

The NNS is efficiently obtained using the $kd$-tree algorithm \citep{Bentley75}, which results in a search tree that can be queried to return neighbours of any order. Here, we use it to obtain the distribution of first nearest-neighbour distances (1-NNDs), from which we directly get $R_{\rm ANN}$, but also indirectly another measure of small-scale clustering - the correlation dimension $d_2$. 

\subsubsection{Small-scale clustering}
A very common diagnostic of small-scale clustering, the correlation dimension $d_2$, can be defined as
\begin{equation}
    d_2 \equiv\lim_{\Delta r\to 0} {d\ln(n_{\rm d})\over d\ln(\Delta r)}
\end{equation}
where $n_{\rm d}(\Delta r)$ is the expected number of dust grains inside a ball of radius $\Delta r$ surrounding a test particle \citep{Pumir16}. The correlation dimension $d_2$ can also be determined from the slope of the power-law end of the distribution of 1-NNDs, derived from the NNS, as mentioned above \citep[see also][]{Mattsson19b}. The $d_2$ tend to show a distinct minimum at a certain $\alpha$ (or $\langle{\rm St}_{\rm int}\rangle$), which normally correlates very well with the minimum of $R_{\rm ANN}$, but filters out the effects of large-scale density variance (compaction) of the carrier (gas).

\subsubsection{Large-scale clustering}
If one wishes to target large-scale clustering, where ``large scale'' means scales comparable to integral flow scale, it is useful to consider the mean logarithmic number density $\mathcal{N}_{\rm d} = \langle\ln n_{\rm d} \rangle$ as a diagnostic. The effects of adding a gravitational force from a self-gravitating carrier are primarily large-scale, so $\langle\ln n_{\rm d} \rangle$ may be of particular importance here. That $\mathcal{N}_{\rm d}$ can be associated with clustering can be surmised from the definition of $d_2$ above. A more stringent way to argue is to say that $\mathcal{N}_{\rm d}$ is the simplest form of a grid-based measure of clustering. Formally, the data space is divided into a grid structure composed of disjoint $d$-dimensional hyper rectangles a.k.a. blocks $B$. Data points are assigned to these blocks and clustering is then determined by a neighbour search algorithm \citep{Aggarwal14}. Grid-based clustering algorithms are typically using the density index $\mathcal{D}_{\rm B} = p_{\rm B}/V_{\rm B}$, where $p_{\rm B}$ is the number of data points in $B$ and $V_{\rm B}$ is the spatial volume of $B$. High values of $\mathcal{D}_{\rm B}$ can be associated with clusters and a cluster is identified by finding adjacent blocks with high $\mathcal{D}_{\rm B}$. The average of $\mathcal{D}_{\rm B}$, taken over the whole domain, is then perhaps the simplest ``overall measure'' of clustering. If $d=3$ and the data space represents the spatial positions of dust particles, we can view $\mathcal{D}_{\rm B}$ as a normalised number density determined with $B$ as the ``binning boxes''.  Thus, $\langle\ln \mathcal{D}_{\rm B} \rangle \approx \mathcal{N}_{\rm d} + \ln\mathcal{D}_0$, where $\mathcal{D}_0$ is a correction constant. In the limit where $V_{\rm B} \to 0$ and the number of blocks $N_{\rm B}\to\infty$, we have $\langle\ln \mathcal{D}_{\rm B} \rangle = \mathcal{N}_{\rm d}$.

\begin{figure*}
      \resizebox{\hsize}{!}{
	\includegraphics[trim=0cm 0cm 2.7cm 0.7cm, clip=true]{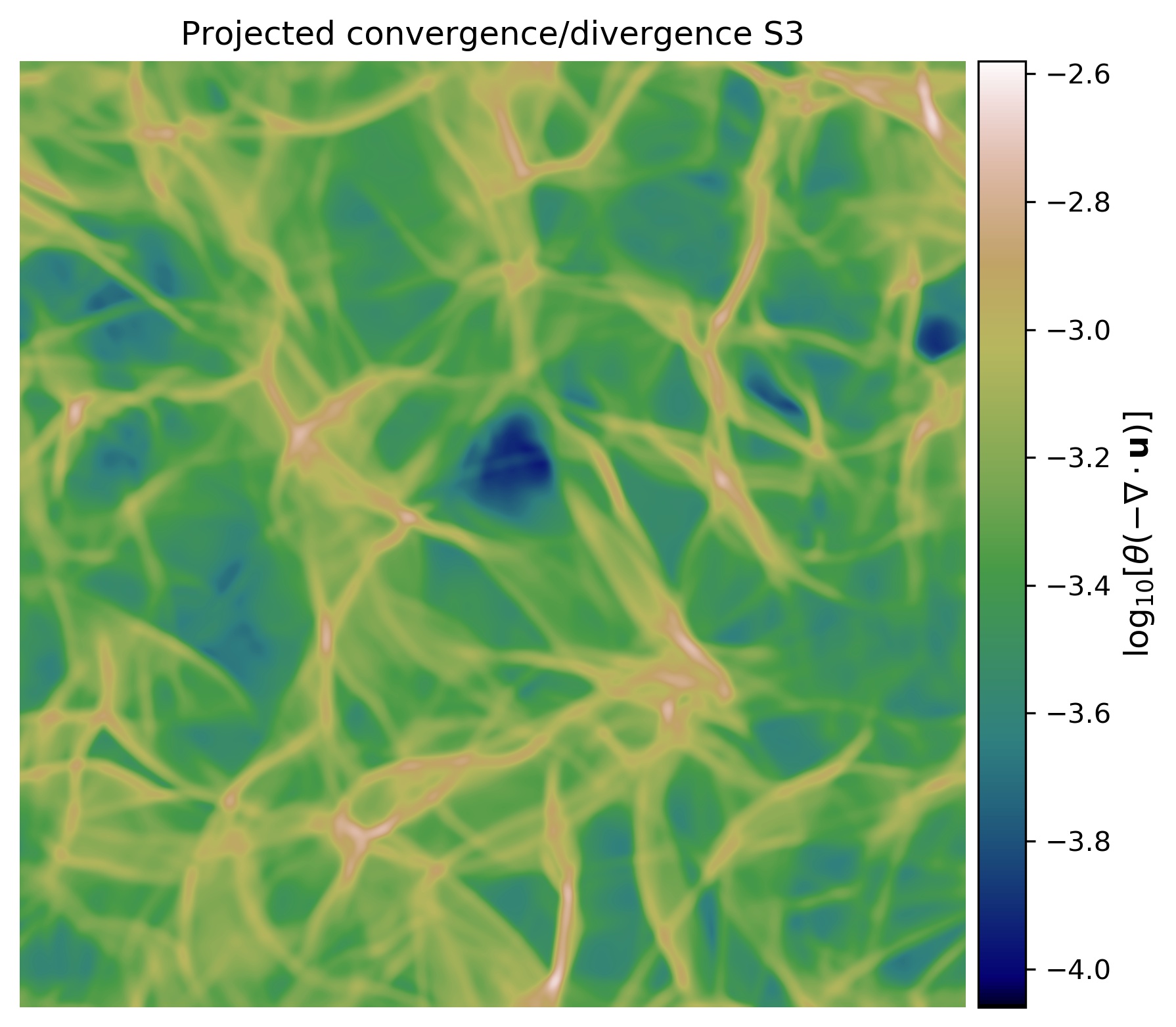}
	\includegraphics[trim=0cm 0cm 0cm 0.7cm, clip=true]{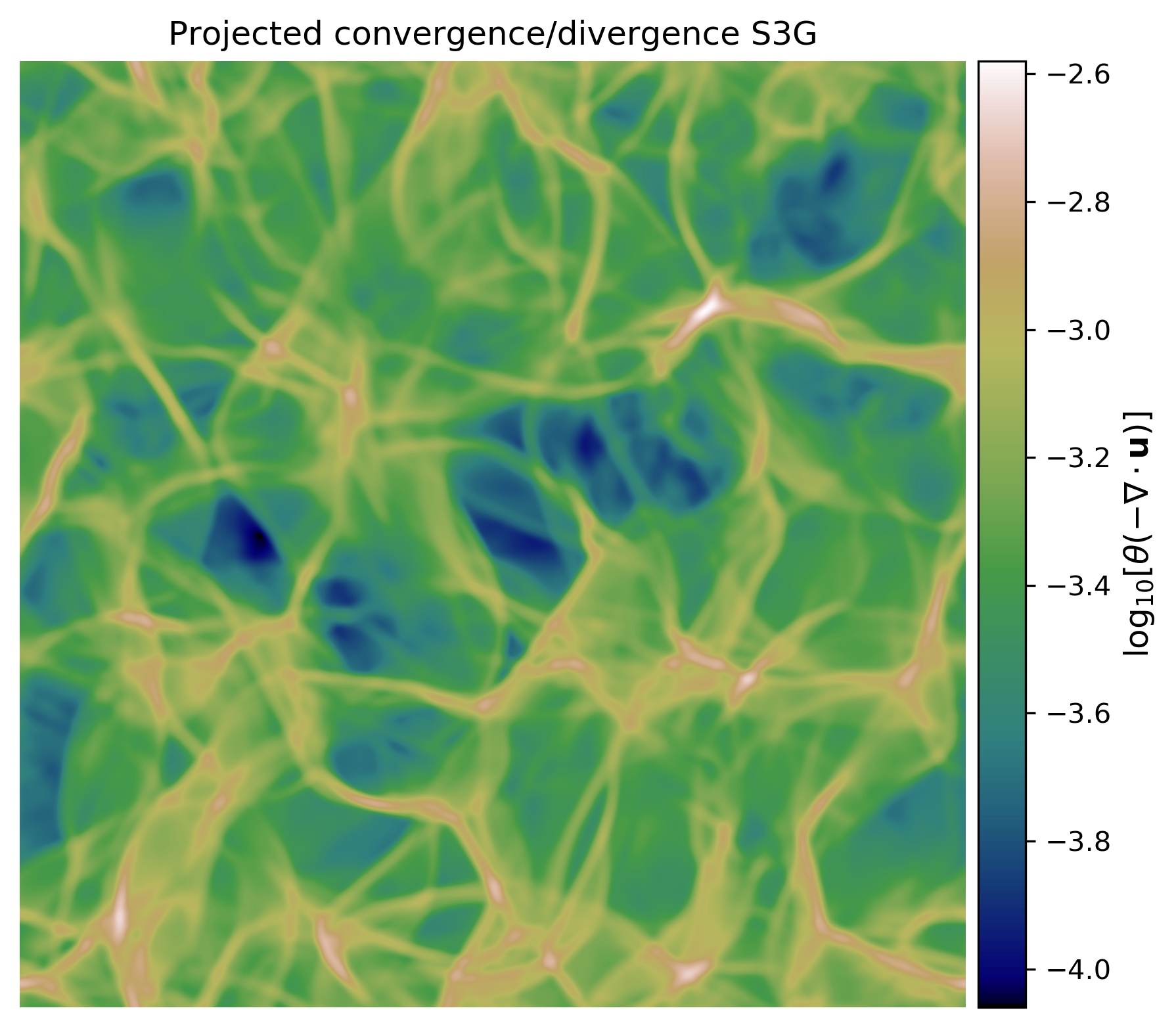}
    }
  \caption{\label{fig:shocks} Projected shock structures as traced by the convergence field $\theta(-\nabla\cdot \mathbfit{u})$. Left panel shows simulation S3 and right panel shows S3G. A small, but notable, difference in shock structure can be seen. }
  \end{figure*}

\section{Results and discussion}
\label{results}
\subsection{Gas flow properties}
\subsubsection{Gas density}
\label{sec:gasdens}
The gas-density variance in the S3 and S3G simulations are very similar at all times. Visually, the projected density fields are indistinguishable in terms of general characteristics. In both cases the gas-density probability distribution (PDF) follows almost exactly a lognormal PDF, as expected for isothermal turbulence with $\mathcal{M}_{\rm rms} \sim 1$, with a log-scale standard deviations $\sigma_s = 0.53$ and $\sigma_s = 0.52$, respectively. The S3G run display a slight tendency of forming a high-density tail in the PDF, which is expected due to the self-gravity force, but the effect is insignificant and have a more or less negligible effect on the higher-order moments of the PDF. Both the S3 and S3G cases can be considered identical in terms of density statistics. It is extremely unlikely that gas-density variance can be the cause of any differences in the clustering and dynamics of the dust.

\subsubsection{Shocks and convergence }
If the gas density fields are statistically almost indistinguishable, the convergence field do perhaps show some qualitative differences in comparison. This observation suggests there is some difference in the shock structure due to the inclusion of self-gravity in S3G, as convergence is tracing shocks in the flow.

Defining {\it convergence} as $\theta(-\nabla\cdot\mathbfit{u})$, where $\theta$ is the Heaviside step function, the S3G simulation show more high-convergence zones and generally somewhat more shock structure (see Fig. \ref{fig:shocks}, which shows the projected convergence fields of the last snapshots in the S3 and S3G simulation time series). We estimate 5-10~\% higher mean convergence in the last snapshot of S3G compared to the last snapshot of S3. But this difference is hardly significant as there is considerable differences between different snapshots of the same simulation. However, convergence patterns are apparently affected by the gravitational field, even if the effect on the mean is small.

  \begin{figure}
      \resizebox{\hsize}{!}{
	\includegraphics{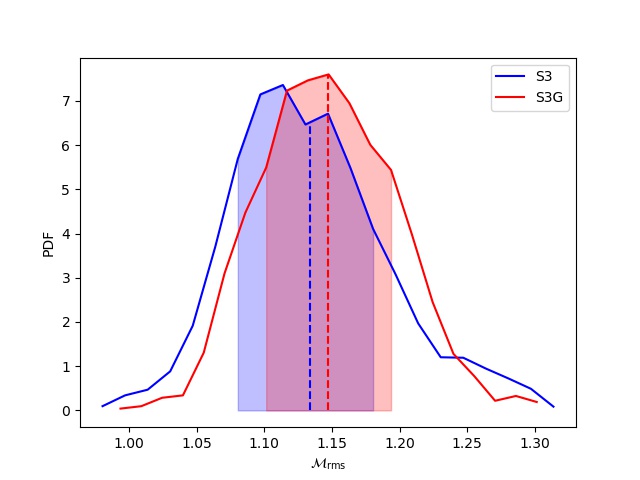}
    }
  \caption{\label{fig:machPDF} Normailised histograms of the mean Mach numbers $\mathcal{M}_{\rm rms}$ obtained from the timeseries of simulations S3 and S3G. Vertical dashed lines mark the means of each distribution and the shaded regions show the standard deviation, which overlap to a large degree. The shape of the histograms are vary similar and it cannot be ruled out that both timeseries belong to the same underlying series.}
  \end{figure} 

\subsubsection{Mach-number statistics}
The most important characteristic of highly compressible turbulence may be the elevated mean Mach number $\mathcal{M}_{\rm rms}$. Simulations of compressible turbulence often show considerable variance in $\mathcal{M}_{\rm rms}$ over time. To some extent this is a consequence of the size $L_{\rm box}$ of the simulation domain being similar to the forcing length $L$. If the domain would be sufficiently large, the change in $\mathcal{M}_{\rm rms}$ from one snapshot to another would be negligible, provided that turbulence can be regarded as an ergodic process.
In Fig. \ref{fig:machPDF} we show the distributions of $\mathcal{M}_{\rm rms}$ obtained as time-series histogram calculated on the interval $t/T= 5\dots 80$ simulation time units ($T$). The shaded regions show the standard deviation, which overlap to a large degree. The shape of the histograms are vary similar and a Kolmogorov-Smirnoff (KS) two-sample test does in fact yield a $p$-value rapidly approaching one. Thus, it cannot be ruled out that both time-series belong to the same underlying series regardless of what the KS statistic is.
The $\mathcal{M}_{\rm rms}$ distributions are similar in both S3 and S3G and can be described as roughly lognormal. This is indicating that both simulations reach similar statistical steady states and can therefore be regarded as having the same overall kinetic-drag effect on the dust, i.e., the acceleration of dust grains due to gas drag is effectively the same. Any difference in grain acceleration can thus very likely be attributed to the inclusion of self-gravity in S3G.

\subsection{Dynamics and distribution of dust}
\subsubsection{Ensemble mean velocities}
\label{sec:enseblemean}
Dust grains in a turbulent gas may have a relatively broad speed distributions, but with a well-defined mean. Root-mean-square velocities $v_{\rm rms}$ of the particles in our simulations are thus an important diagnostic for acceleration of grains and over all kinetic energy of the grains. We will in the following present $v_{\rm rms}$ for different grain sizes calculated as ensemble means, i.e.,
\begin{equation}
\label{eq:ensemblemean}
    v_{\rm rms}^2 = \langle \mathbfit{v}^2\rangle \equiv {1\over N}\sum_{n=1}^N \mathbfit{v}_n^2,
\end{equation}
where $\mathbfit{v}_n$ represents the velocity vectors of the $N$ particles of the ensemble. Provided that $N$ is sufficiently large, this type of mean corresponds to the root-mean-square velocity calculated as a volume average taken of the whole simulation domain. For a tracer particle we then have $v_{\rm rms} = u_{\rm rms}$, which would not be the case for a mass-weighted mean of fluid elements. In the results presented below $v_{\rm rms}$ is always defined according to equation (\ref{eq:ensemblemean}) and $u_{\rm rms}$ is always the straight volume average.

  \begin{figure*}
      \resizebox{\hsize}{!}{
	\includegraphics{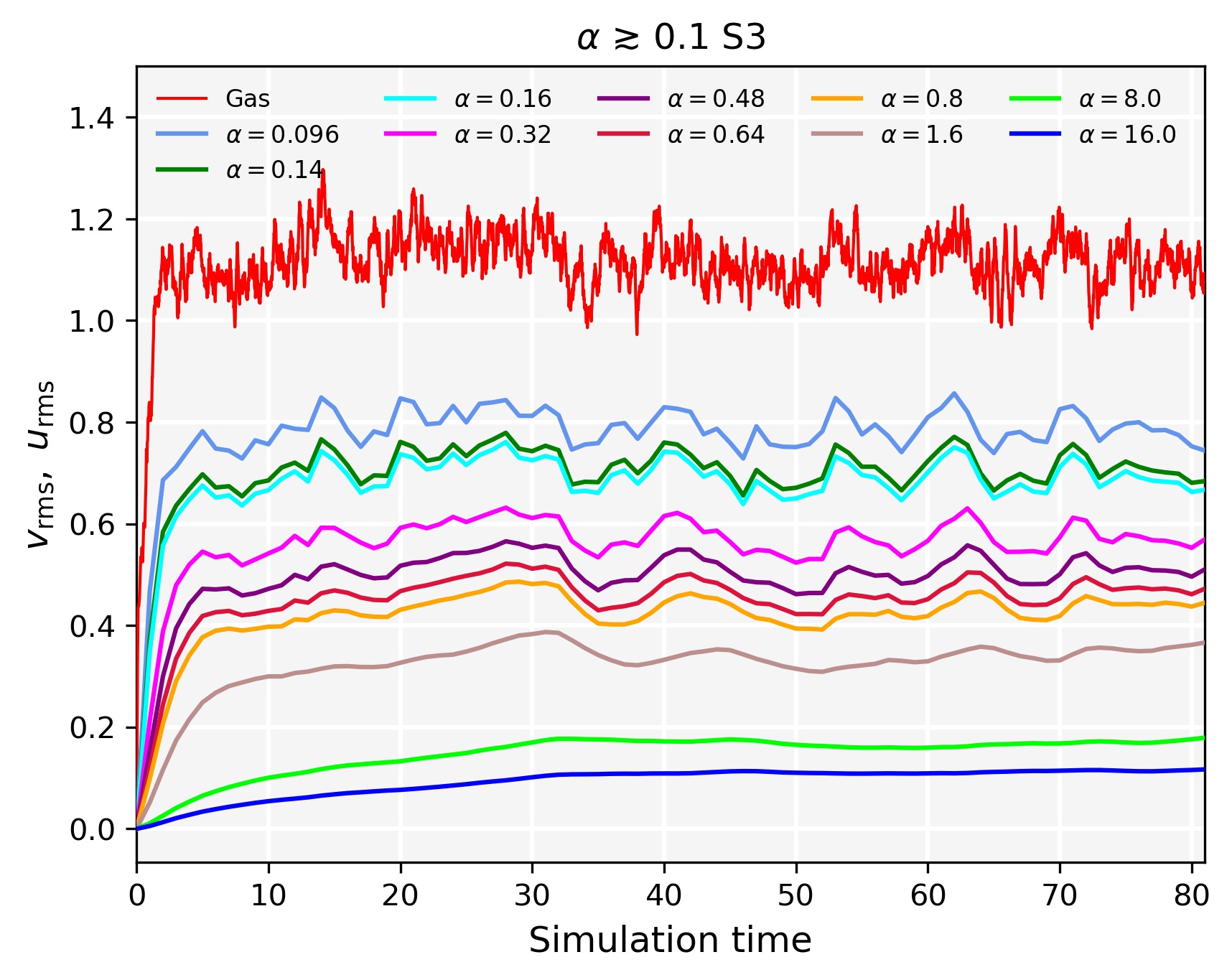}
	\includegraphics[trim=1.45cm 0cm 0.0cm 0.0cm, clip=true]{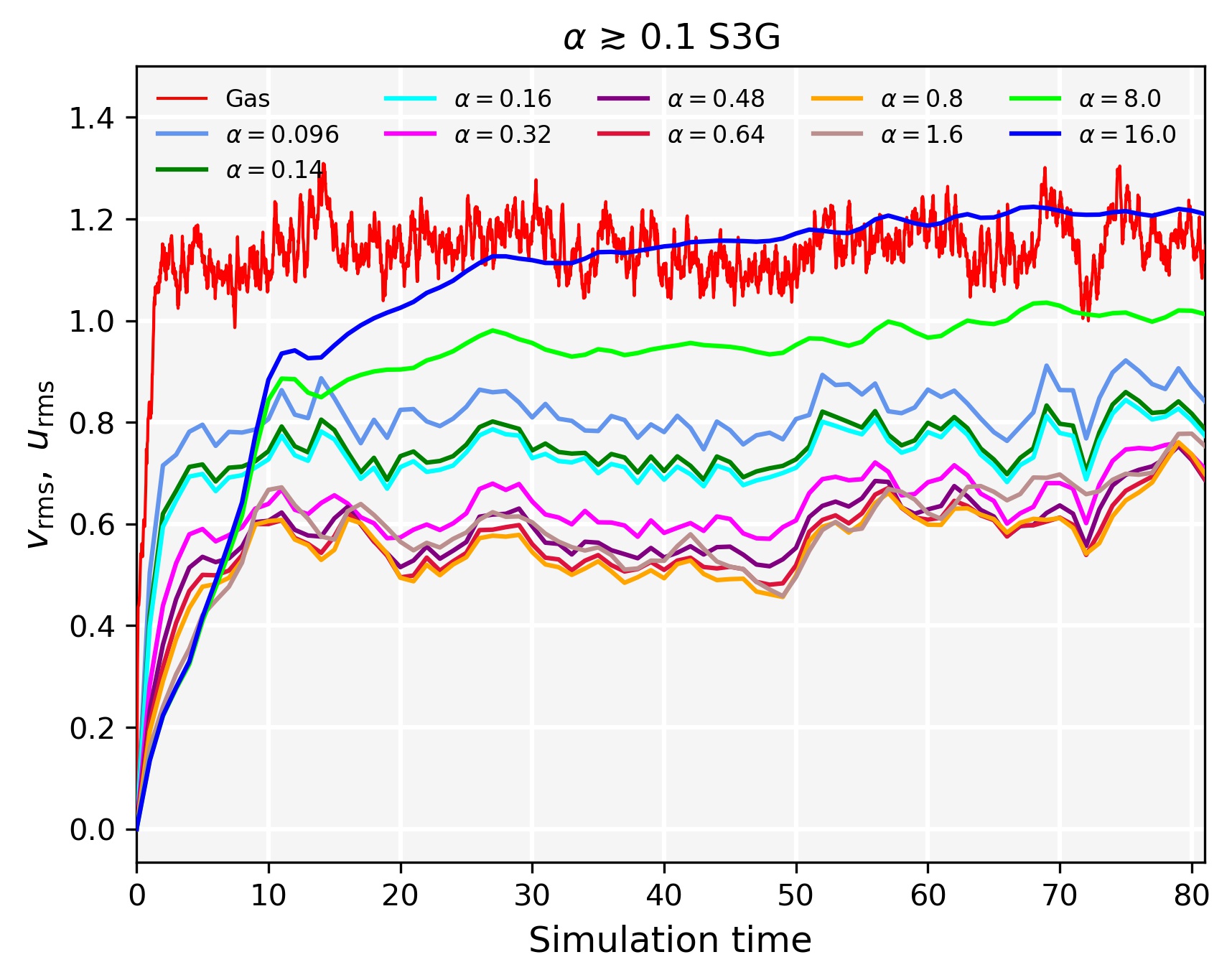}
    }
  \caption{\label{fig:vp_ts} Acceleration of dust grains of various sizes.} 
  \end{figure*}

\subsubsection{Acceleration phase}
Dust particles suspended in a gas of homogeneous isotropic turbulence are expected to reach a statistical steady after an initial phase of acceleration. The time it takes for particles to reach their steady-state mean velocity $v_{\rm rms}$, depends on their sizes/masses. It takes much longer for large-inertia particles to reach their final $v_{\rm rms}$ than it takes for small particles that follow the gas flow and the simulation results follow broadly what we predicted analytically in Section \ref{sec:MEOM} (see Fig. \ref{fig:vp_ts}) and what other simulations have shown \citep[see, e.g.,][]{Pan13}. The reason for this is of course the difference in $\tau_{\rm s}$, which effectively creates a time lag. There is no qualitative difference between S3 and S3G in this regard -- both simulations show acceleration timescales increasing with grain size, but the final $v_{\rm rms}$ can be quite different and at some point gravitational acceleration becomes dominant and breaks this grain-size -- acceleration relation for grains which decouple from the gas.

During the acceleration phase, the simulated dust grains are accelerated by propagating shocks and turbulent eddies via gas drag in both S3 and S3G. In S3G, however, there is also a gravitational force acting on the grains in addition to the drag force. The grains that tend to decouple from the flow (large-inertia particles) gain only a fraction of the momentum that could be transferred from shocks and eddies via gas drag in case of perfect coupling. If there is no other force acting on the grains, as in S3, only very small grains with negligible inertia will be accelerated to the turbulent mean velocity of the gas $u_{\rm rms}$. All larger grains, for which the acceleration is less efficient, will have $v_{\rm rms} < u_{\rm rms}$ when the statistical steady state is reached. But in S3G a stochastic gravitational field is present, created by the self-gravity of the turbulent gas, which accelerate dust grains that decouple from the gas. If the kinetic drag force is comparable to the gravitational force created by the gas, the final $v_{\rm rms}$ is much larger. In Fig. \ref{fig:vp_ts} (right panel) one can clearly see that the largest grains in S3G have much higher $v_{\rm rms}$ than the corresponding grains in S3 (compare the panels of Fig. \ref{fig:vp_ts}). Grains with $\alpha < 0.1$ seem to couple well enough to the gas to be unaffected by gravity, while the gravitational acceleration becomes dominant for $\alpha \gtrsim 1$. 

We also note that really large grains, e.g., $\alpha = 16.0$ in S3G, seem to reach $v_{\rm rms} > u_{\rm rms}$, which is supported by a test simulation we performed with even larger dust particles. In the test simulation we could not ensure that a statistical steady state was reached due to limitations in computational resources, but clearly $v_{\rm rms} > u_{\rm rms}$. However, one must be aware that grains above a certain size (and number density) will cause a non-negligible back-reaction on the gas. That is, the grains will drag the gas along, which will also affect the grain dynamics. As we have omitted this back-reaction in the present simulations, we cannot be sure that large grains are adequately modelled in the context of a self-gravitating carrier.

    \begin{figure}
      \resizebox{\hsize}{!}{
	\includegraphics{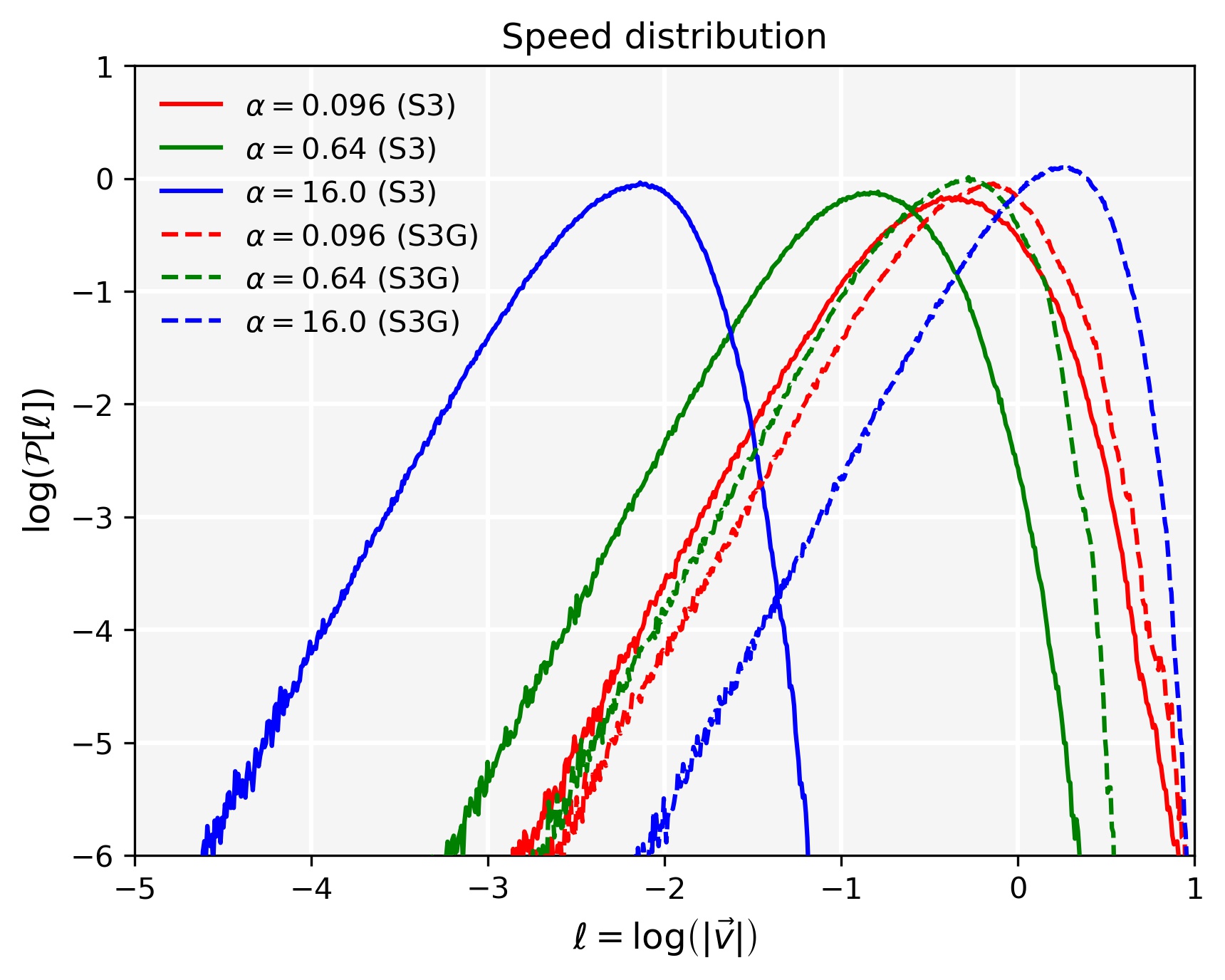}
    }
  \caption{\label{fig:vp_pdf} Probability density functions for dust-particles speeds. } 
  \end{figure}

\subsubsection{Steady-state phase}
After a few simulation time units (corresponding to $\sim 2\tau_{\rm s}$) the acceleration phase ends and the grain dynamics transcends to a statistical steady-state regime. How long this transition takes depends on the stopping time and thus the grain size. We note that after approximately 30 simulation time units all considered grain sizes seem to develop a steady state (see Fig. \ref{fig:vp_ts}). As seen in Fig. \ref{fig:vp_pdf}, showing grain speed distributions for $\alpha = 0.096, 0.64, 16.0$ computed from all saved snapshots at $t> 30$, the grains develop Maxwellian-type velocity distributions regardless of size and whether self-gravity of the gas is included or not. Note, however, that we have plotted the logarithm of the particle speeds, so the linear velocity dispersion varies considerably, although the dispersion of the logarithmic speed is similar for all plotted cases. 

The statistical steady state is stable for the gas -- the time series for $u_{\rm rms}$ is clearly just fluctuating around a temporal mean for $t/T\gtrsim 5$ and, in the case of simulation S3, $v_{\rm rms}$ show a similar behaviour, albeit with progressively smaller variance for larger grains. However, we note a shift in $v_{\rm rms}$ starting at $t/T \sim 50$ in S3G, which is particularly clear for $\alpha \sim 1$. Whether this is a temporary shift, e.g., part of a cyclic variation over a longer time, is impossible to say without a much longer time series, which we cannot obtain due to the computational expense. But we note that $\alpha \sim 1$ is the characteristic grain-size for which kinetic drag and gravitational acceleration tend to be of the same order. This indicates that there is competition between these two forces, which could very well cause cyclic behaviour in the time series of statistical quantities like $v_{\rm rms}$.

\begin{figure*}
      \resizebox{0.9\hsize}{!}{
	\includegraphics[trim=0.0cm 0cm 2.7cm 0.25cm, clip=true]{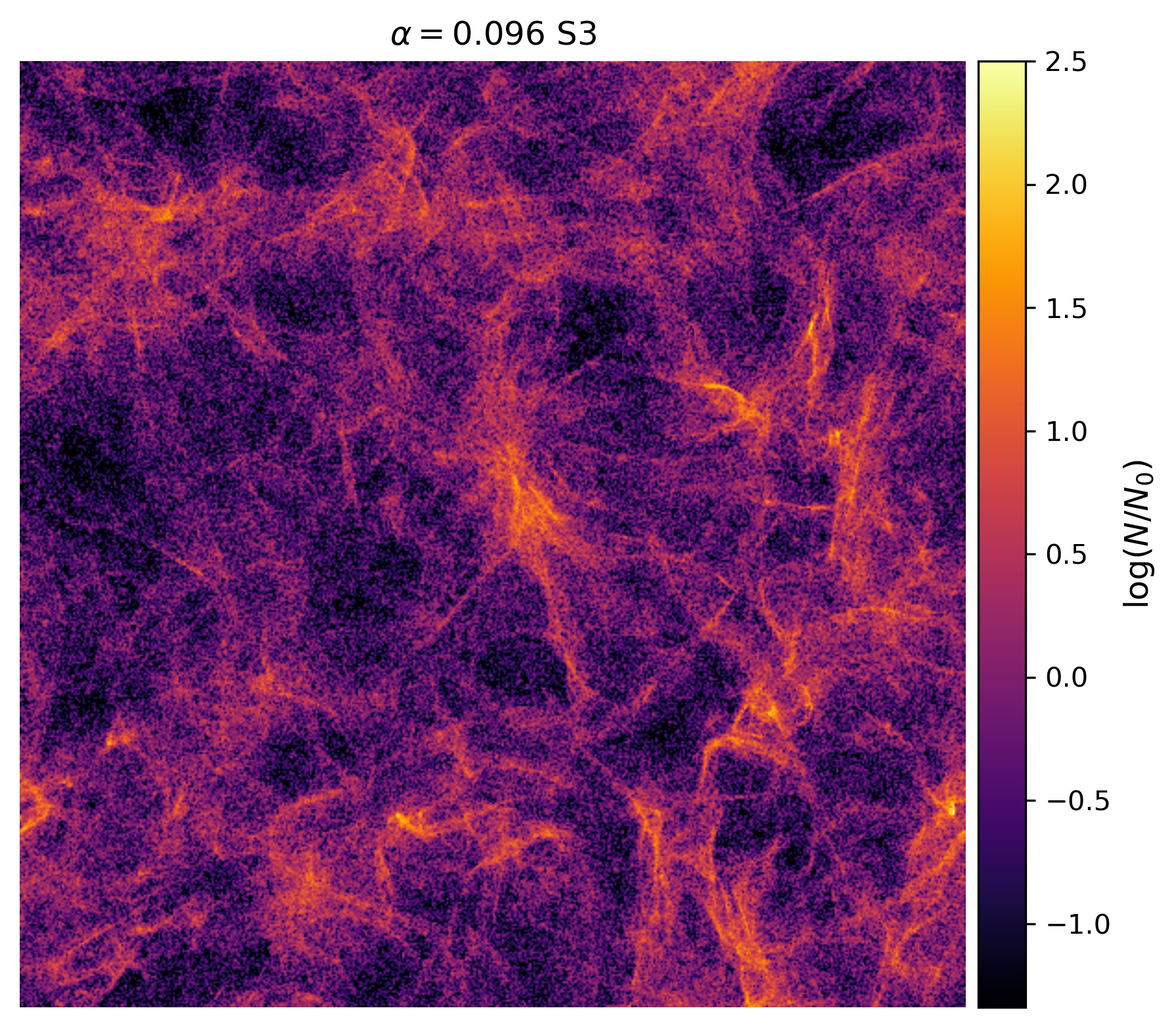}
	\includegraphics[trim=0.0cm 0cm 0.0cm 0.25cm, clip=true]{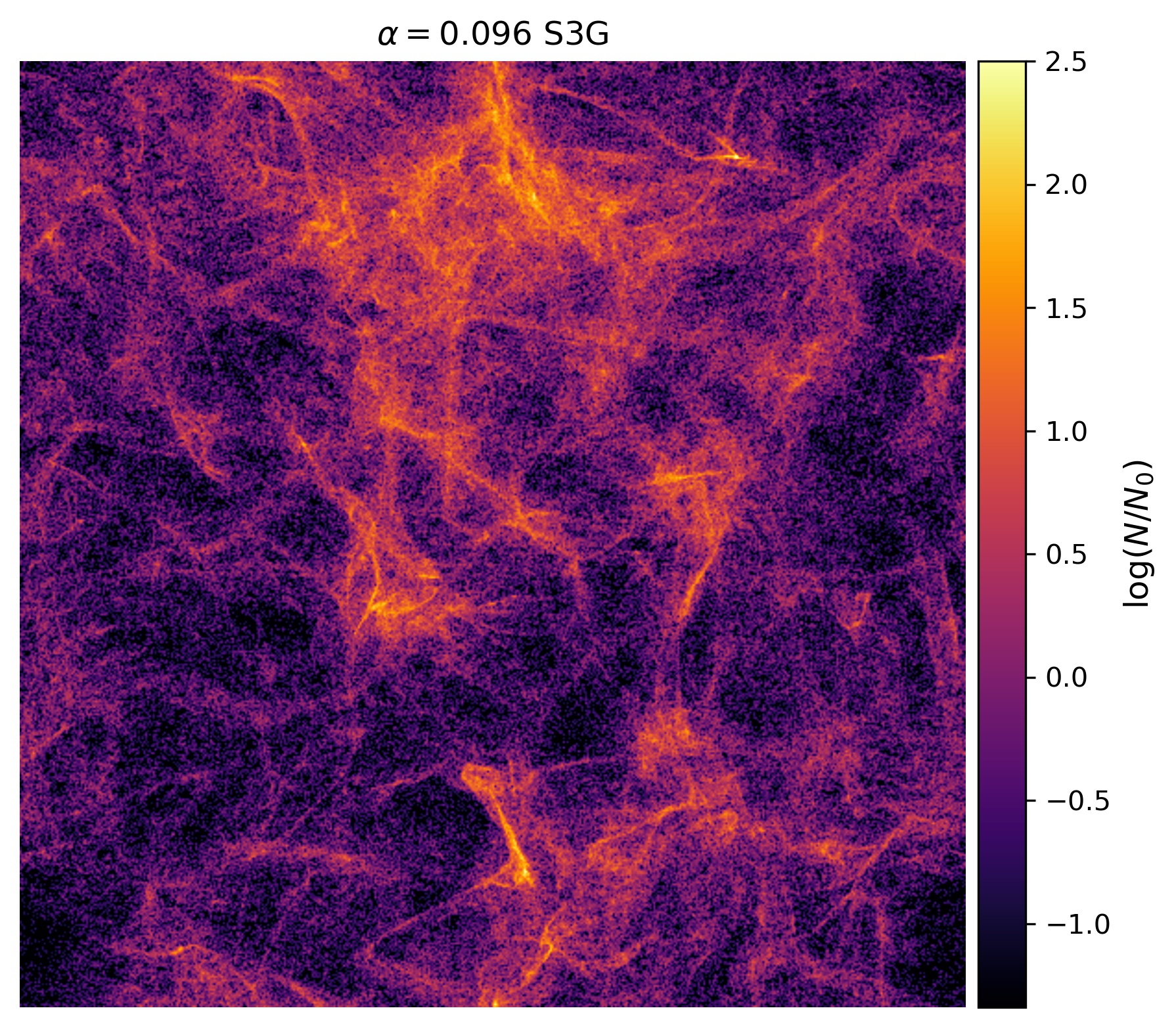}
    }
      \resizebox{0.9\hsize}{!}{
	\includegraphics[trim=0.0cm 0cm 2.7cm 0.25cm, clip=true]{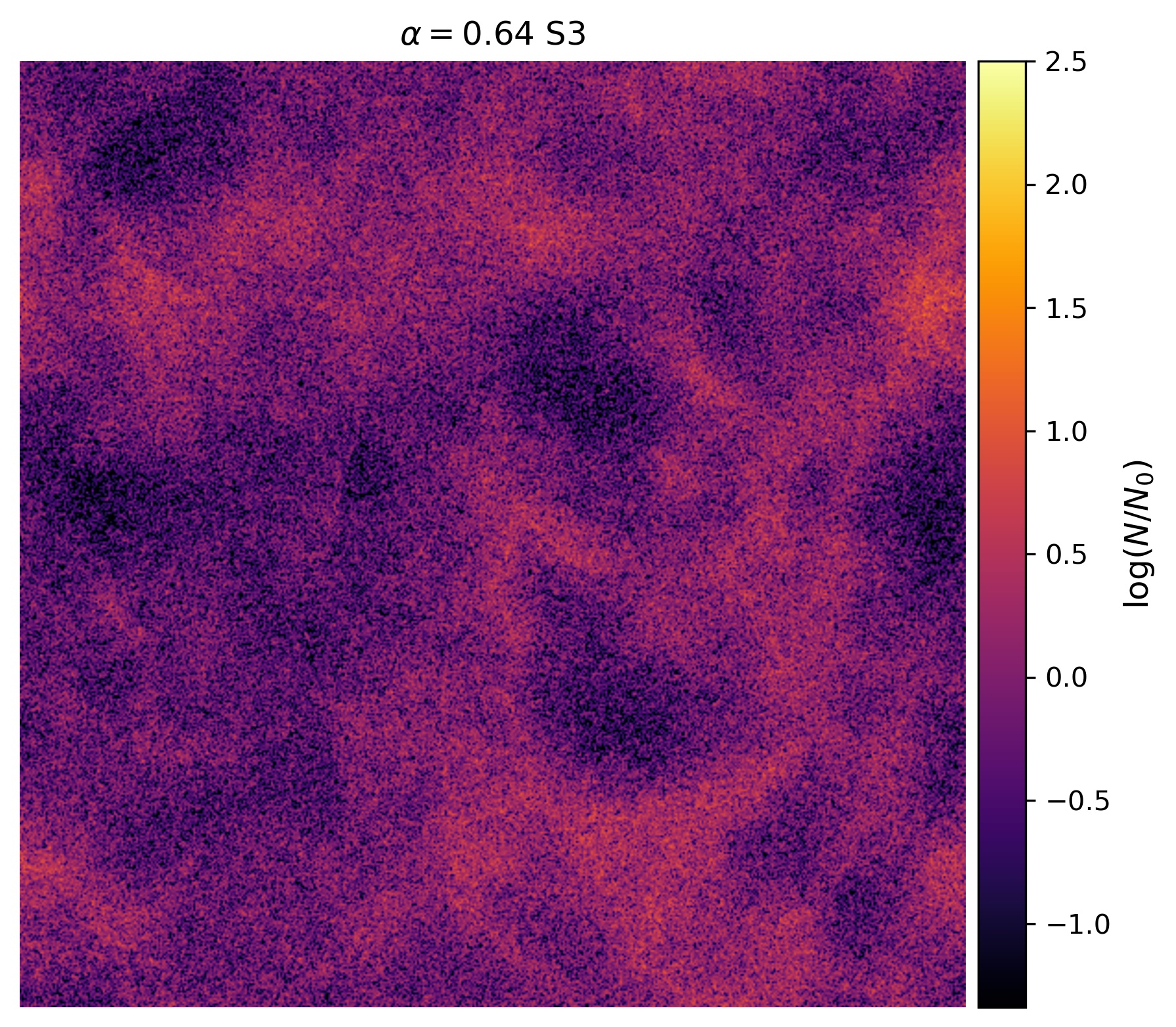}
	\includegraphics[trim=0.0cm 0cm 0.0cm 0.25cm, clip=true]{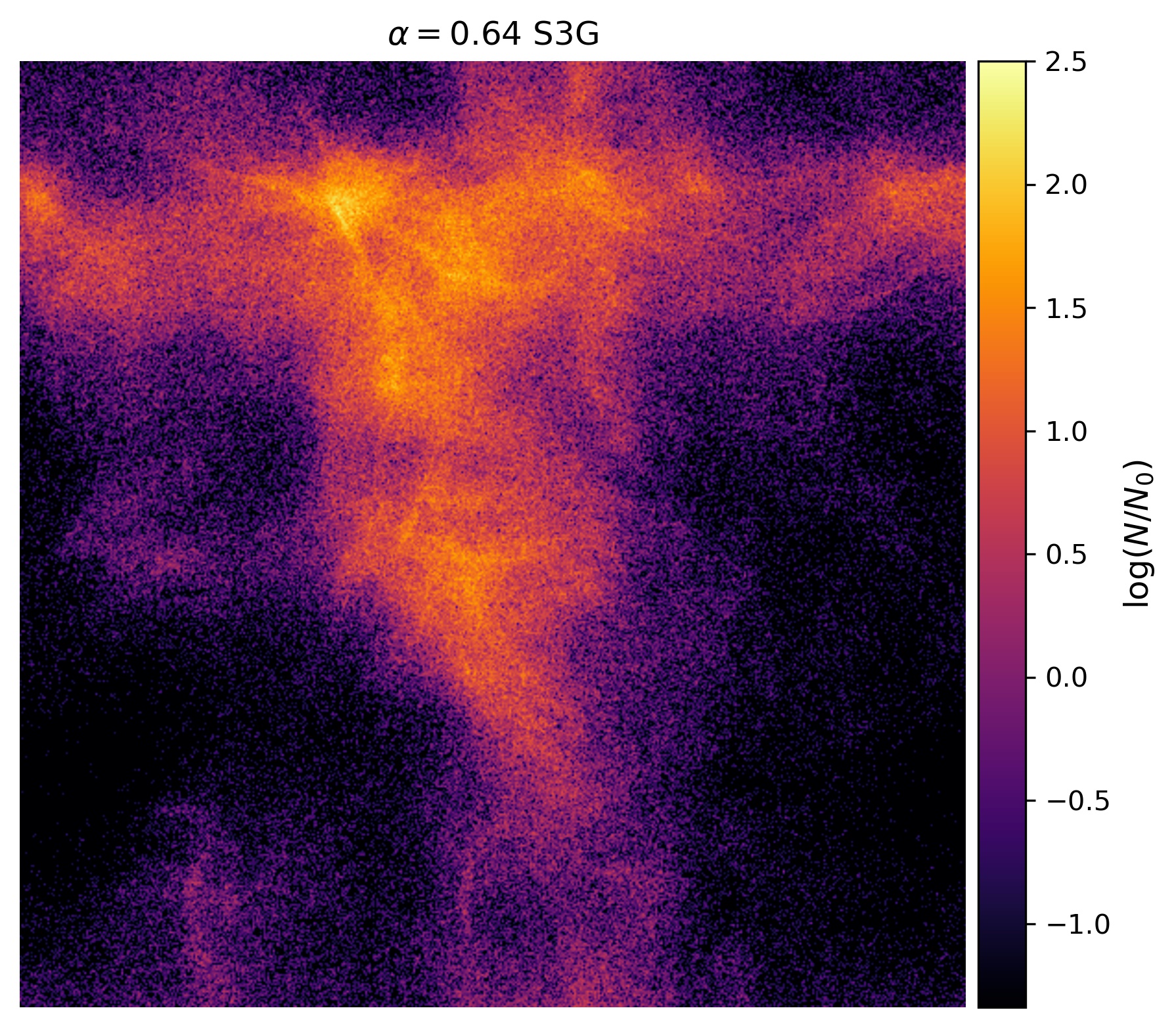}
    }
      \resizebox{0.9\hsize}{!}{
	\includegraphics[trim=0.0cm 0cm 2.7cm 0.25cm, clip=true]{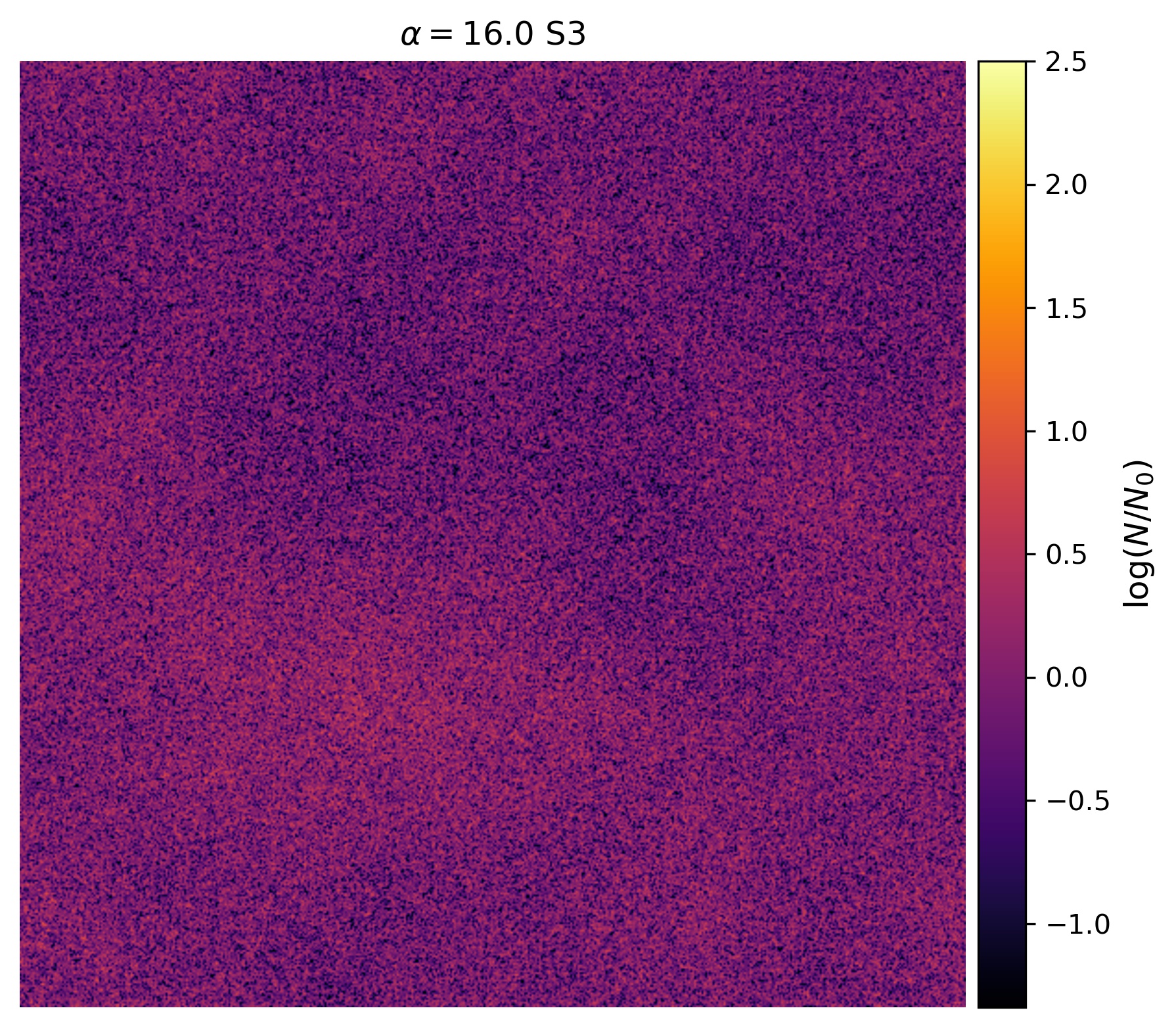}
	\includegraphics[trim=0.0cm 0cm 0.0cm 0.25cm, clip=true]{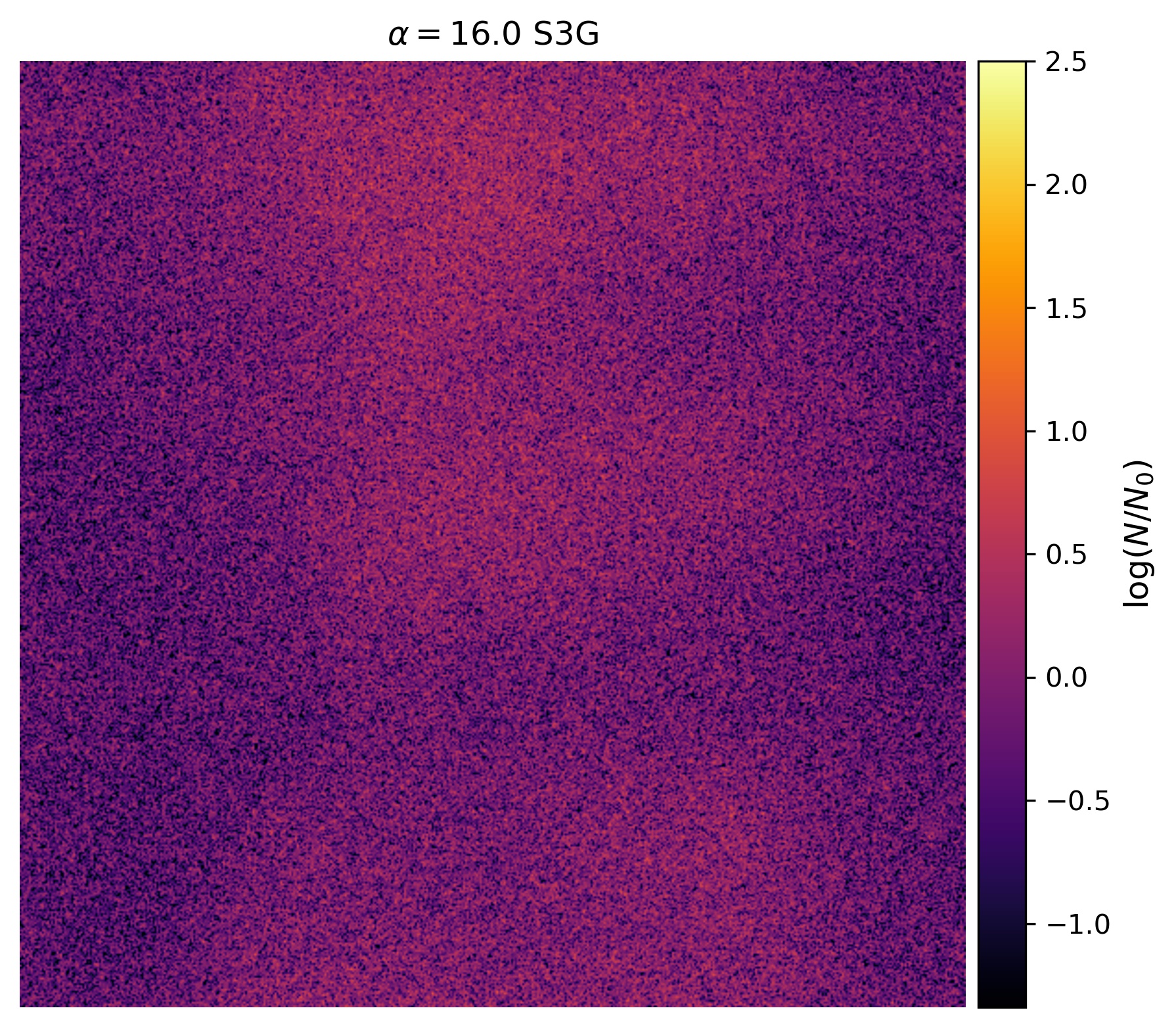}
    }
  \caption{\label{fig:coldens} Projected column (number) density  of particles with size-parameter $\alpha = 0.096$, $0.64$, $16.0$ (top, middle and bottom). Right panels shows the case with self-gravity (right) and the left panels without. Snapshots are taken at the end of the time series ($t/T = 81$).}
  \end{figure*}  
  
  \begin{figure*}
      \resizebox{\hsize}{!}{
	\includegraphics{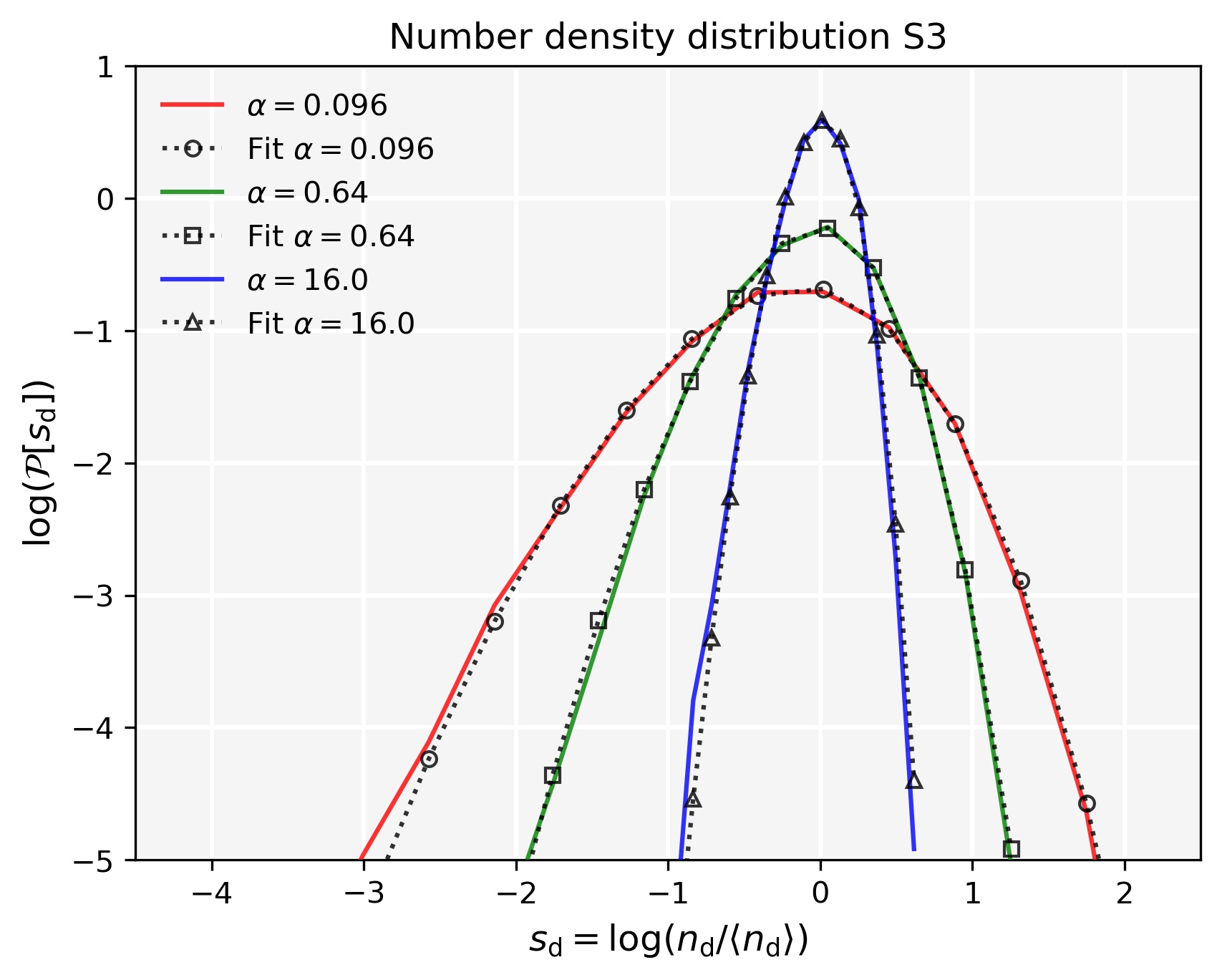}
	\includegraphics[trim=1.45cm 0cm 0cm 0.0cm, clip=true]{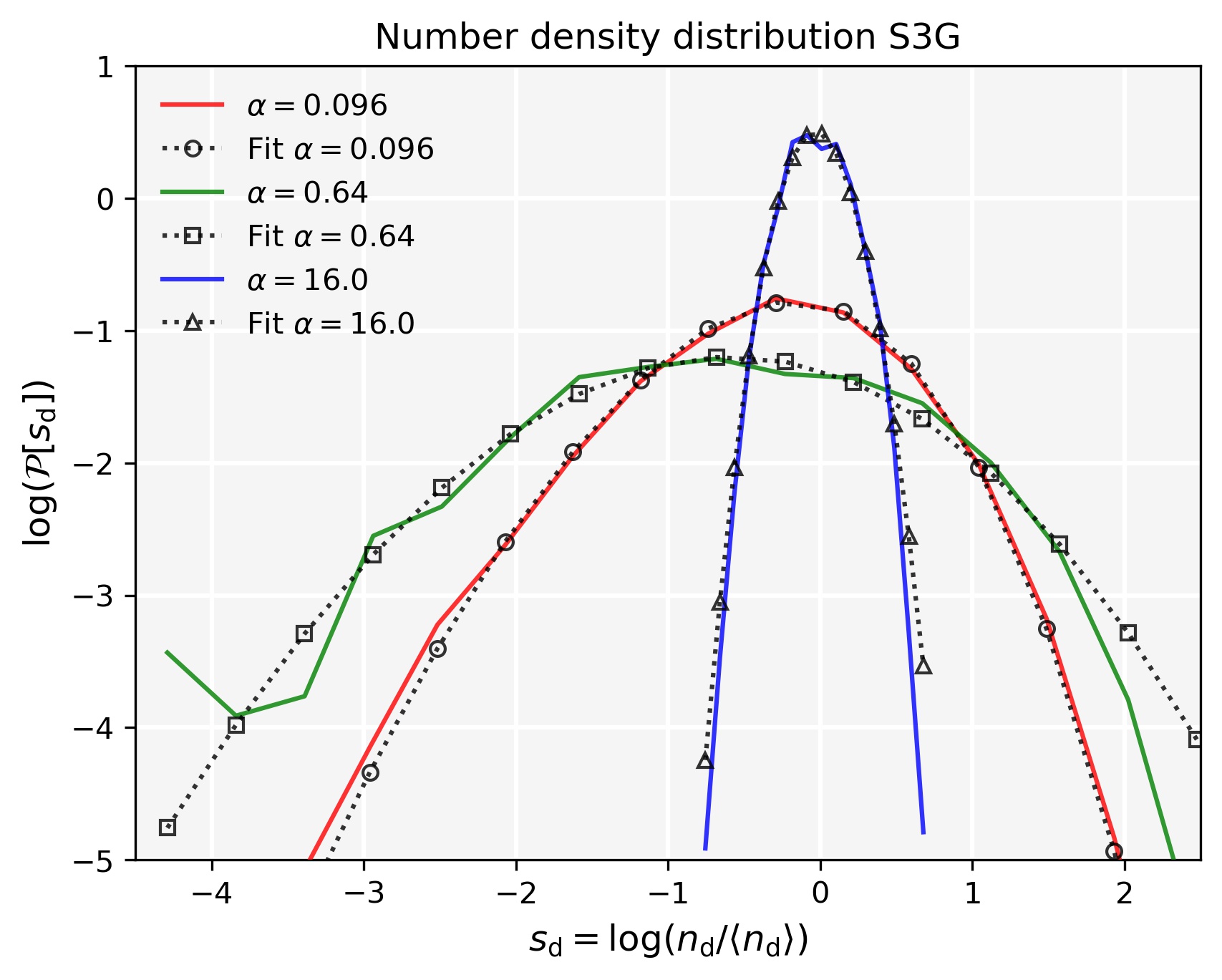}
    }
  \caption{\label{fig:nd_pdf} Probability density functions for three different grain sizes.} 
  \end{figure*} 
  
  \begin{figure*}
      \resizebox{\hsize}{!}{
	\includegraphics[trim=0cm 1.2cm 0.0cm 0.0cm, clip=true]{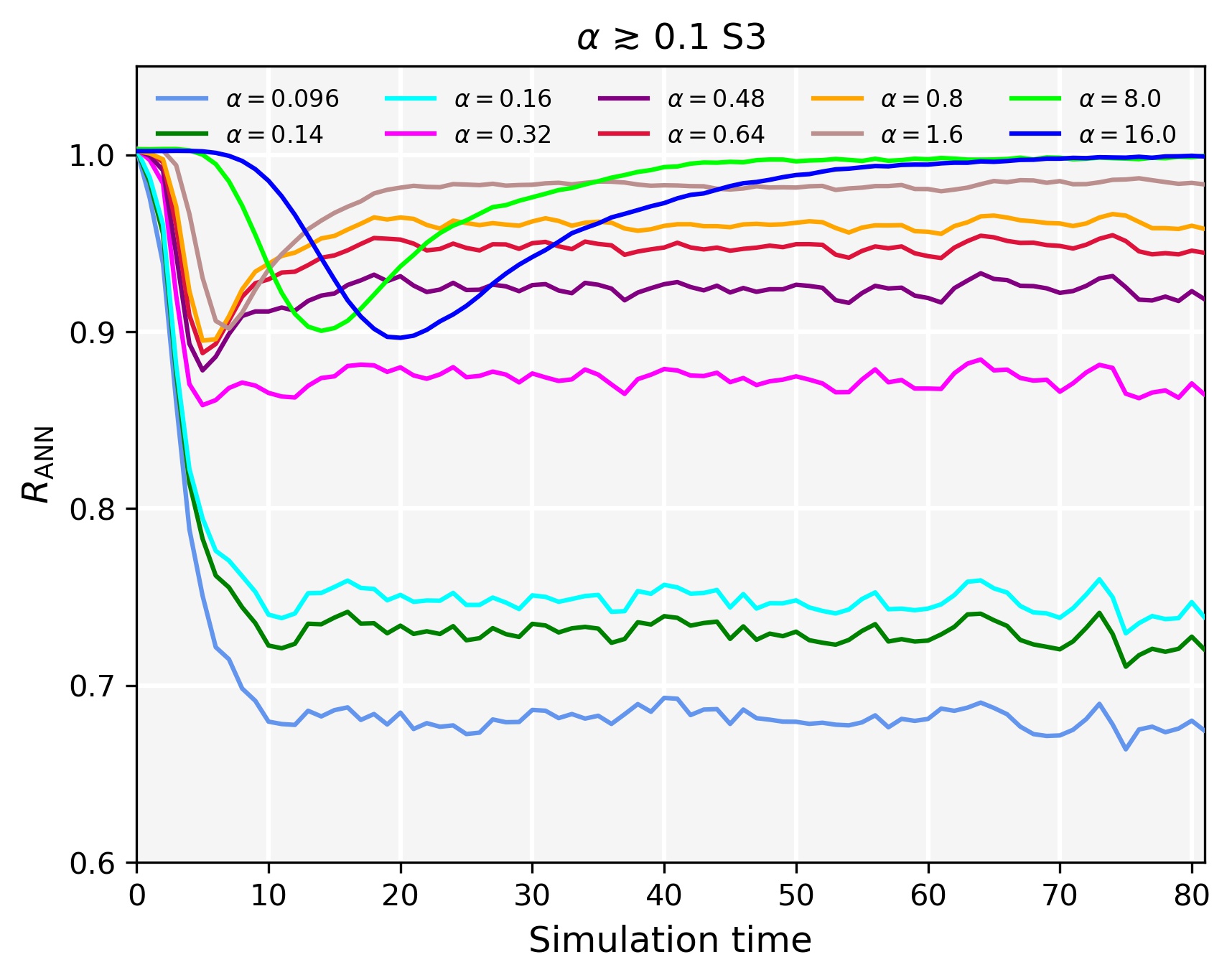}
	\includegraphics[trim=1.45cm 1.2cm 0.0cm 0.0cm, clip=true]{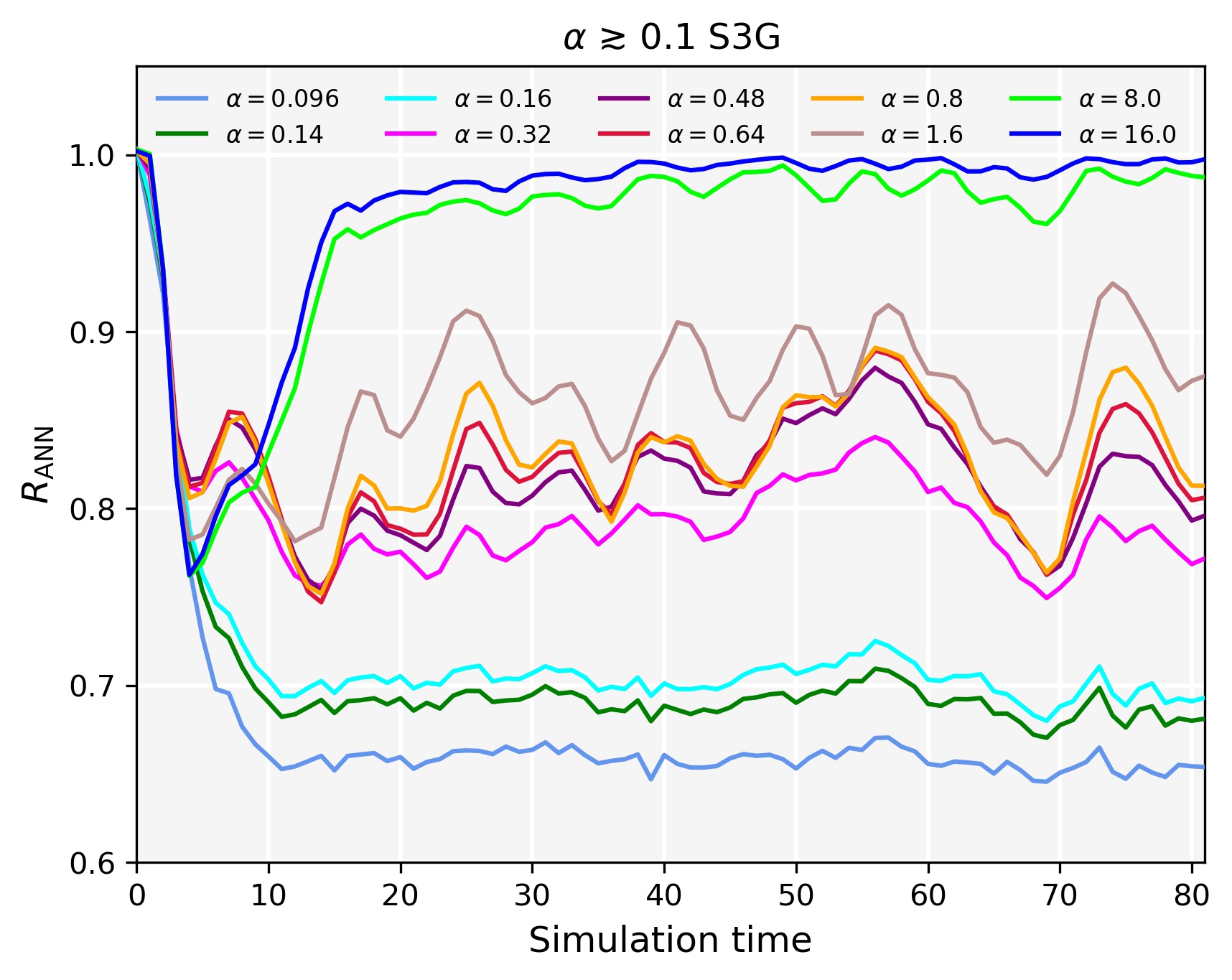}
	}\\[1mm]
    \resizebox{\hsize}{!}{
	\includegraphics[trim=0cm 1.2cm 0.0cm 0.6cm, clip=true]{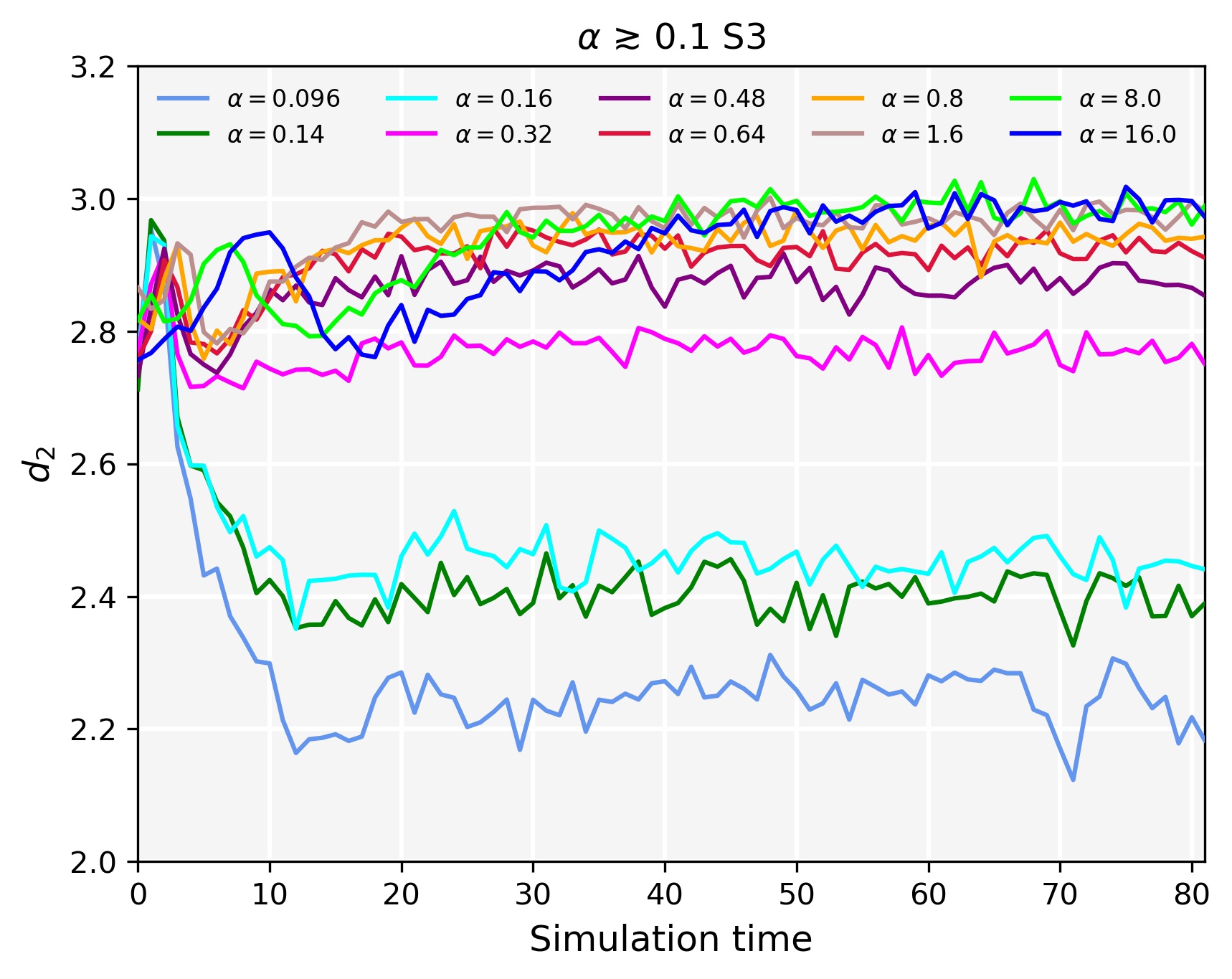}
	\includegraphics[trim=1.45cm 1.2cm 0.0cm 0.6cm, clip=true]{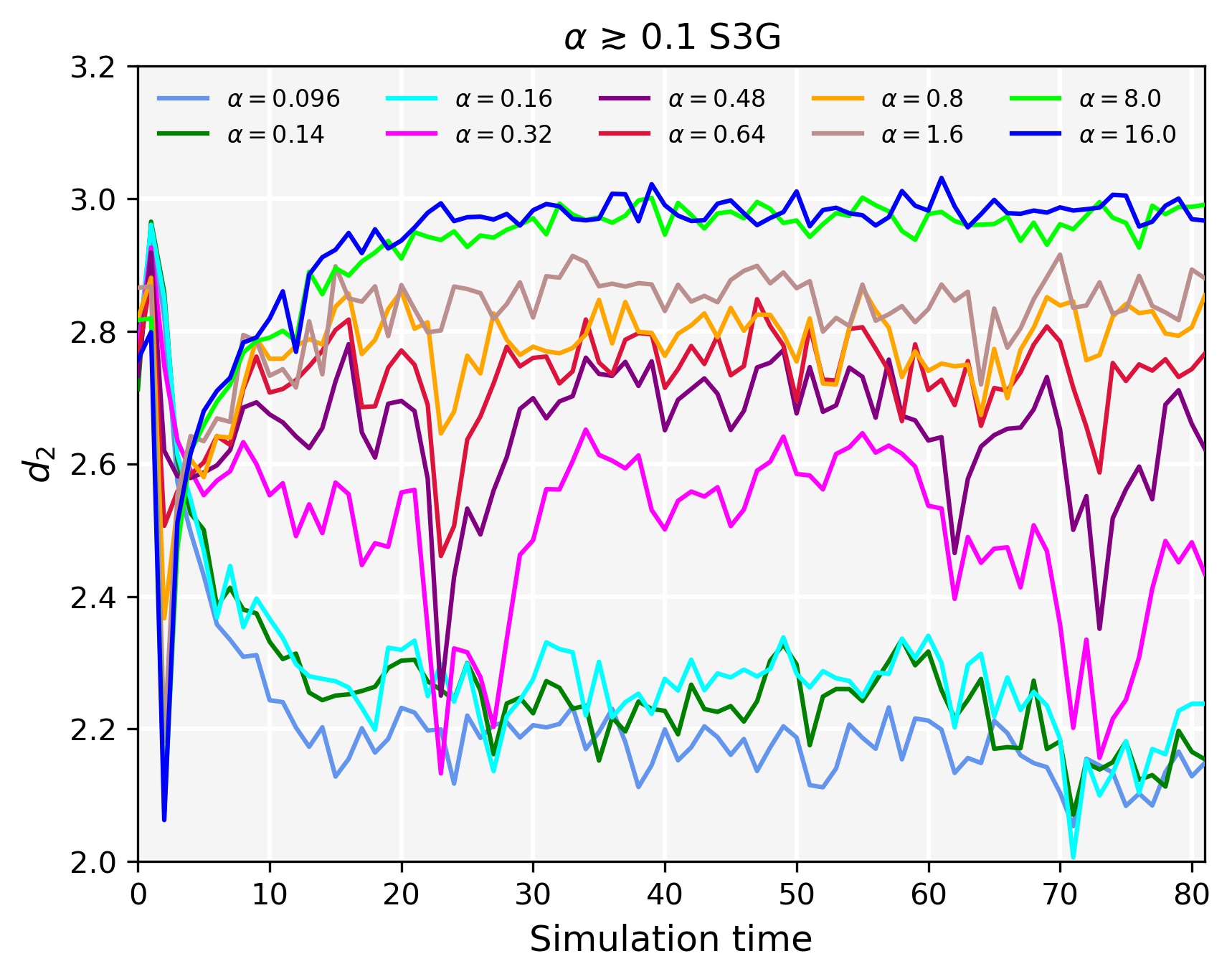}
	}\\[1mm]
	\resizebox{\hsize}{!}{
	\includegraphics[trim=0.0cm 0.33cm 0cm 0.6cm, clip=true]{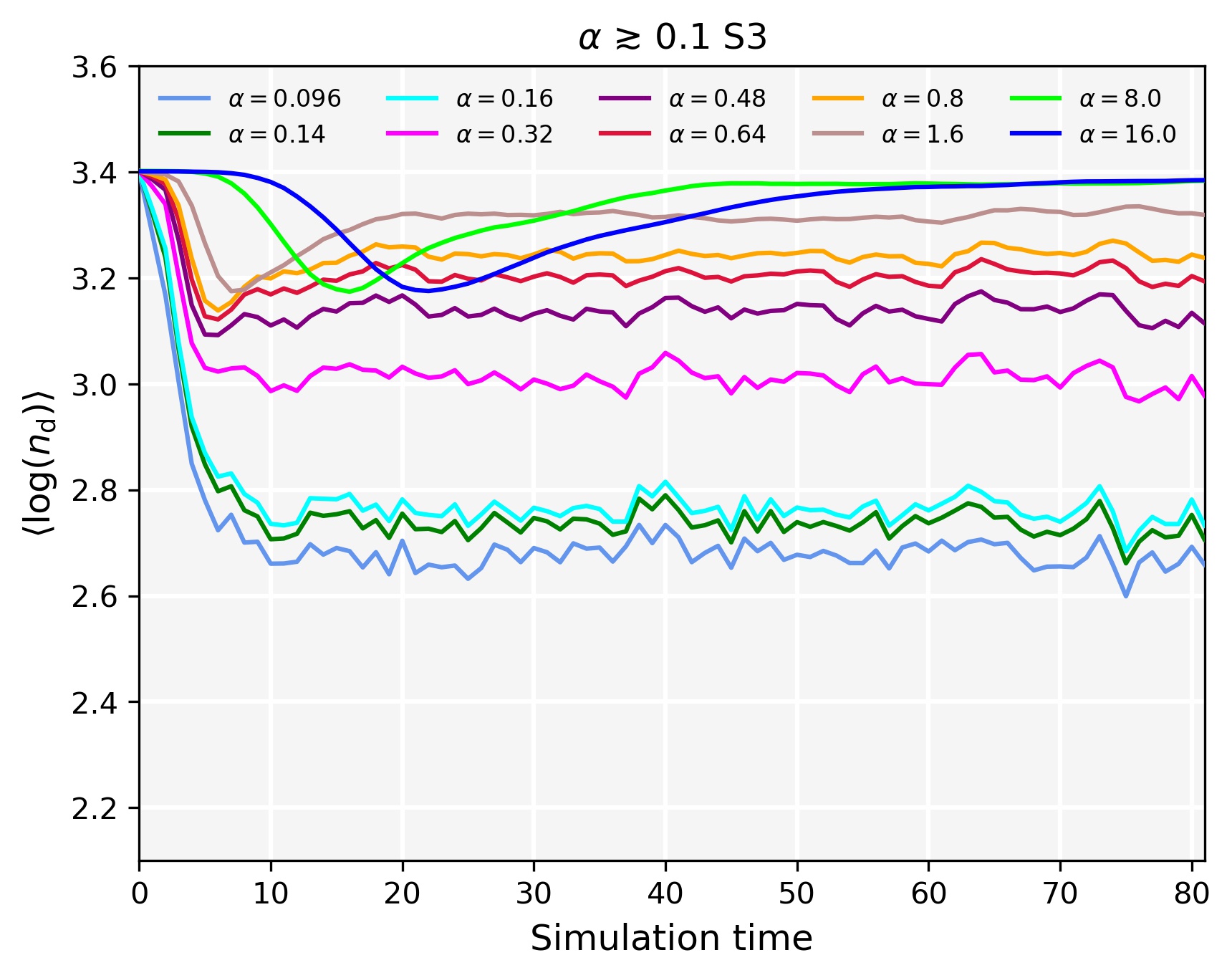}
	\includegraphics[trim=1.45cm 0.33cm 0cm 0.6cm, clip=true]{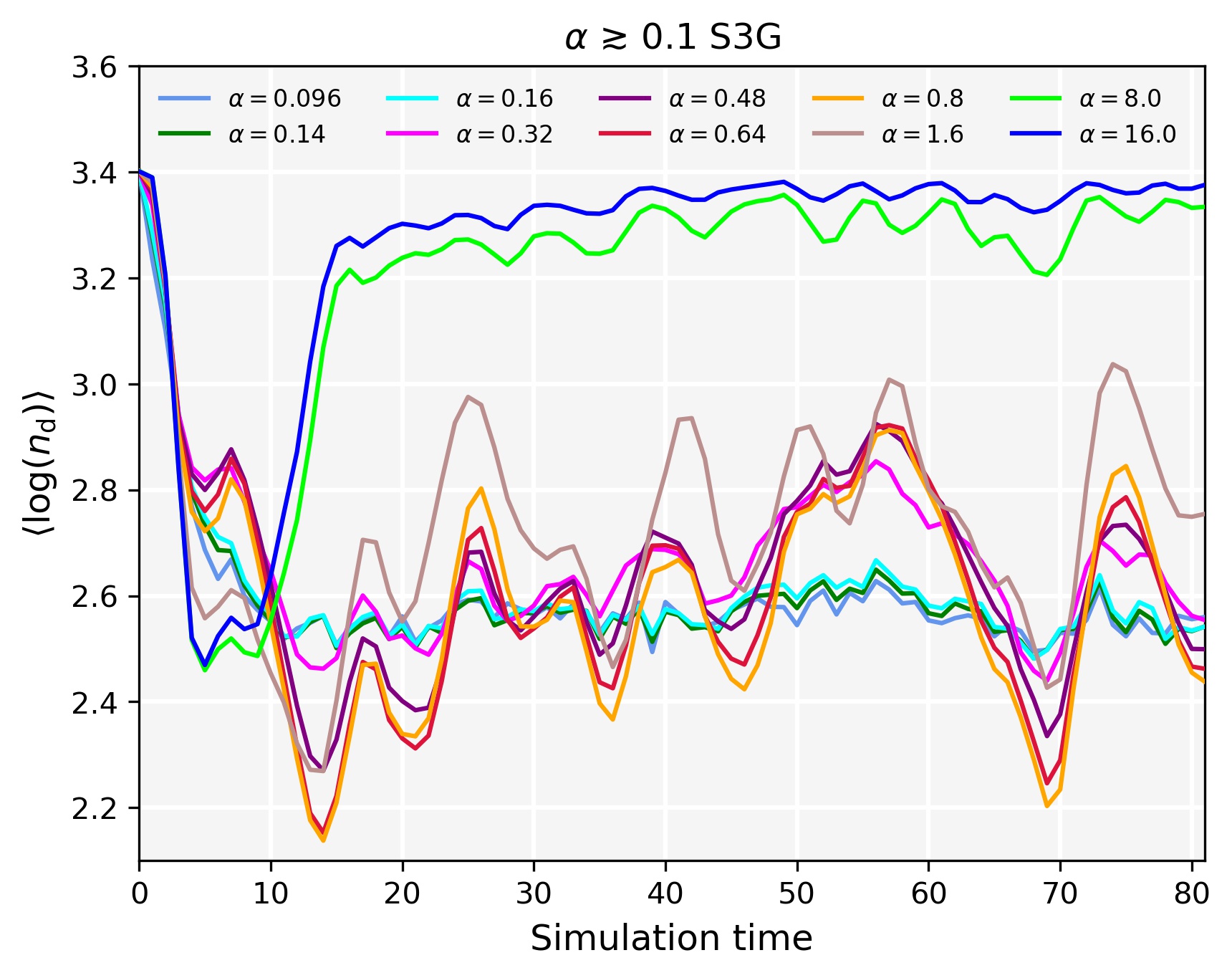}
    }
  \caption{\label{fig:clustering_ts} Average nearest-neighbour distance ratio ($R_{\rm ANN}$, upper panels), correlation dimension ($d_2$, middle panels) and mean logarithmic number density ($\mathcal{N}_{\rm d}=\langle \ln n_{\rm d}\rangle$, lower panels) of particles with size-parameter $\alpha \gtrsim 0.1$. Right panel shows the case with self-gravity and the left panel without.}
  \end{figure*}   
  
\subsubsection{Dust-density variance}
In Fig. \ref{fig:coldens} we show line-of-sight projections for dust with $\alpha = 0.096, 0.64, 16.0$ for S3 (left) and S3G (right). Small dust grains ($\alpha \lesssim 0.1$) tend to couple to the gas and are therefore clustered and distributed in patterns of similar character regardless of whether self-gravity is included or not. Intermediate sized dust grains (e.g., $\alpha = 0.64$) in SG3 shows significantly more clustering on both large and small scales compared to S3. As mentioned previously, $\alpha \sim 1$ corresponds to the grain-size regime where gravity and drag are competing forces. This is why the projected distributions of grains are so different when comparing the $\alpha = 0.64$ cases (middle panels in Fig. \ref{fig:coldens}; see also Appendix \ref{apx:coldens} for other values of $\alpha$). The largest grains in our simulations show generally very little number-density variance and are essentially randomly and homogeneously distributed  over the simulation domain (i.e., their distribution is essentially governed by a Poisson process) once a statistical steady state is reached. 

To quantify the above observations, we measured the PDF of the logarithmic dust-density parameter $s_{\rm d}=\log(n_{\rm d}/\langle n_{\rm d,\,0}\rangle)$ for three different grain sizes ($\alpha = 0.096, 0.64, 16.0$) by means of a simple binning-box method. We divided the simulation domain up in $32^3$ cube segments (binning boxes with size $L_{\rm bin} = 16$ grid points) and then simply counted the number of dust particles in each box to estimate the local number densities $n_{\rm d}$ in the simulation domain. From the resultant number densities we created a histograms, binning $s_{\rm d} = \ln(n_{\rm d}/n_{\rm d,\,0})$ in steps of $\Delta s_{\rm d}$ such that 16 bins were obtained, which are shown in Fig. \ref{fig:nd_pdf}. 

    \begin{table}
  \begin{center}
  \caption{\label{tab:ndPDFfit} Resultant fitting parameters, mean and 1-$\sigma$ deviations from fitting of analytical the distribution functions to the grain number-density PDFs obtained from the simulations.}
  \begin{tabular}{c|rrrrrrr}
  %\hline
  \hline\\[-1mm]
  \rule[-0.2cm]{0mm}{0.2cm}
   & & S3\,\, & & & & S3G \\
   $\alpha$ & 0.096 & 0.64 & 16.0 & & 0.096 & 0.64 & 16.0\\[2mm]
  \hline\\[-1mm]
   $\eta$ & -1.608 & -1.806 & -1.628 & & -1.76 & -0.875 & 0.748\\
   $\xi$ & 0.464 & 0.381 & 0.189 & & 0.505 & 0.166 & -0.157\\
   $\omega$ & 1.083 & 0.716 & 0.309 & & 1.242 & 1.56 & 0.276\\[2mm]
   $\langle s_{\rm d} \rangle$ & -0.27 & -0.119 & -0.021 & & -0.356 & -0.654 & -0.026\\ % Mean from fit
   $\sigma$ & 0.797 & 0.513 & 0.227 & & 0.894 & 1.327 & 0.242\\ % Std from fit
   $\mathcal{S}$ & -0.336 & -0.398 & -0.342 & & -0.383 & -0.101 & 0.069\\
   $\mathcal{K}$ & 0.204 & 0.256 & 0.209 & & 0.244 & 0.041 & 0.025\\[2mm]
  %\hline
  \hline
  \end{tabular}
  \end{center}
  \end{table}

With an exception for the $\alpha = 0.64$ grains in S3G, we can accurately describe the PDFs for $s_{\rm d}$ in Fig. \ref{fig:nd_pdf} as skewed lognormal distributions \citep{Azzalini85},
\begin{equation}
p(s_{\rm d}) = {1\over\sqrt{2\pi}\,\omega}\left\{1+{\rm erf}\left[{\eta(s_{\rm d}-\xi)\over \sqrt{2}\,\omega} \right] \right\}\,\exp\left[ -{(s_{\rm d}-\xi)^2\over 2\,\omega^2}\right],
\end{equation}
where $\eta$, $\xi$ and $\omega$ are fitting parameters. With $\delta = \eta/\sqrt{1+\eta^2}$ we can write the mean and variance as 
\begin{equation}
\langle s_{\rm d} \rangle = \xi + \omega\delta\sqrt{2\over\pi}, \quad 
\sigma^2 = \omega^2 \left(1-{2\over\pi}\delta^2\right),
\end{equation}
and skewness and kurtosis as
\begin{equation}
\mathcal{S} = {4 -\pi\over 2}{(\delta\sqrt{2/\pi})^3\over (1-{2/\pi}\delta^2)^{3/2}}, \quad 
\mathcal{K} = {2(\pi-3)(\delta\sqrt{2/\pi})^4\over (1-{2/\pi}\delta^2)^{2}}.
\end{equation}
The resultant fitting parameters are given in table \ref{tab:ndPDFfit}. The small grains (red lines in Fig. \ref{fig:nd_pdf}) have similar PDFs and thus show similar parameter values. The variance is somewhat larger in S3G compared to S3, but this may be attributed to the slightly larger gas-density variance (see section \ref{sec:gasdens}). Also the large grains ($\alpha = 16.0$) show similar PDFs in both S3 and S3G. The really interesting case is the $\alpha = 0.64$ grains, for which $\sigma$ is 2.6 times larger in S3G compared to S3. The mean value $\langle s_{\rm d}\rangle$ is also more than five times smaller (-0.654 compared to -0.119), which is due to the existence of large voids in the simulation domain with practically no $\alpha = 0.64$ grains.

\subsubsection{Clustering}
One of the main objectives of the present study is to follow up on our  previous work without self-gravity effects \citep{Mattsson19a,Mattsson19c} and determine the maximum degree of clustering and how clustering depends on $\alpha$ (grain size). In the case of incompressible turbulence, centrifuging of particles away from vortex cores leads to accumulation of particles in convergence zones \citep{Squires91,Eaton94,Bec05,Pumir16,Yavuz18}, which is different from the increased density of dust grains that is obviously expected due to compression of the gas and dust. 
Compressible turbulence can display significant shock formation, which may lead to another type of dust clustering: when two shocks interact, dust grains can get ``trapped'' in the high-density sheets and filaments that form via the shock-shock interaction \citep{Haugen21}. This leads to a different type of clustering, which may still show a fractal structure. Examples of such fractal clustering can be seen in the top panels of Fig. \ref{fig:coldens}.

Fig. \ref{fig:clustering_ts} shows how the average nearest-neighbour distance ratio $R_{\rm ANN}$, the correlation dimension $d_2$ and the mean logarithmic number density $\langle\log n_{\rm d} \rangle$ varies over time in S3 (left panels) and S3G (right panels). $R_{\rm ANN}$ for small grains ($\alpha \lesssim 0.1$) show similar evolution for both simulation S3 and S3G, which is due to the fact that such grains couple well to the flow. Initially, $R_{\rm ANN} = 1$, but rapidly declines and reaches its statistical steady-state value ($R_{\rm ANN}\sim 0.7$ for $\alpha \sim 0.1$) after about 10 simulation time units. For $\alpha > 0.1$ we see smaller $R_{\rm ANN}$ values in S3G compared to S3 once a steady state is reached, which is perhaps expected, given how the projected number density of grains differs between the two cases. What is more unexpected, however, is the oscillatory behaviour seen in $R_{\rm ANN}$ for $\alpha \sim 1$ grains in simulation SG3. This phenomenon occurs for the same $\alpha$ values that we previously pointed out as associated with a regime where drag and gravity forces compete. Thus, the ``oscillations'' are likely related to this competition between forces acting upon the grains, but the exact reason for the oscillatory behaviour remains unclear. Large grains ($\alpha \sim 10$) show,  qualitatively, similar behaviour in S3 and S3G. The evolution follows a pattern where $R_{\rm ANN}$ makes a distinct dip and then rises back up to $R_{\rm ANN} = 1$ again. But the timescale is clearly shorter in SG3, compered to S3, which we interpret as an effect of the much faster response to gravitational forcing. Instead of 40-50 time units, it takes only $\sim 15$ time units to get back to $R_{\rm ANN} = 1$ in S3G.

\begin{figure*}
      \resizebox{0.9\hsize}{!}{
      \includegraphics{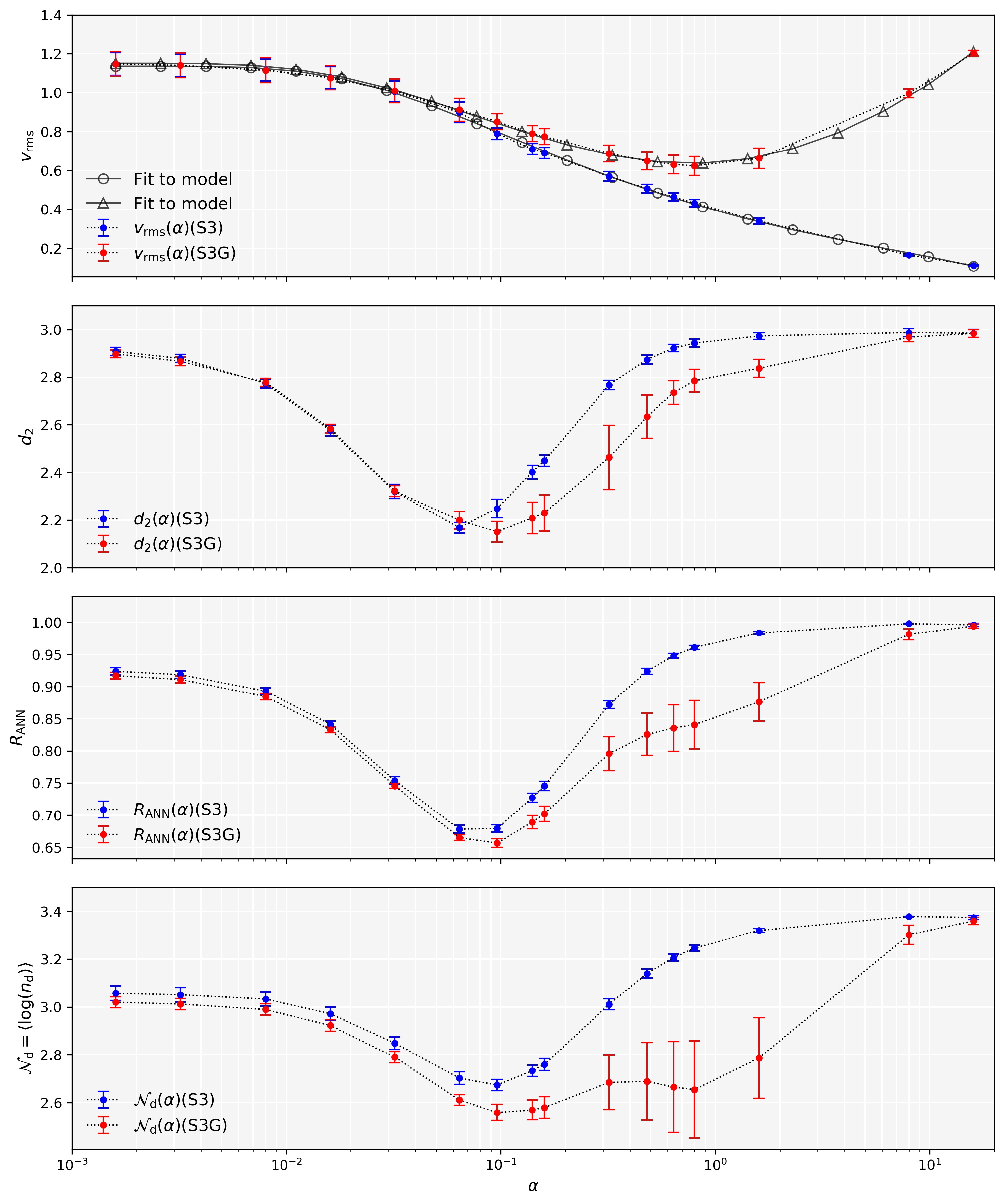}}

  \caption{\label{fig:alphaplot} Root-mean-square velocity ($v_{\rm rms}$), correlation dimension ($d_2$), average nearest-neighbour-distance ratio ($R_{\rm ANN}$) and mean logarithmic number density ($\mathcal{N}_{\rm d}=\langle \ln n_{\rm d}\rangle$) as functions of $\alpha$. The blue curves show the result of simulation S3, while the red curves show S3G. Note how $v_{\rm rms}$ for the two simulations diverge after $\alpha\sim 1$.}
  \end{figure*}

The correlation dimension $d_2$ shows a somewhat different evolution over time, compared to $R_{\rm ANN}$, but also similarities regarding grain-size dependence. A striking difference between S3 and S3G, is the prominent dips in $d_2$ seen for intermediate-size grain in S3G (see middle panels in Fig. \ref{fig:clustering_ts}). $\alpha = 0.32$ grains can temporarily reach $d_2\approx 2.2$, which is otherwise the value of $d_2$ associated with the clustering maximum occurring at $\alpha \approx 0.1$ (see Fig. \ref{fig:alphaplot}). For small grains $d_2$ follow a pattern very similar to that seen in $R_{\rm ANN}$, with an initial phase of decline and then moderate variation around a mean value, which is the case for both S3 and S3G. As expected, large grains ($\alpha \sim 10$) also follow the overall trend seen for $R_{\rm ANN}$ and evolve towards a statistical steady state with small variations around the ``no-clustering value'' $d_2 = 3$. Analogous to the evolution of $R_{\rm ANN}$, $d_2$ returns $d_2 = 3$ after initially showing lower values and the recovery from the initial dips are significantly faster in S3G compared to S3. It is noteworthy that in S3, $d_2$ goes up and down again after an initial decline, which is an artificial result owing to difficulties in measuring of $d_2$ in the early phase. Since calculation of $d_2$ is based on the 1-NND, or the radial distribution function, measurements of $d_2$ on a system too far from either an equilibrium or a statistical steady state will not produce meaningful values because the  distributions computed numerically are too noisy. 

Comparing the middle panels in Fig. \ref{fig:coldens}, one can see that having a diagnostic for large-scale clustering is important. We have therefore computed the mean logarithmic number density $\mathcal{N}_{\rm d}\equiv \langle \log n_{\rm d}\rangle$ for all snapshots we stored while running simulations. The result shows a time evolution of $\mathcal{N}_{\rm d}$ which very much resembles that of  $R_{\rm ANN}$. In particular the oscillatory behaviour for $\alpha \sim 1$ grains is possibly even more pronounced for $\mathcal{N}_{\rm d}$ (see bottom panels of Fig. \ref{fig:clustering_ts}). Since the change of $\mathcal{N}_{\rm d}$ with time is related to the convergence/divergence of a fictitious dust fluid representing the grains, i.e., the averaged continuity equation becomes
\begin{equation}
    {d \mathcal{N}_{\rm d}\over dt} = -\langle \nabla\cdot \mathbfit{v} \rangle,
\end{equation}
which means that $\mathcal{N}_{\rm d}$ can be seen as integrated mean convergence of the dust. For the gas flow we may assume $\langle \nabla\cdot \mathbfit{u} \rangle =0$ in case of a statistical steady state (see Appendix \ref{apx:sssgas}). That is, a compressible turbulent gas flow is divergence-free on average. Clustering of dust, on any scale, will correspond to a process which yields $\langle \nabla\cdot \mathbfit{v} \rangle < 0$ for the simulated domain. The cyclic behaviour of $\mathcal{N}_{\rm d}$ for intermediate sized grains thus suggests a situation similar to a fictitious dust fluid cycling between states of convergence and divergence. We believe this reflects competition between gravitational and drag acceleration, but the exact mechanism is yet to be understood. However, it is noteworthy that for $\alpha\sim 1$, $\mathcal{M}_{\rm rms}\sim 1$, we have that the ratio of the stopping time and the forcing time $\tau_{\rm s}/\tau_{\rm f} \sim 1$ (cf. Eq. \ref{eq:stint} and Fig. \ref{fig:alphaplot}), which may be an indication of a forcing-dependent phenomenon.

\subsubsection{Grain-size dependence}
We have seen in the previous sections how grain size is a key parameter. The velocity statistics as well as clustering properties once the statistical steady-state phase is reached, are distinct functions of $\alpha$. Fig. \ref{fig:alphaplot} shows $v_{\rm rms}$, $d_2$, $R_{\rm ANN}$ and $\mathcal{N}_{\rm d}$ as functions of $\alpha$, averaged over several snapshots\footnote{The averaging for different $\alpha$ is not done over the exact same snapshots. First, we need to ensure that a steady state has been reached in the dust dynamics, which happens after a significantly longer time for large grains. Second, we need to ensure that the time-averaged $u_{\rm rms}$ is essentially the same for the chosen intervals, which means we have adjusted the beginning and the end of each interval so that the time-averaged $u_{\rm rms}$ is as similar as possible. Without this careful choice of intervals there may otherwise be artificial ``kinks'' showing up in $v_{\rm rms}$ as a function of $\alpha$. In principle this is also the case for $d_2$, $R_{\rm ANN}$ and $\mathcal{N}_{\rm d} = \langle \ln n_{\rm d} \angle$, but for these quantities the effect is much smaller.}. In Fig. \ref{fig:alphaplot} we also include data from corresponding simulations with $\alpha < 0.1$, in which the grains couple well to the gas flow. The error bars in Fig. \ref{fig:alphaplot} display the $1 \sigma$ deviation from the mean, which is calculated only for the part of the time series in which the simulations have reached a statistical steady state.

The perhaps most striking difference between S3 and S3G is the upturn in $v_{\rm rms}$ for $\alpha\gtrsim 1$ (see, Fig. \ref{fig:alphaplot}, first panel from the top). When dust grains decouple from the gas, they are in our simulations only affected by gravitational pull from gas clumps (over-densities). The energy available for accelerating the grains is set by the gravitational potential energy, which by far exceeds the energy that can be transferred to dust grains via gas drag. More precisely, the maximum mean kinetic energy of dust which is coupled to the gas flow is $\langle E^{\rm d}_{\rm kin}\rangle \approx {1\over 2} \rho_{\rm d} \,u_{\rm rms}^2$, where $\rho_{\rm d}$ is the mass-density of dust. For decoupled grains, we have instead $\langle E^{\rm d}_{\rm kin}\rangle \sim \rho_{\rm d}\, \langle \nabla\Phi\rangle$, i.e., scaling with the mean gravitational potential energy. Thus, acceleration to mean velocities above $v_{\rm rms } = u_{\rm rms }$ should be possible, but we stress again that we need to add also the back-reaction of the dust, i.e., the drag exerted on the gas by large dust grains. 

Inspection of the bottom panel of Fig. \ref{fig:alphaplot} reveals another striking difference: the overall ($\mathcal{N}_{\rm d}$) clustering of grains in the size range $\alpha = 0.1\dots 10$ is significantly stronger in SG3 compared to S3. This increased clustering is also reflected in in $R_{\rm ANN}$ (third panel from the top) and, to a lesser degree, in $d_2$ (second panel from the top). Since $d_2$ primarily measures the fractal small-scale clustering, we conclude that a self-gravitating gas mainly causes large-scale clustering, although all considered diagnostics for clustering are affected to some degree. The error bars ($1\sigma$ deviation) of $d_2$, $R_{\rm ANN}$ and $\mathcal{N}_{\rm d}$ near $\alpha \sim 1$ is considerably larger in S3G than in S3. This is an effect of the oscillatory behaviour seen Fig. \ref{fig:clustering_ts}. Although the variance is too large to say anything conclusive, there seems to be a second minimum in $\mathcal{N}_{\rm d}$ near $\alpha \sim 1$, which coincides with the minimum of $v_{\rm rms}$ (upper panel of Fig. \ref{fig:alphaplot}).

The stochasticity of the gravitational potential in a turbulent gas, where the energy injected by the driving force is sufficient to avoid collapsing regions, is considerable and becomes a random forcing on the dust.
Consequently, the largest grains ($\alpha = 16$), which are almost completely decoupled from the gas flow, show essentially only random motion and very little clustering. This means that $d_2 = 3$ and $R_{\rm ANN} = 1$ in S3G, just as in S3. It also means that grain-grain collisions in case of decoupling are not enhanced by clustering, but the relative velocities of interacting grains. We shall therefore have a closer look at how the collision kernel is affected. 

\subsubsection{Effects of self-gravity of the gas on the grain-grain collision rate}
A mono-dispersed population of dust grains in a vigorously turbulent gas has a local collision kernel
$C = 2\upi\, a^2\,g(a)\,\Delta\mathbfit{v}$,
where $a$ is the radius of the grains, $g$ is the radial distribution function (of grains surrounding a reference grain) and $\Delta\mathbfit{v}$ is the relative velocity of the colliding grains. As shown by, e.g., \citet{Li21}, the ensemble average of such a kernel is 
\begin{equation}
\langle C \rangle = \sqrt{16\upi\over 3} a^2\,v_{\rm rms}(a), 
\end{equation} 
which implies that ${\langle C \rangle_{\rm S3G} / \langle C\rangle_{\rm S3}} \approx 5$ if $a$ and the scaling of the simulations are chosen such that $\alpha \approx 10$ (see Fig. \ref{fig:alphaplot}). That is, the large-particle collision rate can be a few times higher with self-gravity included. 

The mean collision rate (per unit volume) can be estimated as $\langle R\rangle \sim \langle n_{\rm d}^2\rangle \langle C\rangle$. For $\alpha \sim 1$ grains we have a situation where $\langle n_{\rm d}^2\rangle > \langle n_{\rm d}\rangle^2$, while ${\langle C \rangle_{\rm S3G} / \langle C\rangle_{\rm S3}} \approx 1.5$. The net effect is that intermediate sized grains ($\alpha \sim 1$) may have collision rates similar to those of large grains ($\alpha \sim 10$). The effects on $\langle C\rangle$ suggests that addition of self-gravity in simulation of coagulation in compressible turbulence such as those by \citet{Li21} will lead to more efficient coagulation, but likely also an increased rate of grain shattering as large grain may collide with significantly higher energies.

\section{Summary and conclusions}
We have studied the dynamics and clustering of dust grains in an ISM context using 3D periodic-boundary box simulations of stochastically forced, transonic steady-state turbulence, with and without self-gravity (S3G and S3, respectively). The S3G simulation is nearly Jeans-unstable with a box-side length $L_{\rm box}=\lambda_{\rm J}$, where $\lambda_{\rm J}$ is the Jeans wavelength, which enables a statistical steady state without irreversible gravitational collapses. From the simulations we draw the following conclusions:
\begin{itemize}
\item Large grains with high inertia, which more or less decouple from the gas, will be accelerated to much higher mean velocities $v_{\rm rms}$ if suspended in a compressible self-gravitating turbulent gas. The difference is entirely due to the fact that when the kinetic drag is inefficient, gravity is the main force acting on the particles. 
\item The stochasticity of the potential gives rise to a random acceleration of dust grains and thus negligible correlation with the velocity field of the gas.
\item Maximal clustering occurs at similar grain sizes in both S3 and S3G ($\alpha \sim 0.1$). At somewhat larger grain sizes ($\alpha \sim 1$), S3G displays more clustering than S3. In particular, there seems to be a sign of a second minimum in the mean logarithmic number density $\mathcal{N}_{\rm d}$. 
\item The divergence of $v_{\rm rms}$ vs. $\alpha$ for S3 and S3G appears to begin at the point of maximal clustering (minimum $d_2$, which occurs at $\alpha \sim 0.1$) and becomes tangible in the size range where S3G display a clustering excess compared to S3 ($\alpha \sim 1$).
\item In the transonic regime considered here, the collision rate for large-inertia particles increase at least an order of magnitude. This is of particular importance for the formation of large dust aggregates.
\end{itemize}
The results presented here suggest the need for future studies on simulating aggregation/coagulation including self-gravity in the nearly Jeans-unstable regime. This is a rather delicate task, which will require a considerable computational effort, however. Less computationally expensive extensions of the present work would be to study a minimally forced system as well as the Jeans-unstable regime.

\section*{Acknowledgments}
The authors thank the anonymous reviewer for his/her concise and constructive report.
This project is supported by the Swedish Research Council (Vetenskapsrådet), grant no. 2015-04505. 
Nordita funding is shared between the Nordic Council of Ministers, the Swedish Research Council, the two host universities KTH Royal Institute of Technology and Stockholm University, and Uppsala University. Our simulations were performed using computational resources provided by the Swedish National Infrastructure for Computing (SNIC) at the PDC Center for High Performance Computing, KTH Royal Institute of Technology in Stockholm.

\section*{Data Availability}

The data underlying this article will be made available upon request. The \textsc{Pencil Code} is open source and available at \href{https://github.com/pencil-code/pencil-code}{\texttt{https://github.com/pencil-code/pencil-code}}.

\bibliographystyle{mnras}
\bibliography{refs_dust}
%~\\[9cm]
%\newpage
\appendix
\section{Implications of a statistical steady state for gravoturbulent gas}
\label{apx:sssgas}
Averaging the equation of continuity for the gas leads to
\begin{equation}
    {{\rm d}\langle \ln\rho\rangle\over{\rm d}t} = {\partial \langle\rho\rangle\over \partial t} + \langle\mathbfit{u}\cdot\nabla \ln\rho\rangle = -\langle \nabla\cdot \mathbfit{u}\rangle = 0,
\end{equation}
where the fact that $\langle \ln\rho\rangle$ is a constant (mass is conserved) requires that $\langle \nabla\cdot \mathbfit{u}\rangle = 0$ and $\langle\mathbfit{u}\cdot\nabla \ln\rho\rangle = 0$ in case of a statistical steady state. 

For the equation of motion, assuming negligible viscosity (high Re) and statistical steady state, we have
\begin{equation}
    {{\rm d}\langle \mathbfit{u}\rangle\over{\rm d}t} = -\left\langle{\nabla p\over \rho}\right\rangle -\langle \nabla\Phi\rangle + \left\langle {\mathbfit{F}_{\rm force}\over \rho}\right\rangle = 0.
\end{equation}
The turbulence forcing ($\mathbfit{F}_{\rm force}$) is conservative and has a vanishing mean. Thus, we have a quasi-hydrostatic (statistical) equilibrium,
\begin{equation}
\left\langle{\nabla p\over \rho}\right\rangle = -\langle \nabla\Phi\rangle,
\end{equation}
which in case of an isothermal condition implies that
\begin{equation}
c_{\rm s}^2\,{\nabla \langle\ln\rho}\rangle = -\langle \nabla\Phi\rangle = 0.
\end{equation}
That is, in our simulations, we can assume there is essentially no net acceleration of the gas due to the inclusion of self-gravity.

Scalar multiplication of the (isothermal) equation of motion with $\mathbfit{u}$ before averaging yields
\begin{equation}
\label{eq:scalarmult_u_gas}
    \left\langle\mathbfit{u}\cdot{{\rm d} \mathbfit{u}\over{\rm d}t}\right\rangle = -c_{\rm s}^2\,{\langle\mathbfit{u}\cdot\nabla \ln\rho}\rangle -\langle \mathbfit{u}\cdot\nabla\Phi\rangle + \left\langle {\mathbfit{u}\cdot\mathbfit{F}_{\rm force}\over \rho}\right\rangle = 0,
\end{equation}
where we note that the second term on the right-hand-side is zero (see the paragraph about the continuity equation above). Thus, equation (\ref{eq:scalarmult_u_gas}) redices to
\begin{equation}
 \langle \mathbfit{u}\cdot\nabla\Phi\rangle = \left\langle {\mathbfit{u}\cdot\mathbfit{F}_{\rm force}\over \rho}\right\rangle.
\end{equation}
Without self-gravity ($\nabla\Phi = 0$) it is clear that the forcing term must also be vanishing, i.e., $\langle {\mathbfit{u}\cdot\mathbfit{F}_{\rm force}/ \rho}\rangle = 0$. Since the velocity statistics of the our simulations with (S3G) and without (S3) self-gravity is similar (see Fig. \ref{fig:machPDF}) we have good reasons to assume that $\langle {\mathbfit{u}\cdot\mathbfit{F}_{\rm force}/ \rho}\rangle \approx 0$ is a fair assumption in general, as long as the self-gravity effects are not altering the mean-flow properties significantly. Moreover, we would also have $\langle \mathbfit{u}\cdot\nabla\Phi\rangle \approx 0$, which is assumed throughout our analysis in the present paper.

\section{Derivation of equation (8)}
\label{apx:dustacceq}
Starting from the EOM for the dust, we can derive a ``dust acceleration equation'', which describes how the kinetic drag and the self-gravity of the gas determine the steady-state mean velocity of the dust. In order for the derivation below to make sense, both $\mathbfit{u}$ and $\mathbfit{v}$ must be regarded as vector fields, which more or less requires that we treat the dust as a fluid, as mentioned in section \ref{sec:dustfluid}.

Scalar multiplication of equation (\ref{eom_dust2}) with $\mathbfit{v}$ and $\mathbfit{u}$ yields, respectively,
\begin{equation}
\label{eq:scalarmult1}
\mathbfit{v}\cdot{{\rm d}\mathbfit{v}\over{\rm d}t} = {1\over \tau_{\rm s}}\,(\mathbfit{v}\cdot\mathbfit{u}-\mathbfit{v}^2) - \mathbfit{v}\cdot \nabla\Phi,
\end{equation}
\begin{equation}
\label{eq:scalarmult2}
\mathbfit{u}\cdot{{\rm d}\mathbfit{v}\over{\rm d}t} = {1\over \tau_{\rm s}}\,(\mathbfit{u}^2-\mathbfit{v}\cdot\mathbfit{u}) - \mathbfit{u}\cdot \nabla\Phi.
\end{equation}
By combination of eqns. (\ref{eq:scalarmult1}) and (\ref{eq:scalarmult2}) above, we obtain
\begin{equation}
\label{eq:scalarmult_comb}
\mathbfit{v}\cdot{{\rm d}\mathbfit{v}\over{\rm d}t} + \mathbfit{u}\cdot{{\rm d}\mathbfit{v}\over{\rm d}t}= {\mathbfit{u}^2 -   \mathbfit{v}^2\over \tau_{\rm s}} - \mathbfit{v}\cdot \nabla\Phi - \mathbfit{u}\cdot \nabla\Phi.
\end{equation}
Assuming equilibrium drag/drift, we have
\begin{equation}
    {{\rm d}\mathbfit{v}\over{\rm d}t} = {{\rm d}\mathbfit{u}\over{\rm d}t} \quad\rightarrow\quad \mathbfit{u}\cdot{{\rm d}\mathbfit{v}\over{\rm d}t} = \mathbfit{u}\cdot{{\rm d}\mathbfit{u}\over{\rm d}t},
\end{equation}
which leaves us with 
\begin{equation}
\label{eq:scalarmult_comb2}
\mathbfit{v}\cdot{{\rm d}\mathbfit{v}\over{\rm d}t} + \mathbfit{u}\cdot{{\rm d}\mathbfit{u}\over{\rm d}t}= {\mathbfit{u}^2 -   \mathbfit{v}^2\over \tau_{\rm s}} - \mathbfit{v}\cdot \nabla\Phi - \mathbfit{u}\cdot \nabla\Phi.
\end{equation}
Consider now the EOM for the (isothermal) gas and multiply by $\mathbfit{u}$,
\begin{equation}
\label{eq:scalarmult_gas}
 \mathbfit{u}\cdot{{\rm d}\mathbfit{u}\over{\rm d}t}= -c_{\rm s}^2 \,\mathbfit{u}\cdot\nabla\ln\rho - \mathbfit{u}\cdot \nabla\Phi + \mathbfit{u}\cdot\mathbfit{f}_{\rm force},
\end{equation}
where the last term is due to forcing. We not that the mean of this term must be vanishing. Inserting eq. (\ref{eq:scalarmult_gas}) into eq. (\ref{eq:scalarmult_comb2}) we find that
\begin{equation}
\label{eq:scalarmult_comb3}
\mathbfit{v}\cdot{{\rm d}\mathbfit{v}\over{\rm d}t} = {\mathbfit{u}^2 -   \mathbfit{v}^2\over \tau_{\rm s}} - \mathbfit{v}\cdot \nabla\Phi + c_{\rm s}^2 \,\mathbfit{u}\cdot\nabla\ln\rho -  {\mathbfit{u}\cdot\mathbfit{F}_{\rm force}\over\rho}.
\end{equation}
Averaging this equation results in a ``dust-acceleration equation'' of the form
\begin{equation}
\label{eq:DAE}
    {1\over 2}{{\rm d} v_{\rm rms}^2\over {\rm d}t} = {u_{\rm rms}^2 - v_{\rm rms}^2 \over \tau_{\rm s}} - \langle\mathbfit{v}\cdot \nabla\Phi \rangle + c_{\rm s}^2 \,\langle \mathbfit{u}\cdot\nabla\ln\rho\rangle,
\end{equation}
since the forcing term vanishes by definition. The last term, which arises from gas pressure, will in fact also vanish (see Appendix \ref{apx:sssgas}). 
With $\langle \mathbfit{u}\cdot\nabla\ln\rho \rangle=0$ we then have
\begin{equation}
    {1\over 2}{{\rm d} v_{\rm rms}^2\over {\rm d}t} = {u_{\rm rms}^2 - v_{\rm rms}^2 \over \tau_{\rm s}} - \langle\mathbfit{v}\cdot \nabla\Phi \rangle,
\end{equation}
which is equivalent to eq. (\ref{acc_eq}) introduced in section \ref{sec:MEOM}.

\section{Derivation of equation (15)}
Equation (\ref{eq:HMeq}) in Section \ref{sec:addselfgrav} can be rewritten as
\begin{equation}
\label{eq:hm19}
\left\langle\,\left({{\rm d} \mathbfit{v}\over{\rm d}t}\right)^2\right\rangle - \langle(\nabla\Phi)^2\rangle = {\langle\mathbfit{u}^2\rangle -  \langle\mathbfit{v}^2\rangle \over \tau_{\rm s}^2}.
\end{equation}
We will here show how this relationship arises from the equation of motion of the dust after some algebra.
Scalar multiplication and averaging of equation (\ref{eom_dust2}) with $\nabla\Phi$ yields
\begin{equation}
\label{eq:scalarmult_nphi}
\left\langle\nabla\Phi\cdot{{\rm d}\mathbfit{v}\over{\rm d}t}\right\rangle = {1\over \tau_{\rm s}}\,(\langle\mathbfit{u}\cdot\nabla\Phi\rangle-\langle\mathbfit{v}\cdot\nabla\Phi\rangle) - \langle(\nabla\Phi)^2\rangle,
\end{equation}
where $\langle\mathbfit{u}\cdot\nabla\Phi\rangle = 0$ as shown in Appendix \ref{apx:sssgas}.
Similarly, scalar multiplication and averaging of equation (\ref{eom_dust2}) with $\mathbfit{v}$ yields
\begin{equation}
\label{eq:scalarmult_vvec}
\left\langle\mathbfit{v}\cdot{{\rm d}\mathbfit{v}\over{\rm d}t}\right\rangle = {1\over \tau_{\rm s}}\,[\langle\mathbfit{u}\cdot\mathbfit{v}\rangle -\langle(\mathbfit{v})^2\rangle] - \langle\mathbfit{v}\cdot\nabla\Phi\rangle = 0,
\end{equation}
where we note that the right-hand-side will vanish in case of a statistical steady state. ``Squaring'' and averaging of equation (\ref{eom_dust2}) results in
\begin{equation}
\label{eq:squared}
\left\langle\,\left({{\rm d} \mathbfit{v}\over{\rm d}t}\right)^2\right\rangle + 2\,\left\langle\nabla\Phi\cdot {{\rm d} \mathbfit{v}\over{\rm d}t}\right\rangle + \langle(\nabla\Phi)^2\rangle = 
{\langle\mathbfit{u}^2\rangle - 2\,\langle\mathbfit{v}\,\mathbfit{u}\rangle +  \langle\mathbfit{v}^2\rangle \over \tau_{\rm s}^2}.
\end{equation}
It is now easy to see that combination equations (\ref{eq:scalarmult_nphi}), (is \ref{eq:scalarmult_vvec}) and (is \ref{eq:squared}) results in equation (\ref{eq:hm19}).

\section{Grain-velocity fit to model}
As suggested in Section \ref{sec:genformula}, the mean velocity of grains of a certain size $\alpha$ can be modelled by the following formula
\begin{equation}
 \label{eq:fitting}
 \mathcal{V}(\mathcal{G}_0,b,\alpha_0;\alpha) =
%    {v_{\rm rms}(a)\over u_{\rm rms}} = 
    {\bigg[1 + \mathcal{G}_0\,\left({\alpha\over \alpha_0} \right)^{2b}\bigg] \,\bigg[1-\exp\left\{- \left({\alpha_0\over \alpha}\right)^b\right\}\bigg]},
\end{equation}
where $b$, $\alpha_0$ and $\mathcal{G}$ are free parameters. Using these as fitting parameters we can obtain very accurate least-square fits to our simulation results (see Fig. \ref{fig:alphaplot}). We have used the \texttt{curve\_fit} routine from the \textsc{Python} module \texttt{scipy}, with the $1 \sigma$ deviations from the $v_{\rm rms}$ values as ``errors''. The resultant parameter values are given in Table \ref{tab:vfit}. A good fit to the S3 data can be obtained by simply setting $\mathcal{G}_0=0$ in the model fit to the S3G data. This is indicating that the simple analytic model is sound despite the many assumptions made.

  \begin{table}
  \begin{center}
  \caption{\label{tab:vfit} Resultant parameters from fitting equation (\ref{eq:fitting}) to the $v_{\rm rms}$ values obtained from the simulations. Note that the $u_{\rm rms}$ values are fixed by the simulations and not varied in the fitting process. }
  \begin{tabular}{l|cccc}
  %\hline
  \hline
  \rule[-0.2cm]{0mm}{0.6cm}
       & $u_{\rm rms}/ c_{\rm s}$ & $\mathcal{G}_0$ & $\alpha_0$ & $b$ \\
  \hline\\[-1mm]
   S3G & $1.15$ & $0.03$    & $0.06$ & $0.7$ \\
   S3  & $1.14$ & $0.00$ & $0.06$ & $0.7$ \\[2mm]
  %\hline
  \hline
  \end{tabular}
  \end{center}
  \end{table}

\section{Projected dust-number densities}
\label{apx:coldens}
\begin{figure}
      \resizebox{0.98\hsize}{!}{
	\includegraphics[trim=0.0cm 0cm 2.7cm 0.25cm, clip=true]{Bilder/column_density_log_PVAR81_01_0.096_S3_inferno.jpeg}
	\includegraphics[trim=0.0cm 0cm 0.0cm 0.25cm, clip=true]{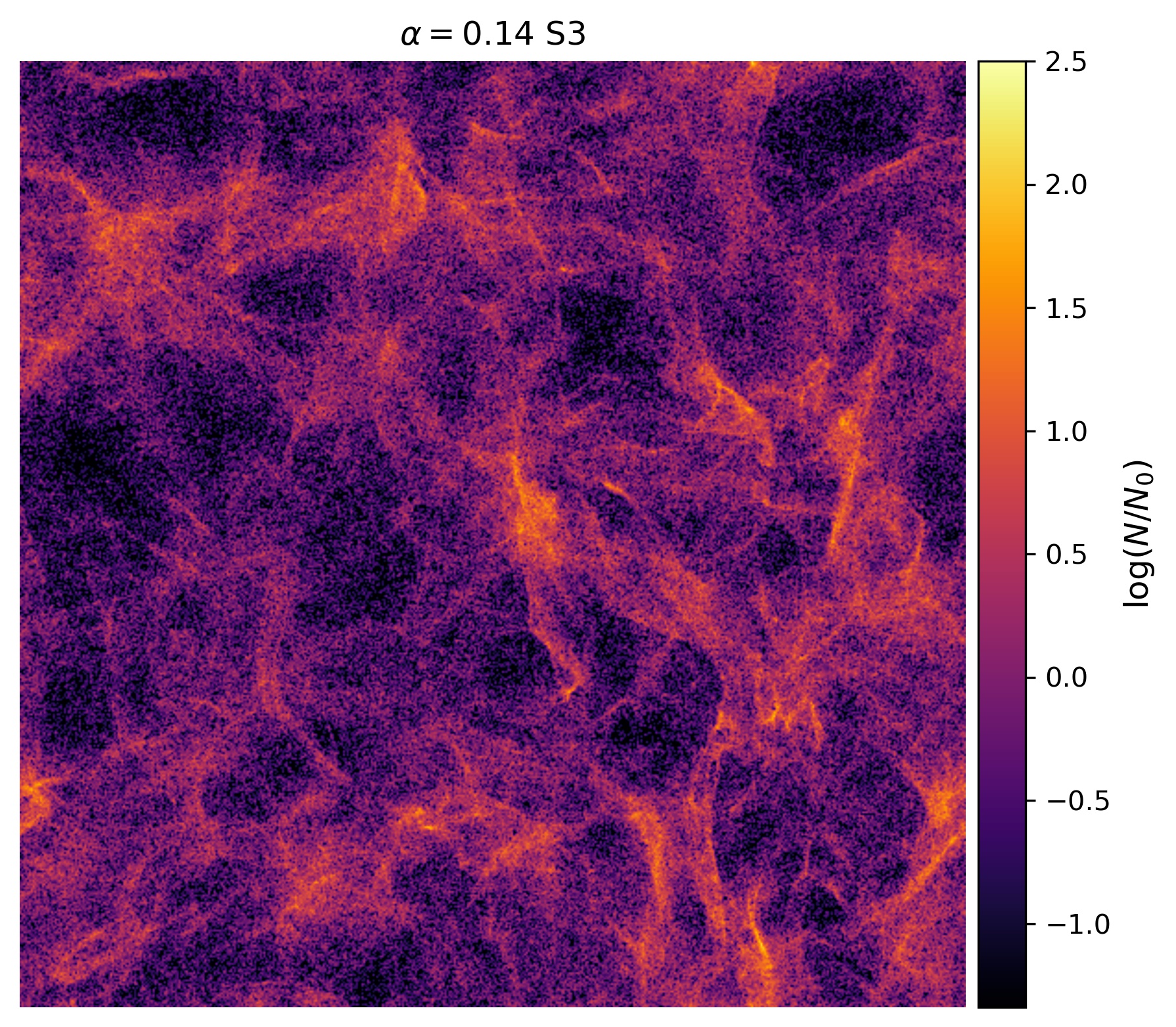}
    }
      \resizebox{0.98\hsize}{!}{
	\includegraphics[trim=0.0cm 0cm 2.7cm 0.25cm, clip=true]{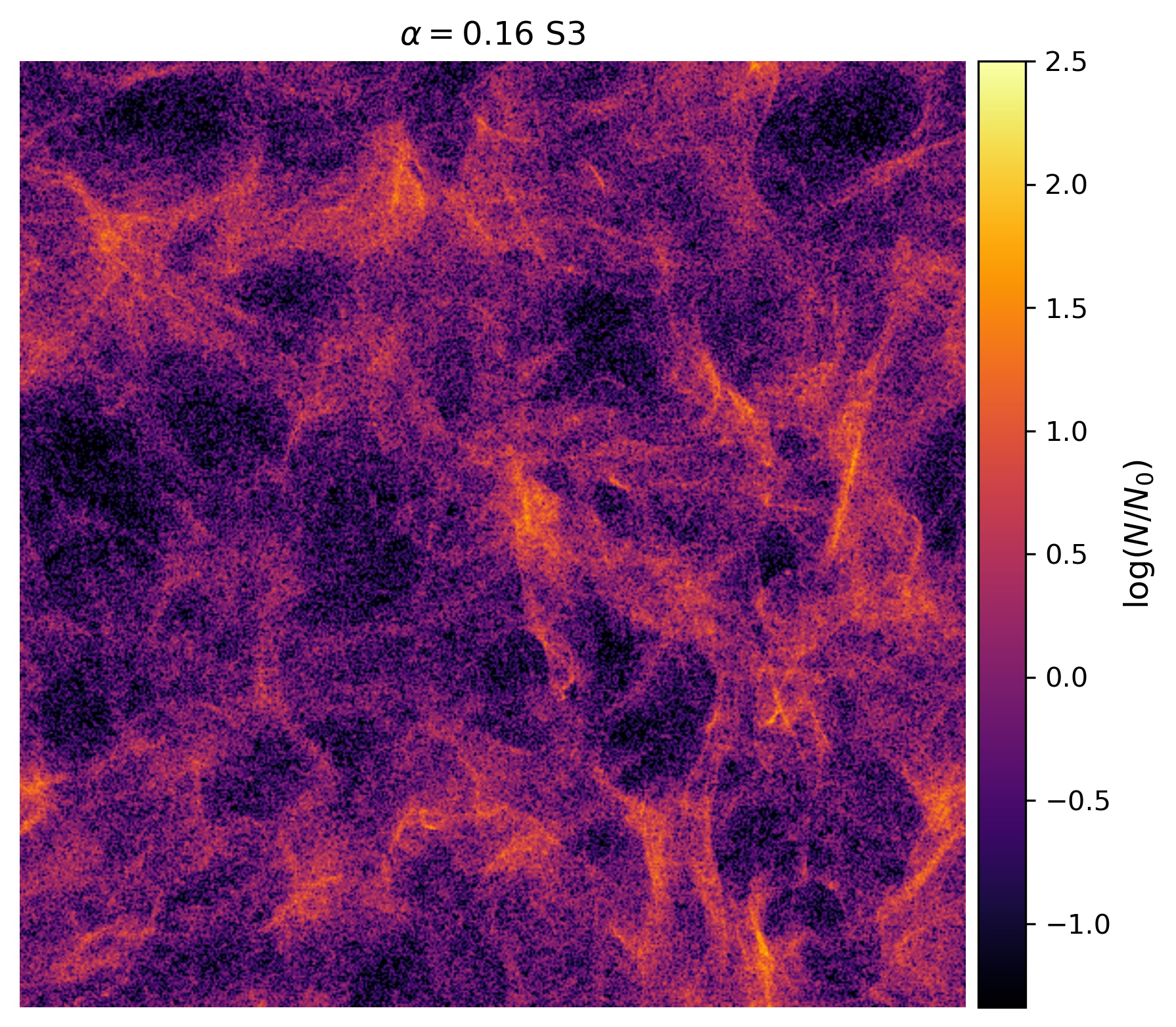}
	\includegraphics[trim=0.0cm 0cm 0.0cm 0.25cm, clip=true]{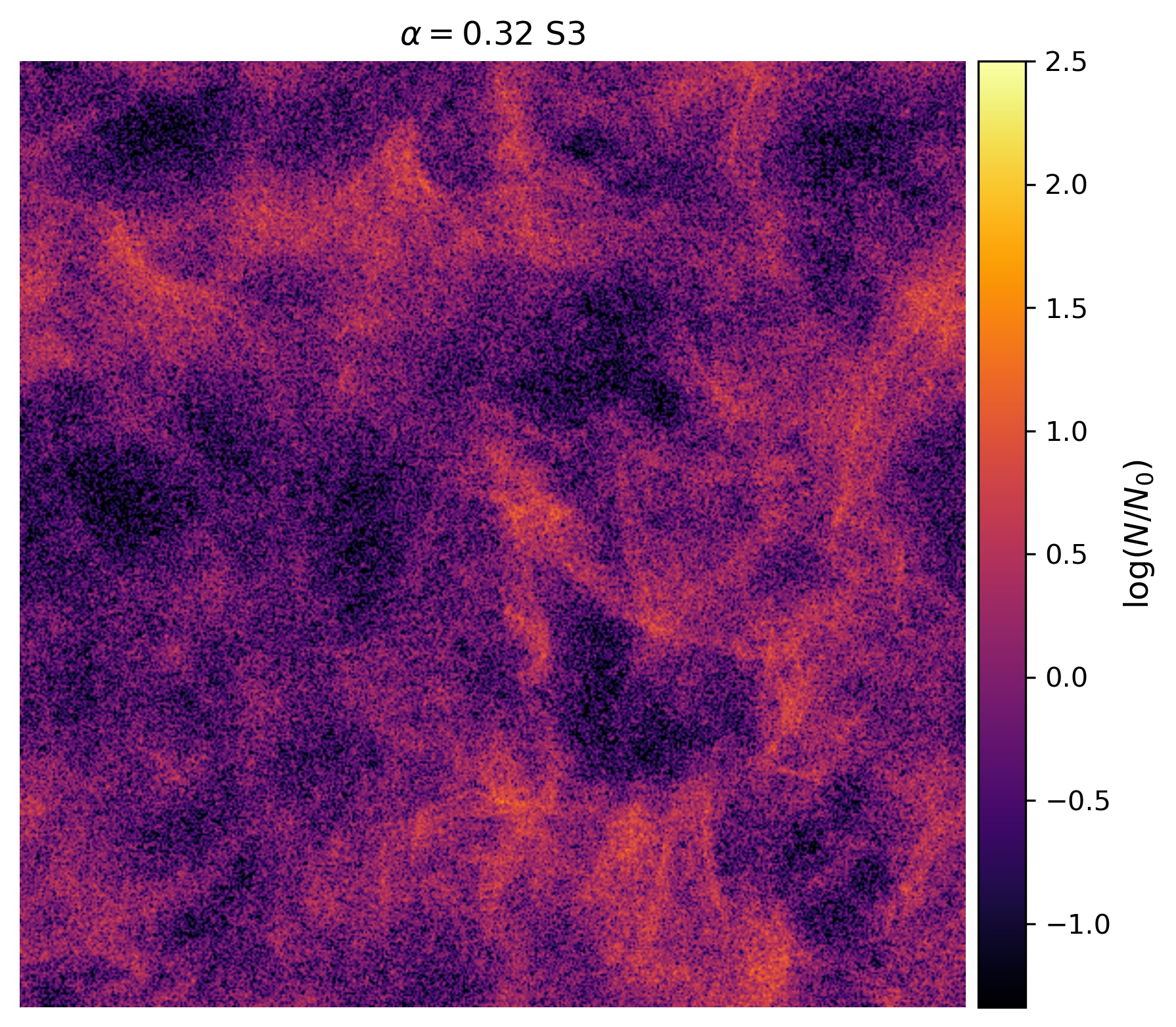}
    }
      \resizebox{0.98\hsize}{!}{
	\includegraphics[trim=0.0cm 0cm 2.7cm 0.25cm, clip=true]{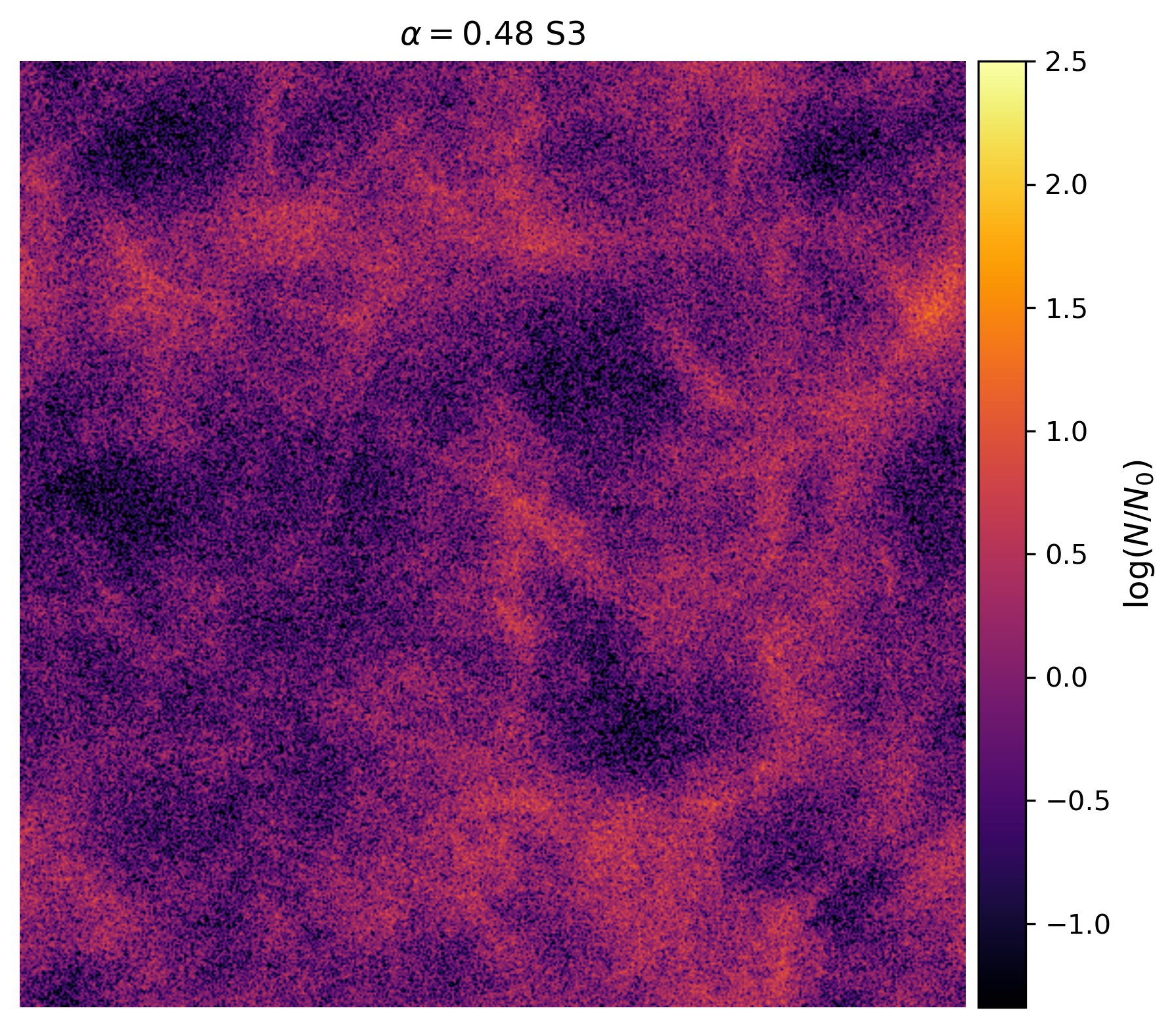}
	\includegraphics[trim=0.0cm 0cm 0.0cm 0.25cm, clip=true]{Bilder/column_density_log_PVAR81_06_0.64_S3_inferno.jpeg}
    }
      \resizebox{0.98\hsize}{!}{
	\includegraphics[trim=0.0cm 0cm 2.7cm 0.25cm, clip=true]{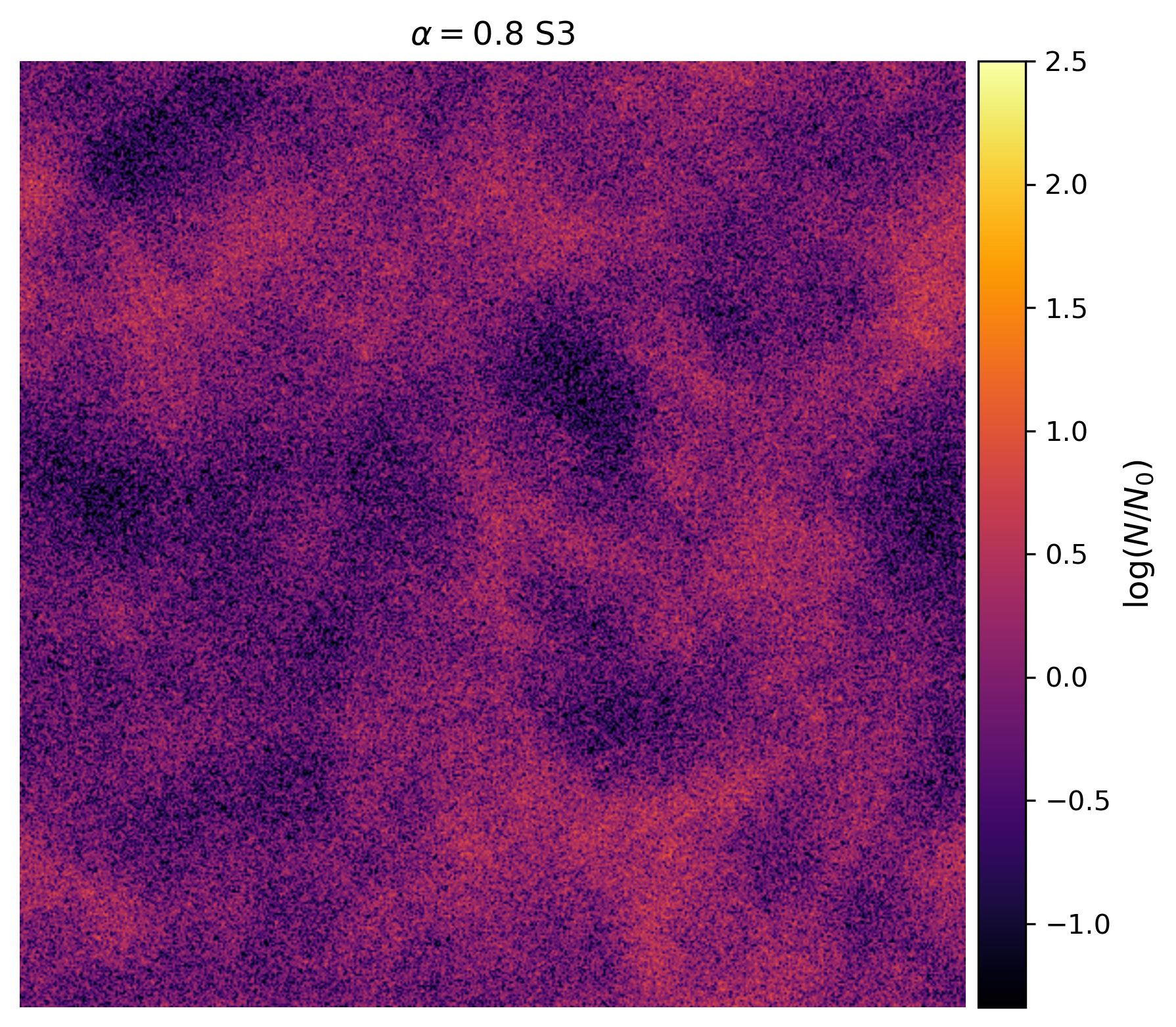}
	\includegraphics[trim=0.0cm 0cm 0.0cm 0.25cm, clip=true]{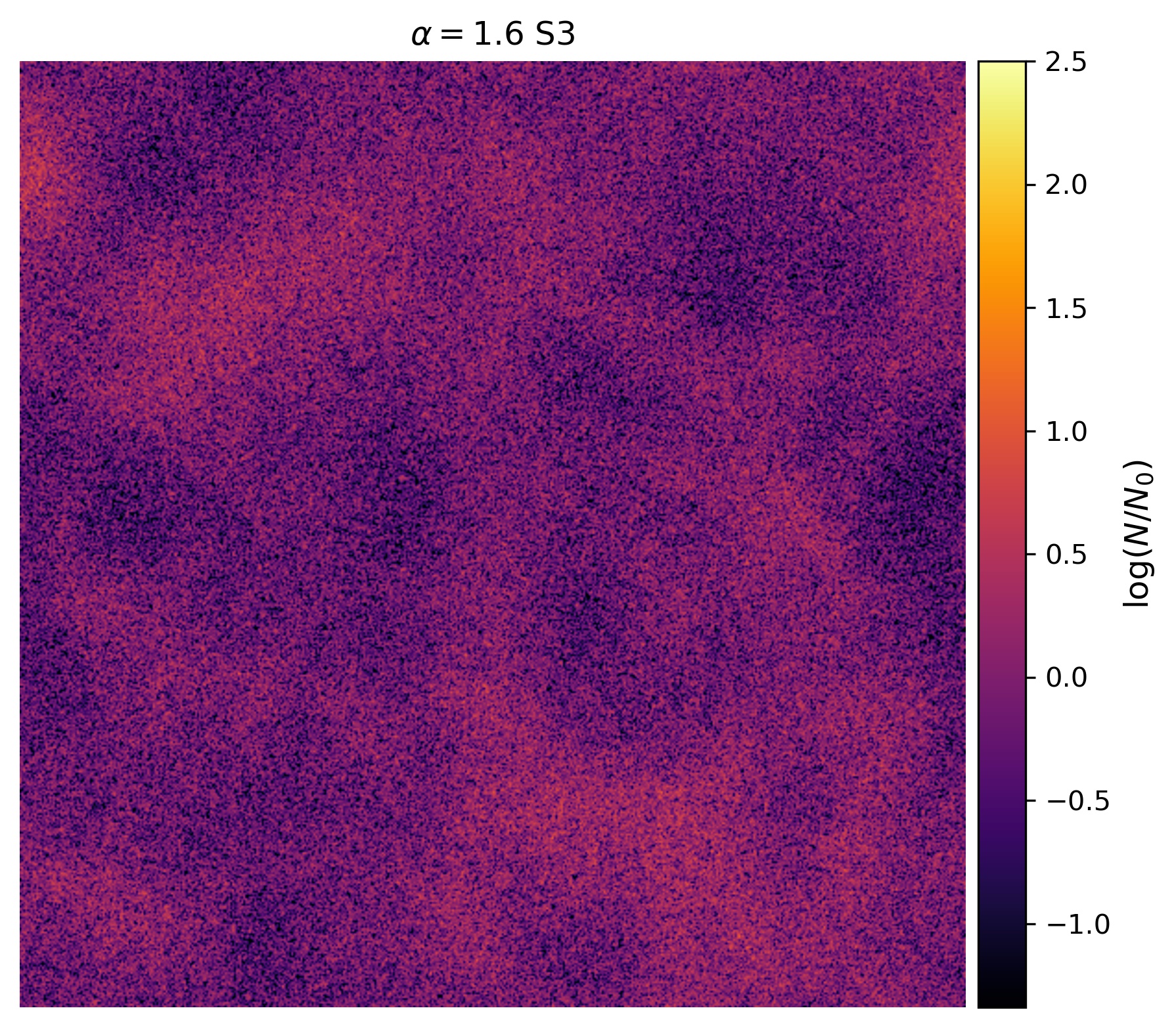}
    }
      \resizebox{0.98\hsize}{!}{
	\includegraphics[trim=0.0cm 0cm 2.7cm 0.25cm, clip=true]{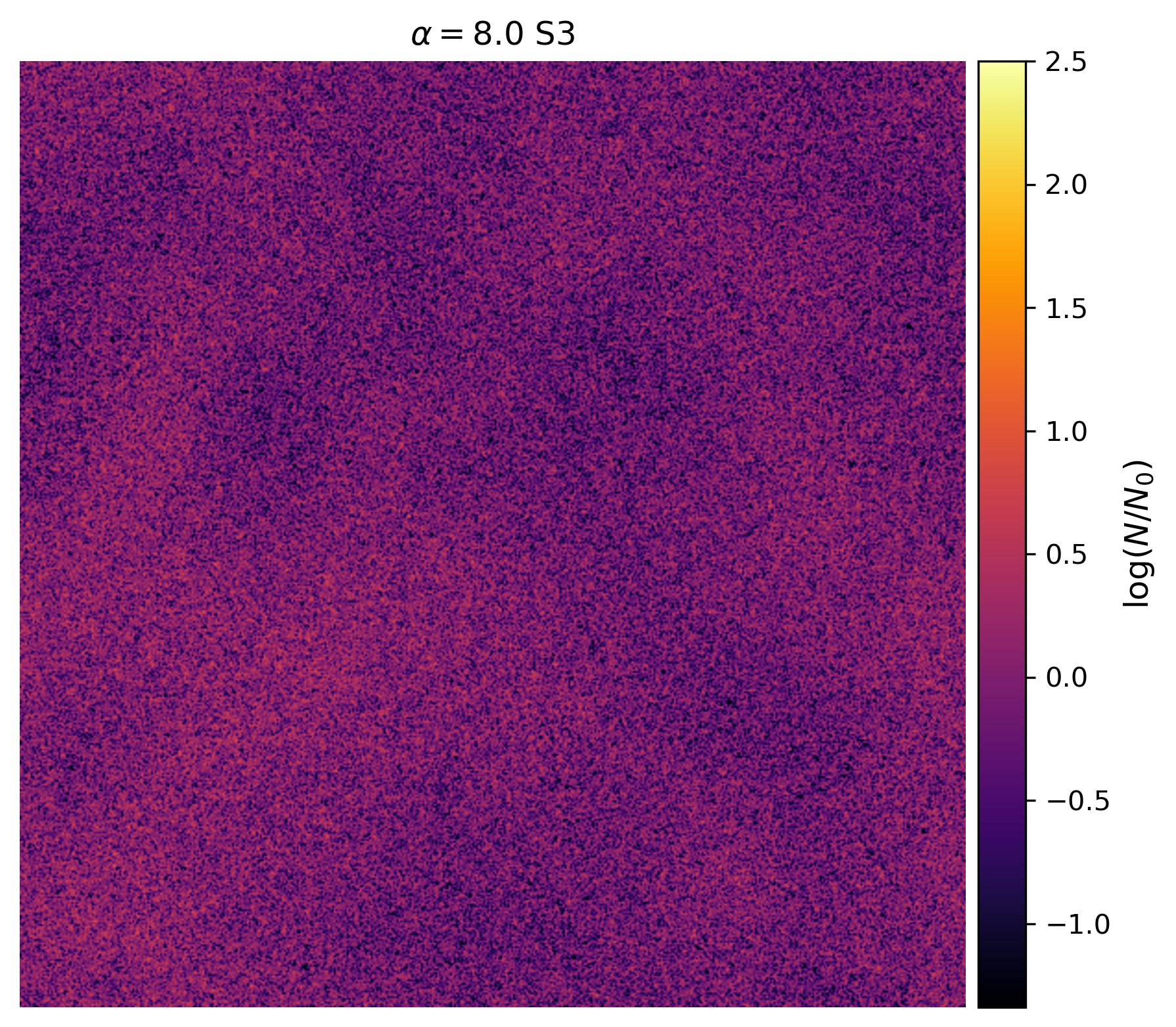}
	\includegraphics[trim=0.0cm 0cm 0.0cm 0.25cm, clip=true]{Bilder/column_density_log_PVAR81_10_16.0_S3_inferno.jpeg}
    }
  \caption{\label{fig:coldens_S3_apx} Projected column (number) density of dust particles for different size-parameters $\alpha$ for simulation S3. Snapshots are taken at the end of the time series ($t/T = 81$).}
  \end{figure}

\begin{figure}
      \resizebox{0.98\hsize}{!}{
	\includegraphics[trim=0.0cm 0cm 2.7cm 0.25cm, clip=true]{Bilder/column_density_log_PVAR81_01_0.096_S3G_inferno.jpeg}
	\includegraphics[trim=0.0cm 0cm 0.0cm 0.25cm, clip=true]{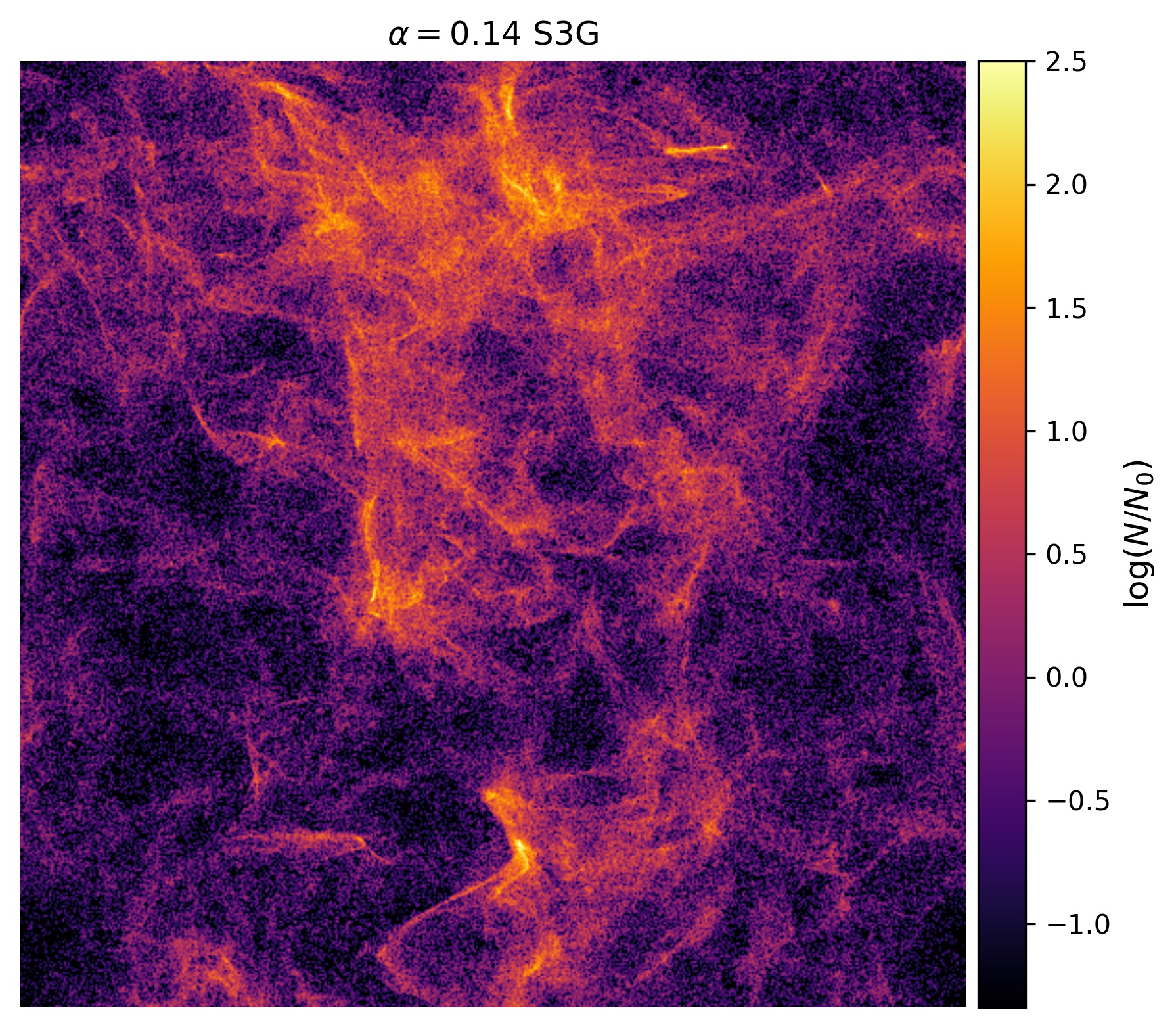}
    }
      \resizebox{0.98\hsize}{!}{
	\includegraphics[trim=0.0cm 0cm 2.7cm 0.25cm, clip=true]{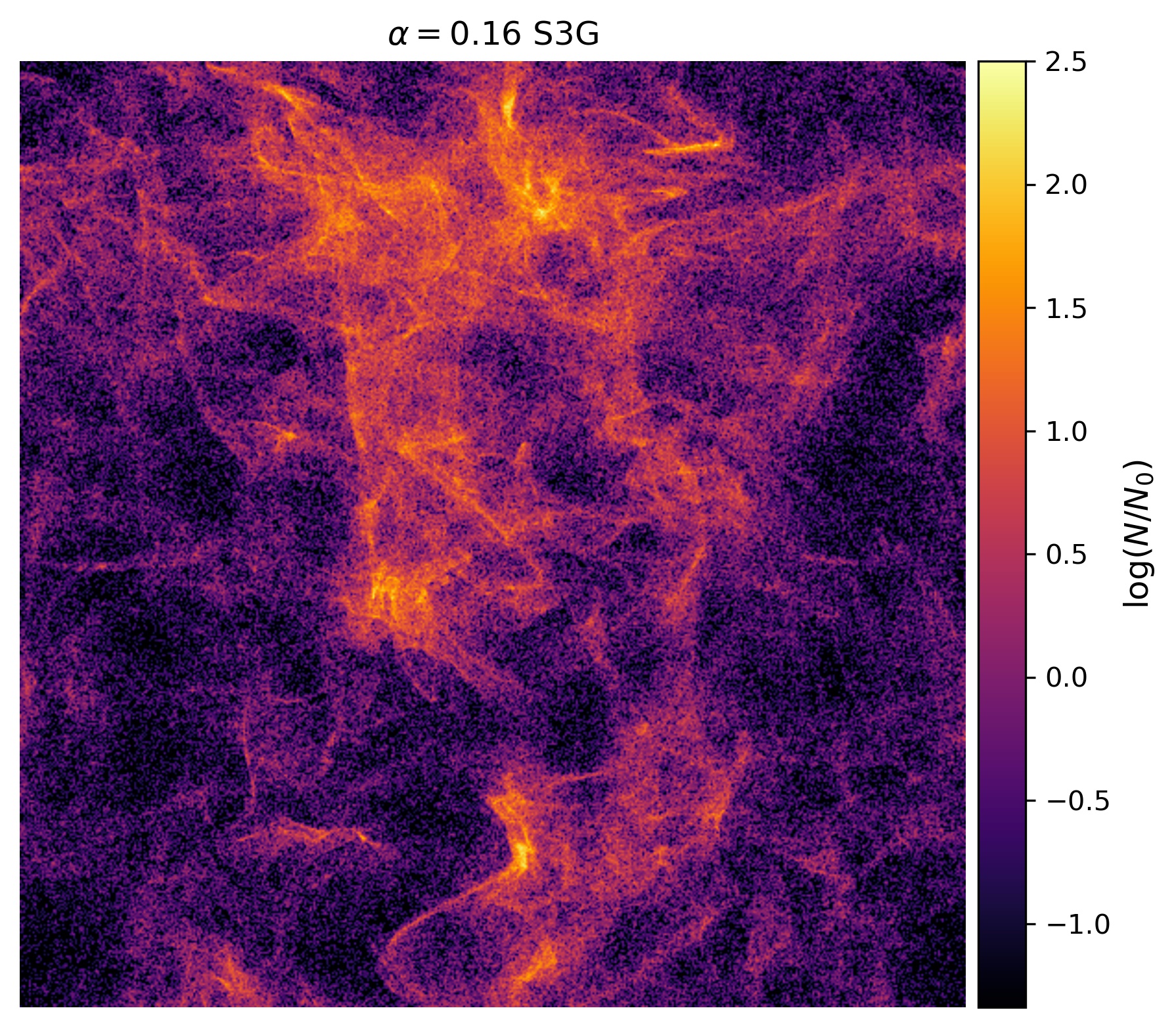}
	\includegraphics[trim=0.0cm 0cm 0.0cm 0.25cm, clip=true]{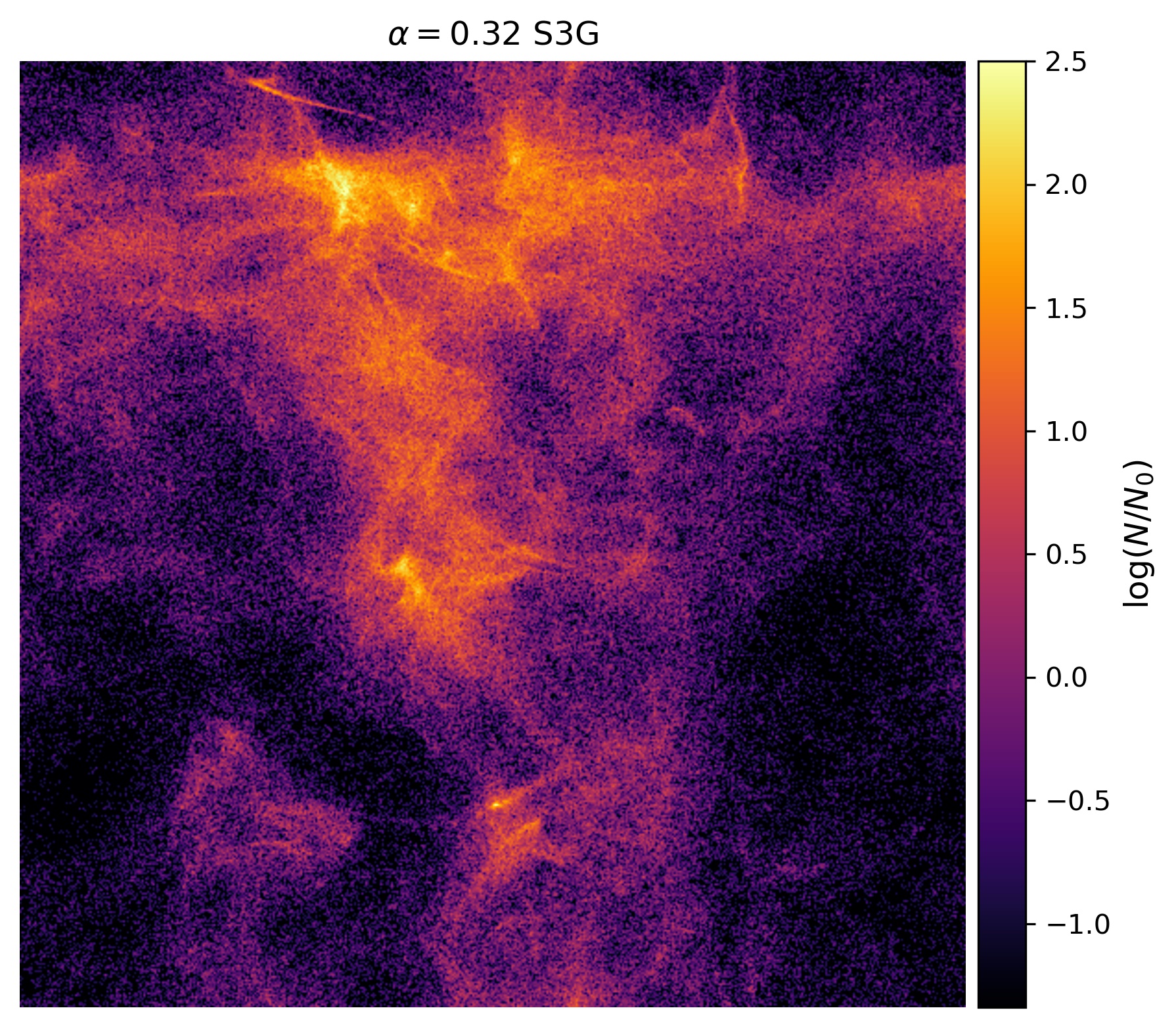}
    }
      \resizebox{0.98\hsize}{!}{
	\includegraphics[trim=0.0cm 0cm 2.7cm 0.25cm, clip=true]{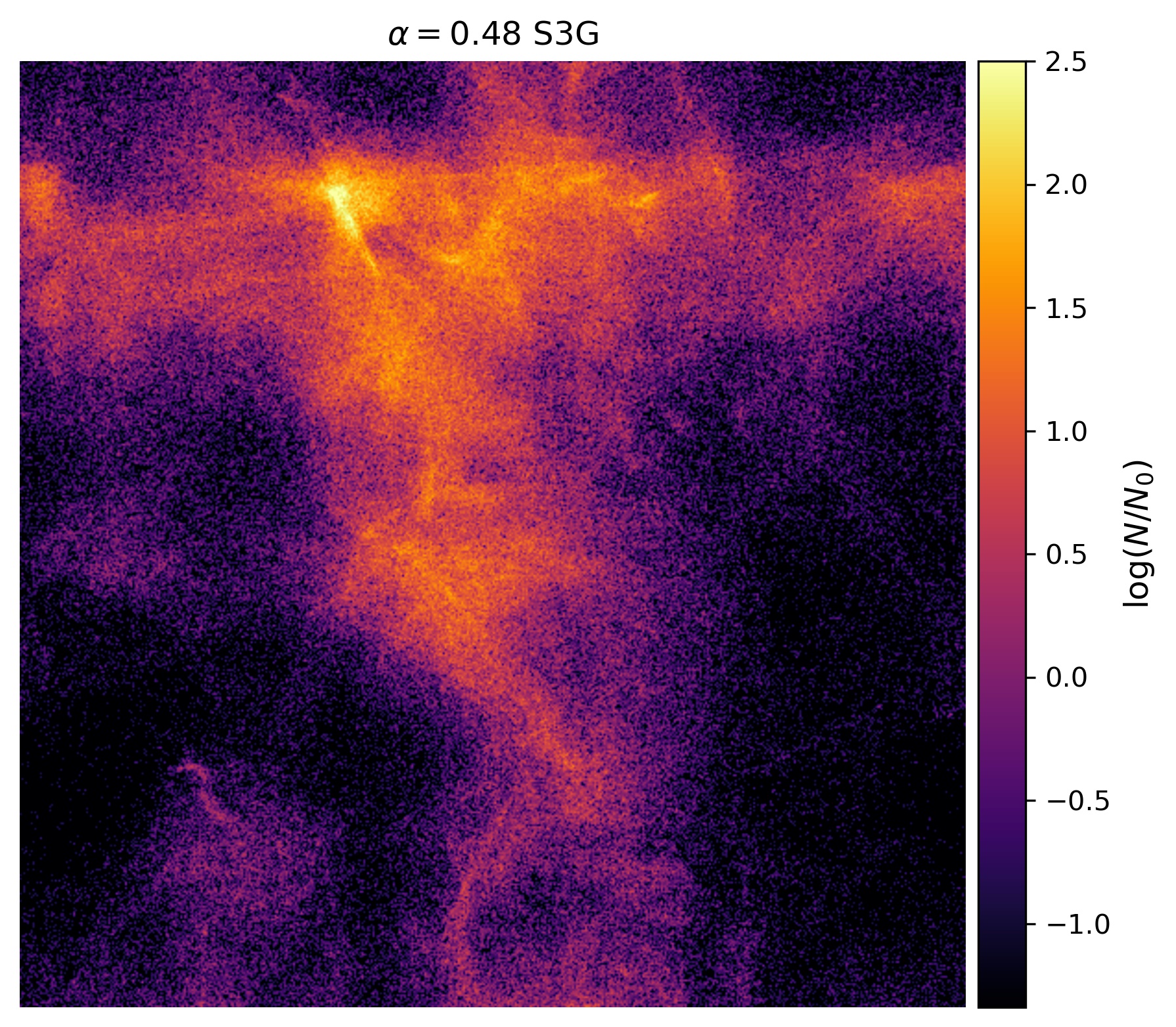}
	\includegraphics[trim=0.0cm 0cm 0.0cm 0.25cm, clip=true]{Bilder/column_density_log_PVAR81_06_0.64_S3G_inferno.jpeg}
    }
      \resizebox{0.98\hsize}{!}{
	\includegraphics[trim=0.0cm 0cm 2.7cm 0.25cm, clip=true]{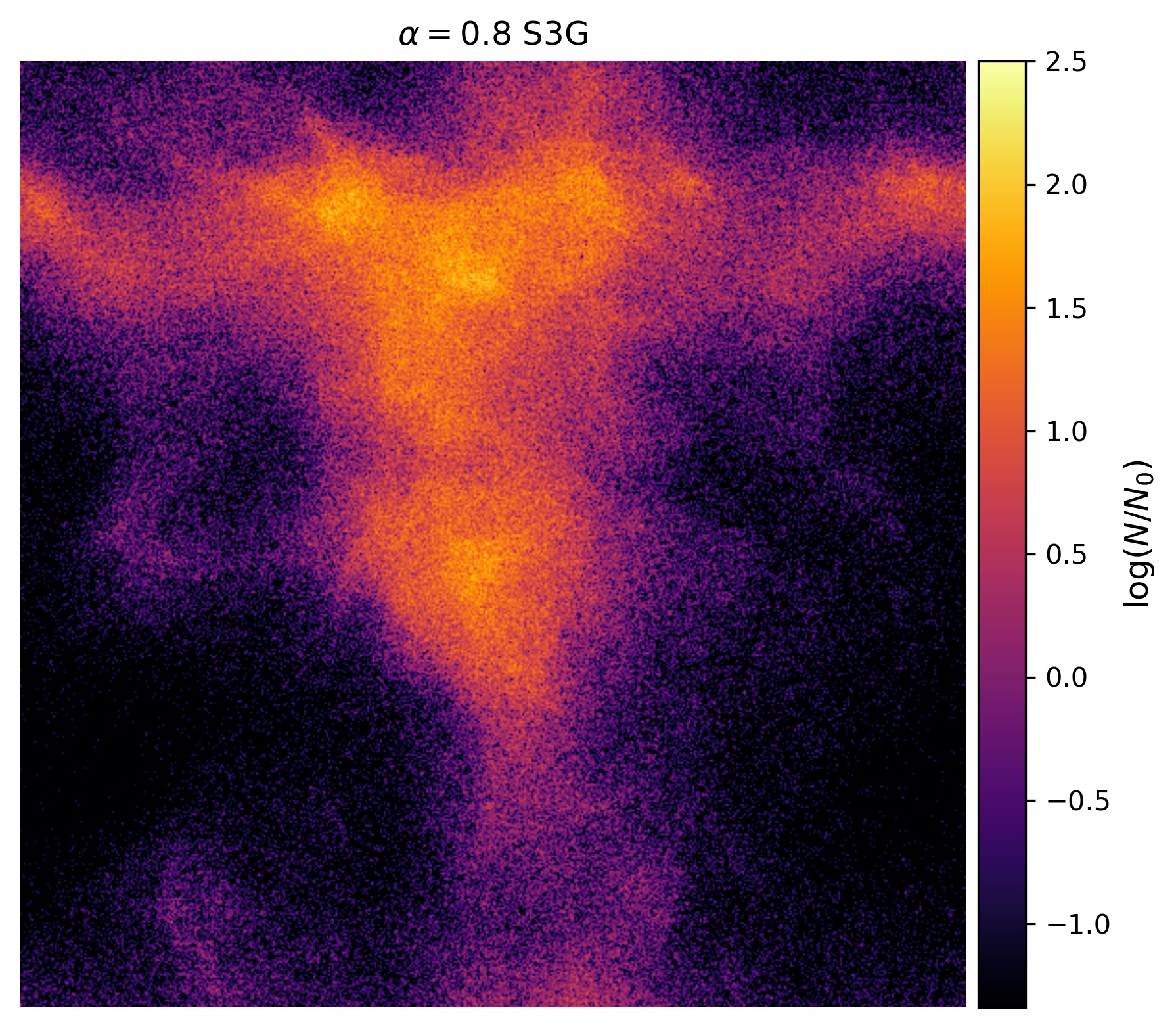}
	\includegraphics[trim=0.0cm 0cm 0.0cm 0.25cm, clip=true]{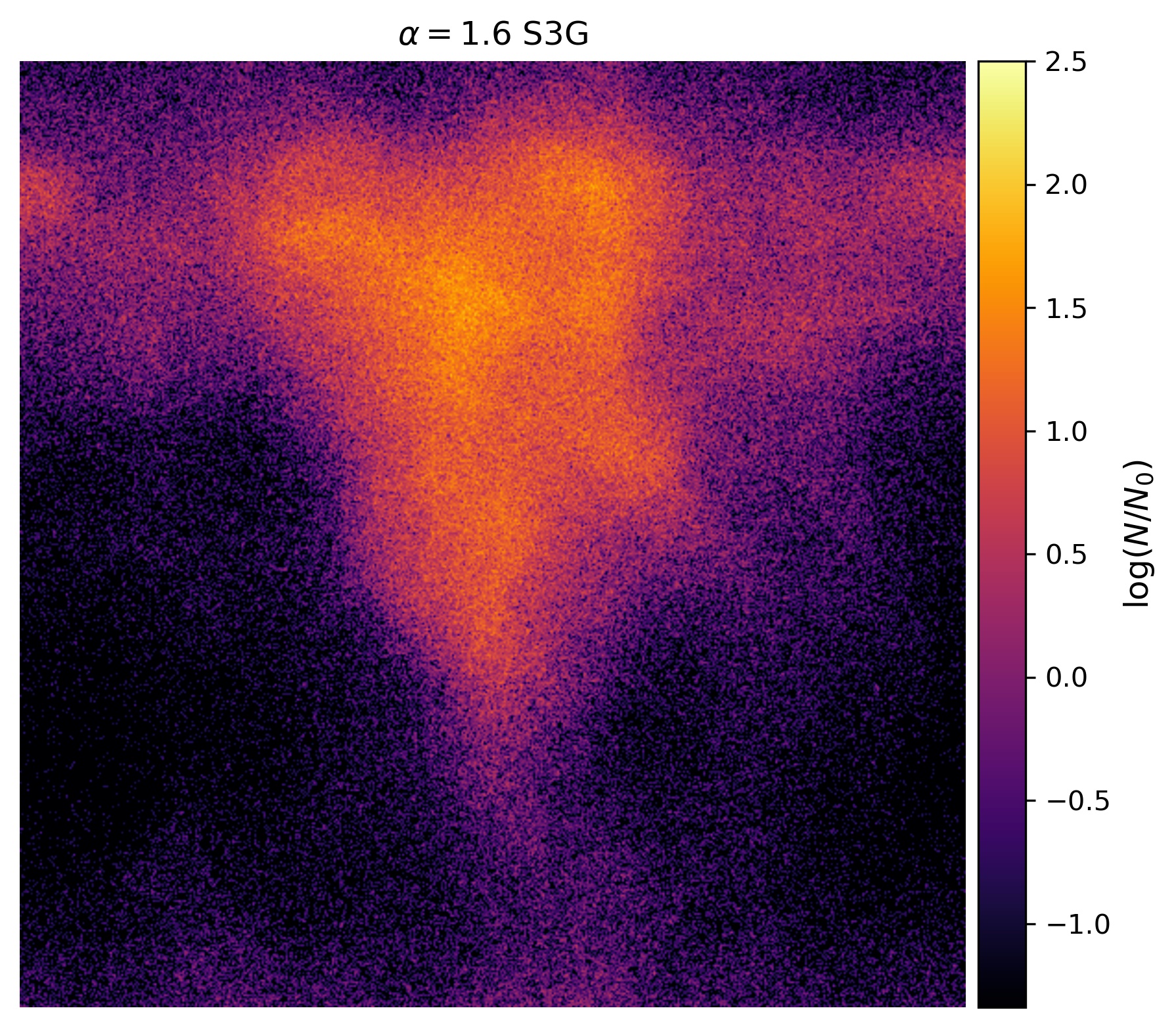}
    }
      \resizebox{0.98\hsize}{!}{
	\includegraphics[trim=0.0cm 0cm 2.7cm 0.25cm, clip=true]{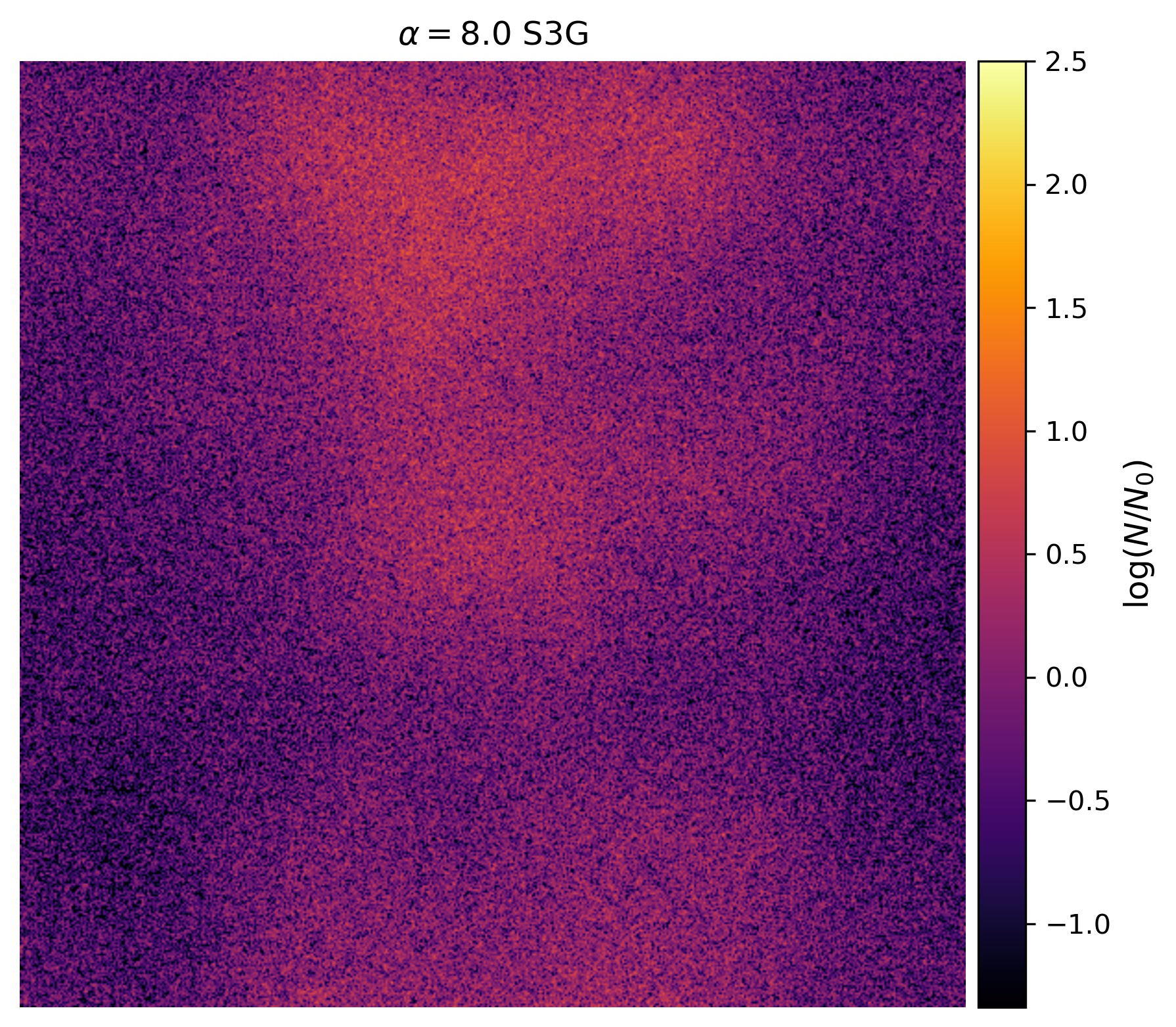}
	\includegraphics[trim=0.0cm 0cm 0.0cm 0.25cm, clip=true]{Bilder/column_density_log_PVAR81_10_16.0_S3G_inferno.jpeg}
    }
  \caption{\label{fig:coldens_S3G_apx} Same as Fig. \ref{fig:coldens_S3_apx} but for simulation S3G, including self-gravity of the gas.}
  \end{figure}

In Section \ref{results} we only present projected number densities for three grain sizes (see Fig. \ref{fig:coldens}). For completeness, we supply in this appendix projections for all grain sizes. Fig. \ref{fig:coldens_S3_apx} shows the line-of-sight projections for simulation S3, while Fig. \ref{fig:coldens_S3G_apx} shows the same for S3G.

\onecolumn
\newpage
\twocolumn

\label{lastpage}

\end{document}